\documentclass[twocolumn,secnumarabic,amssymb, nobibnotes, aps,prx,showpacs]{revtex4}
\usepackage{amsmath,amsthm,amssymb}
\usepackage[dvips]{graphicx,color}
\usepackage{color}
\usepackage{comment}

\begin{document}

\title{Ultraslow diffusion in language: Dynamics of appearance of already popular adjectives on Japanese blogs} 
\author{Hayafumi Watanabe$^{1,2,3}$}\email[E-mail: ]{hayafumi.watanabe@gmail.com}
\affiliation{$^1$Risk Analysis Research Center, The Institute of Statistical Mathematics, 10-3 Midori-cho, Tachikawa, Tokyo 190-8562, Japan}
\affiliation{$^2$Hottolink,Inc., 6 Yonbancho Chiyoda-ku, Tokyo 102-0081, Japan}
\affiliation{$^3$Joint Support-Center for Data Science Research, The Research Organization of Information and Systems, 10-3 Midori-cho, Tachikawa, Tokyo 190-8562, Japan}
\begin{abstract}
 What dynamics govern a time series representing the appearance of words in social media data? In this paper,  
 we investigate an elementary dynamics, from which word-dependent special effects are segregated, such as breaking news, increasing (or decreasing) concerns, or seasonality.
 To elucidate this problem, we investigated approximately \textcolor{black}{three billion Japanese blog articles} over a period of six years, and analysed some corresponding solvable mathematical models. From the analysis, we found that a word appearance can be explained by the random diffusion model based on the power-law forgetting process, which is a type of long memory point process related to ARFIMA(0,0.5,0). 
In particular, we confirmed that ultraslow diffusion (where the mean squared displacement grows logarithmically), which the model predicts, reproduces the actual data in an approximate manner.
  In addition, we also show that the model can reproduce other statistical properties of a time series: (i) the fluctuation scaling, (ii) spectrum density, and (iii) shapes of the probability density functions. 
\end{abstract}
\maketitle
\section{Introduction}
Languages are constantly changing. 
In linguistics, language changes have been observed and analysed for many years.
For example, in a large time scale (i.e., on the order of 100 years), a historical relationship of languages, called “language trees” was found \cite{hock2009language}.
Moreover, in a short (i.e., monthly or daily) or middle (i.e., yearly) time scale, many studies investigating how newly emerging words spread both quantitatively and qualitatively have been conducted \cite{wurschinger2016using}. 
 In contrast, the changes in usage of many already popularized words (i.e, not new words), which seem to be ``mostly unchanged'' in a daily time scale, have not been studied quantitatively in spite of the core words of languages and one of most basic dynamic state of words.  
Although there have previously been difficulties in distinguishing from a stationary time series through precision limits, these words should be changing gradually from day to day.
In this study, we investigate the dynamics of these ``mostly unchanged'' words precisely by applying two concepts developed in statistical physics: “anomalous diffusion” and “fluctuation scaling” to large-scale nation-wide blog data for an improvement in accuracy. It can also be said that the purpose of our study is to clarify the elementary process of the time variation of a word appearance in a``normal'' state in which special effects such as breaking news, and increasing (or decreasing) concerns or recognitions, are segregated. \par
A time series representing the appearance of considered keywords, that is, a sequence of daily counts of the appearance of a considered word within a large social media dataset, is used in our investigation. 
This quantity is mostly used to measure temporal changes in social concerns related to the considered word in both practical applications (such as marketing, television shows, politics, and finance) and basic sciences (such as sociology, physics, psychology, and information science). \cite{preis2012quantifying, ugander2011anatomy, ceron2014every, ginsberg2009detecting, sakaki2010earthquake, grajales2014social, yu2012survey}.  
Therefore, it is expected that information regarding the basic dynamics of time series of keywords, \textcolor{black}{which we investigate}, will help us to precisely observe human behaviours using social media data, and in particular, will be important to extracting essential information from noisy time series data in a practical manner.
\par
In statistical mechanics or complex systems science, a diffusion analysis based on the mean squared displacement(MSD)is a commonly used technique to characterize the dynamics of a non-stationary time series.
The (time average) MSD, which is the average squared displacement of a time series $\{X(t)\}$ as a function of the lag time $L$,  is defined as 
\begin{equation}
\Pi(L; \{X\}) =E[(X(t+L)-X(t))^2]=\sum^{T-L}_{t=1}\frac{(X(t+L)-X(t))^2}{T-L}, 
\end{equation}
where $E[A]$ is the temporal mean of a time series $\{A(t)\}$.
%
\par
In many empirical observations, the power law MSD,  
\begin{equation}
\Pi(L;\{X\}) \propto L^{\alpha} \quad (\alpha>0), 
\end{equation}
is observed \cite{bouchaud1990anomalous,metzler2000random,da2014ultraslow}. The diffusion with the power law MSD is classified using a scaling exponent $\alpha$, and this value provides insight into the dynamics. 
For $\alpha=1$, the diffusion corresponds to a normal diffusion, such as particles in water, which is modelled using a random walk, $X(t+1)=X(t)+\eta(t)$. 
Under this situation, $\eta(t)$ is independent, identically distributed, and finite variant noise.  
In contrast, the diffusion for $\alpha \neq 1$ is called ``anomalous diffusion''.  
 \textcolor{black}{Anomalous diffusion has been known since 1926, in which anomalous diffusion of turbulence was discovered \cite{metzler2000random}.}  Nowadays, many systems have been shown to exhibit anomalous diffusion in diverse areas, such as physics, chemistry, geophysics, biology, and economy \cite{metzler2000random, da2014ultraslow}.
\textcolor{black}{Anomalous diffusion is explained through a correlation of random noise (e.g., random walk in disordered media) \cite{bouchaud1990anomalous}, finite-variance (e.g., a Levy flight) \cite{bouchaud1990anomalous, metzler2000random}, the power-law wait time (e.g., continuous random walk) \cite{bouchaud1990anomalous, metzler2000random}, and long memory (e.g., fractional random walk) \cite{lowen2005fractal}.}
\par
It has also been known that, with logarithmic type diffusion,  
\begin{equation}
\Pi(L;\{X\})=\log(L)^\alpha.
\end{equation}
This type of diffusion is called ``ultraslow diffusion'', which has been mainly studied theoretically. One of best known examples is the diffusion in a disordered medium (in the case of $\alpha=4$, it is called Sinai diffusion \cite{sinai1983limiting}). 
Empirically, it has also been reported that the mobility of humans and monkeys obeys an ultra-slow diffusion-like behaviour \cite{song2010modelling,boyer2011non}． \par 
\textcolor{black}{The other concept that we employed to analyse the data precisely is ``fluctuation scaling'' (in \cite{RD_base}, and we intensively studied the fluctuation scaling of Japanese blogs for a daily time scale.)}
\textcolor{black}{Fluctuation scaling (FS), which is also known as ``Taylor's law'' \cite{taylor1961aggregation} in ecology, is a power law relation between the system size (e.g., mean) and the magnitude of fluctuation (e.g., standard deviation). }
FS has been observed in various complex systems, such as random work in a complex network \cite{PhysRevLett.100.208701}, Internet traffic \cite{argollo2004separating}, river flows \cite{argollo2004separating}, animal populations \cite{xu2015taylor}, insect numbers \cite{xu2015taylor, eisler2008fluctuation}, cell numbers \cite{eisler2008fluctuation}, foreign exchange markets \cite{sato2010fluctuation}, the numbers of Facebook application downloads \cite{onnela2010spontaneous}, word counts of Wikipedia \cite{gerlach2014scaling}, academic papers \cite{gerlach2014scaling}, old books \cite{gerlach2014scaling}, crime \cite{10.1371/journal.pone.0109004}, and Japanese blogs \cite{sano2010macroscopic}. \par 
\par
 A certain type of FS can be explained through the random diffusion (RD) model \cite{PhysRevLett.100.208701}.
The RD model, which has been introduced as a mean field approximation for a random walk on a complex network, is described by a Poisson process 
with a random variable Poisson parameter.
It can be demonstrated that the fluctuation of the RD model obeys the FS with an exponent of $0.5$ for a small system size (i.e., a small mean), or $1.0$ for a large system size (i.e., a large mean). 
Because this model is based only on a Poisson process, it is not only applicable to random walks on complex networks, but also to a wide variety of phenomena related to random processes. For instance, this model can reproduce a type of FS regarding the appearance of words in Japanese blogs \cite{sano2009, PhysRevE.87.012805}. 
 \par
 Note that physicists have studied the linguistic phenomena using concepts of complex systems \cite{link1}, such as competitive dynamics \cite{abrams2003linguistics}, statistical laws \cite{altmann2015statistical}, and complex networks \cite{cong2014approaching}. Our study can also be positioned within this context, that is, we study the properties of the time series of word counts in nationwide blogs (a linguistic phenomenon) using a diffusion analysis and the FS, which are concepts of complex science or statistical physics.
\par
In this study, we tried to clarify the elementary process of the time variation of a word appearance that is not affected by a special effect, such as breaking news, an increase (or decrease) of concern (recognition), or seasonality.
First, we investigate the FSs of the word appearance time series for various time-scales using 5 billion Japanese blog articles from 2007 in order to obtain a clue of their dynamics.
Second, we introduce the random diffusion model based on a long memory stocastic process, and the model can reproduce the empirical FSs.
Third, we show that the model can also reproduce other statistical properties of word count time series data: (i) mean squared displacement, (ii)spectrum density, and (iii) shapes of the probability density functions. 
In this part, we also show that the empirical data have the properties of ultraslow diffusion.
Finally, we conclude with a discussion. \par
\begin{figure*}
\begin{minipage}{0.48\hsize}
\centering
\includegraphics[width=7cm,height=4cm]{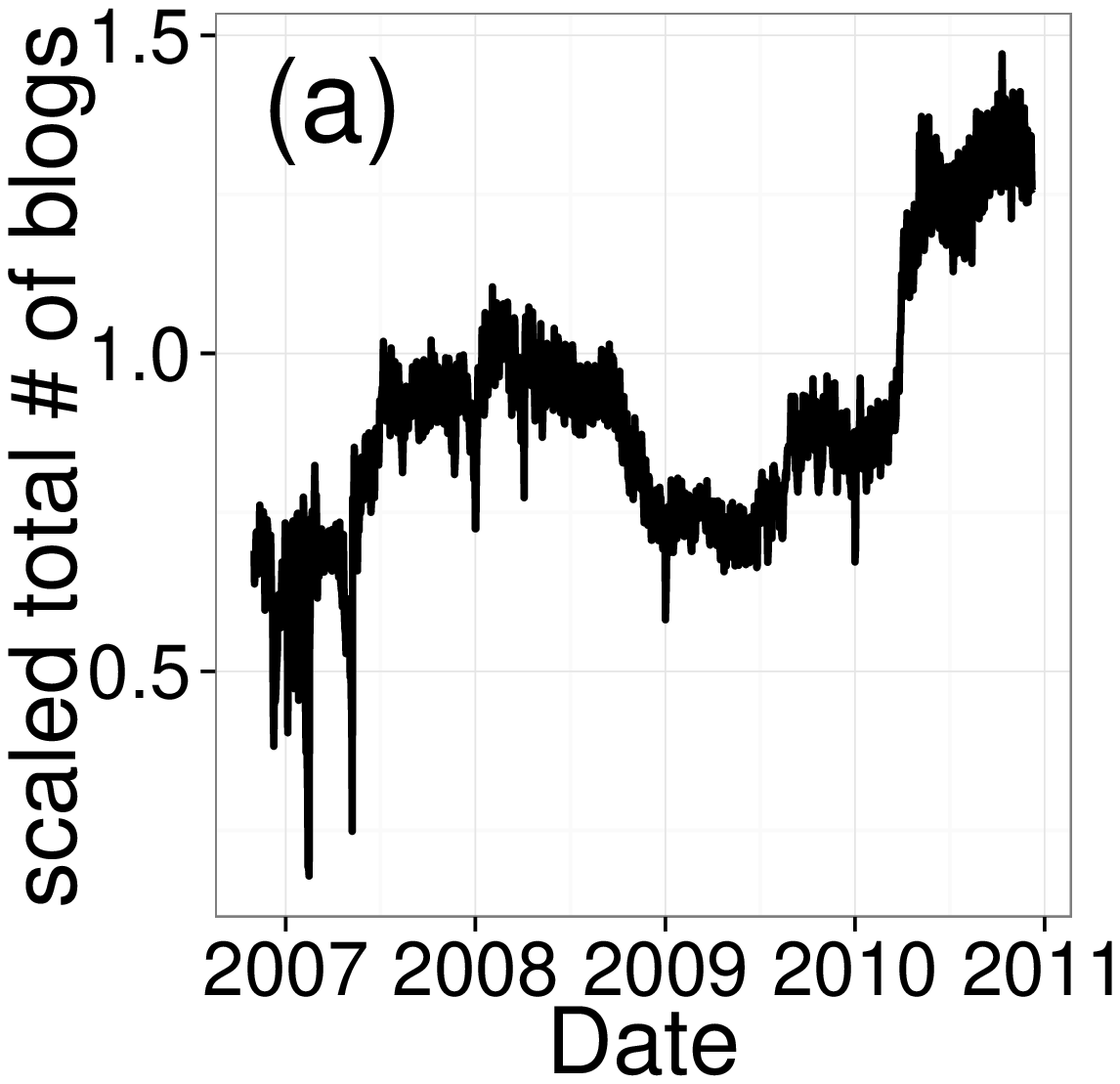}
\end{minipage}
\begin{minipage}{0.48\hsize}
\centering
\includegraphics[width=7cm,height=4cm]{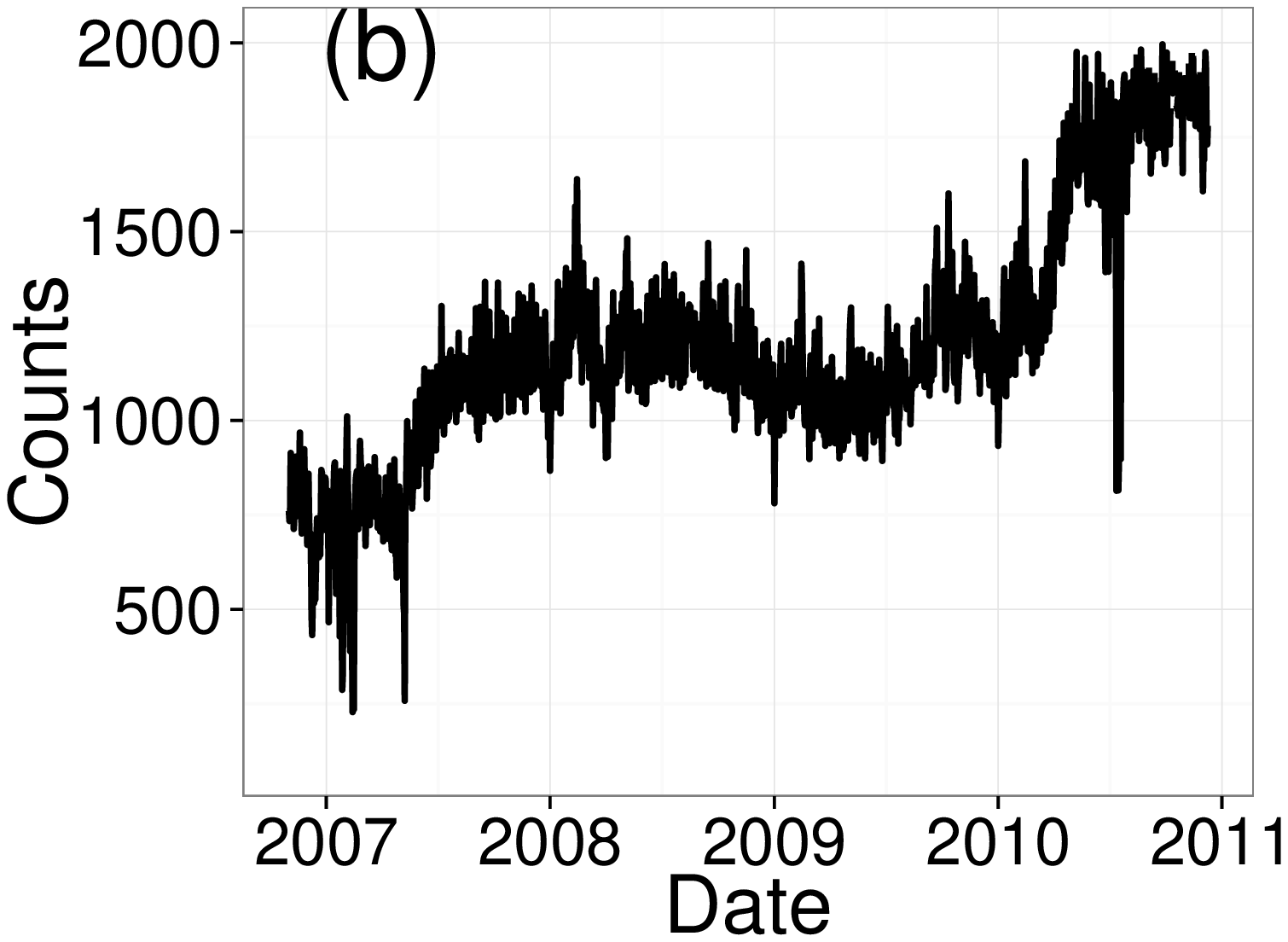}
\end{minipage}
\begin{minipage}{0.48\hsize}
\centering
\includegraphics[width=7cm,height=4cm]{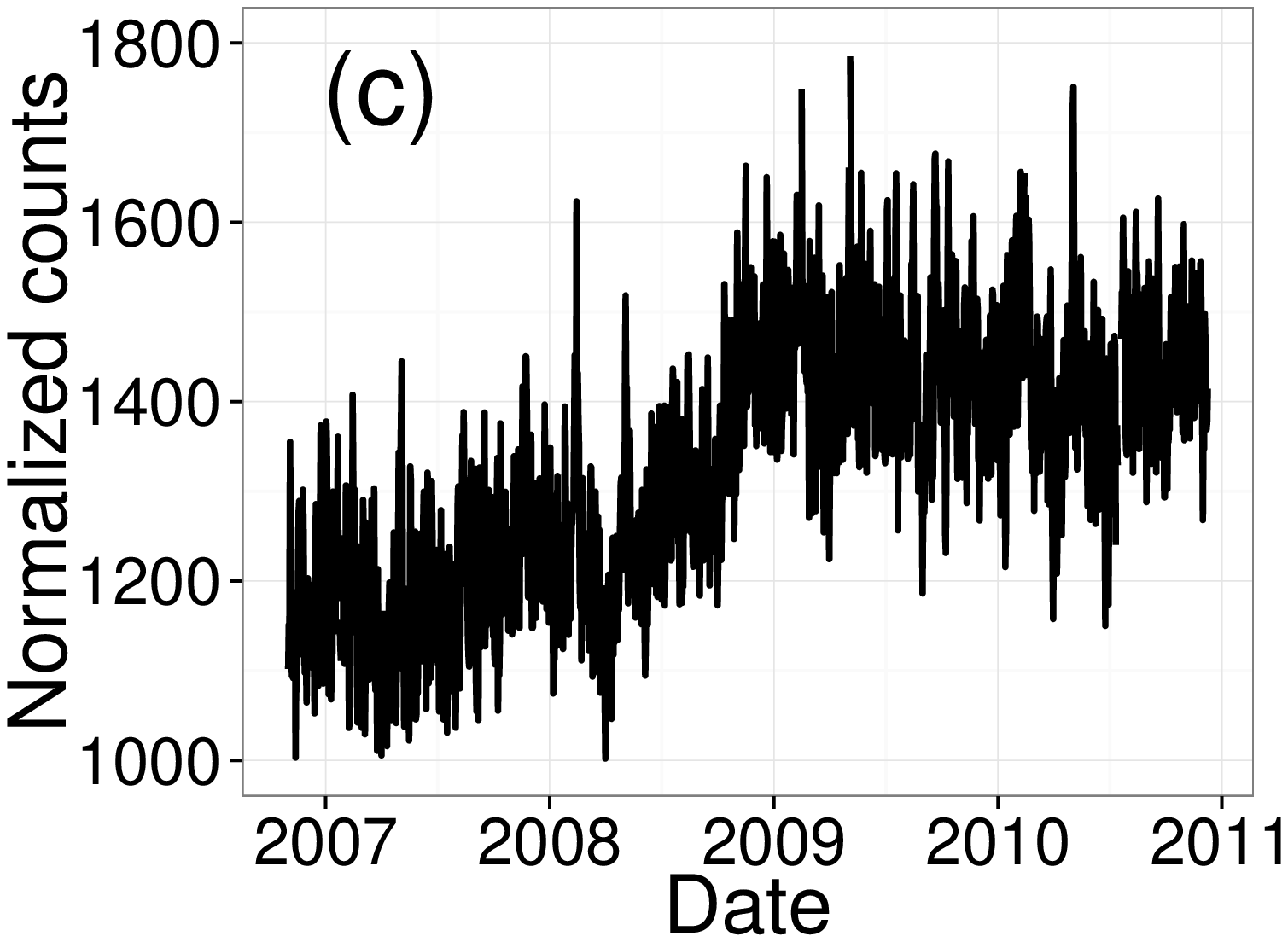}
\end{minipage}
\begin{minipage}{0.48\hsize}
\centering
\includegraphics[width=7cm,height=4cm]{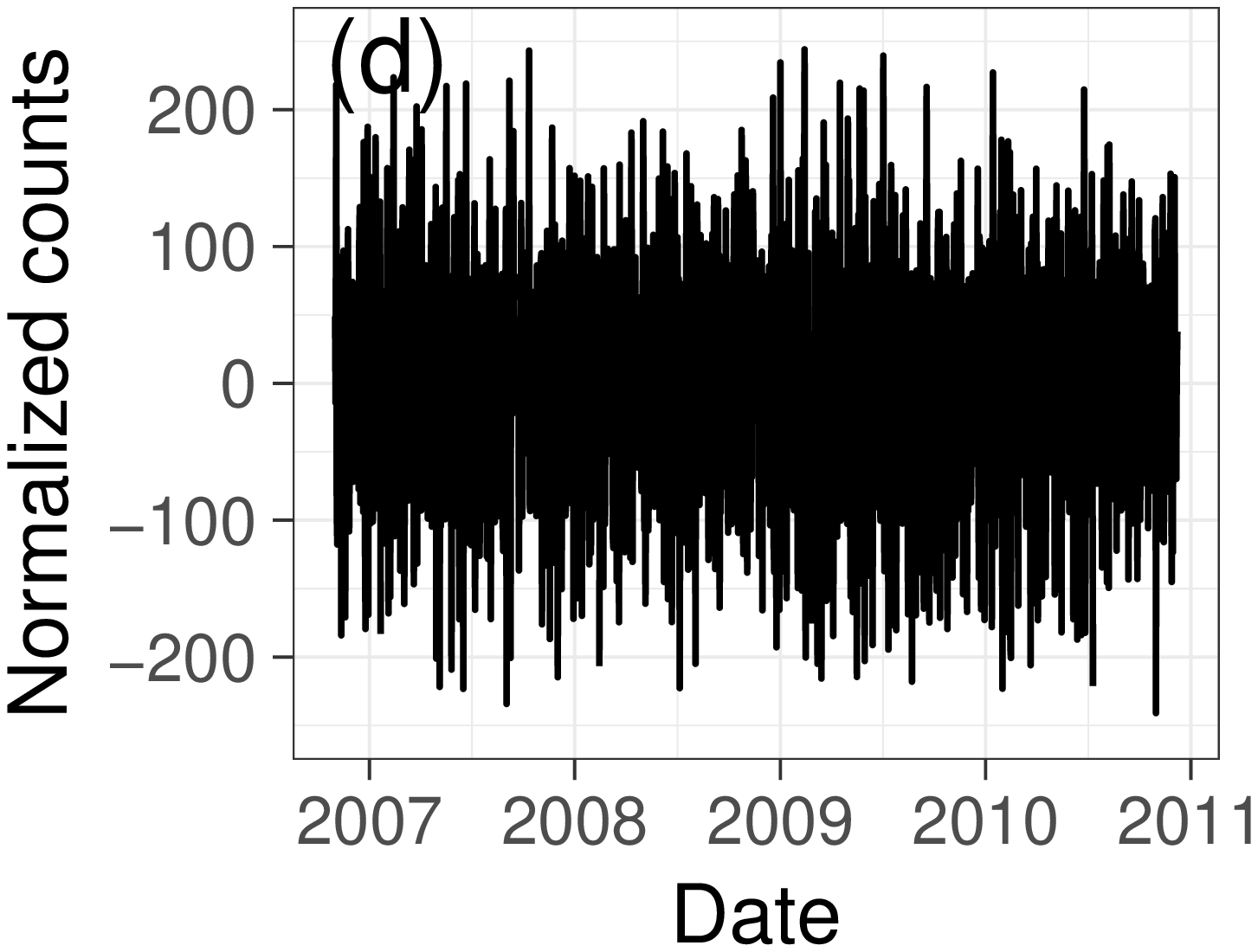}
\end{minipage}
\begin{minipage}{0.48\hsize}
\centering
\includegraphics[width=7cm,height=4cm]{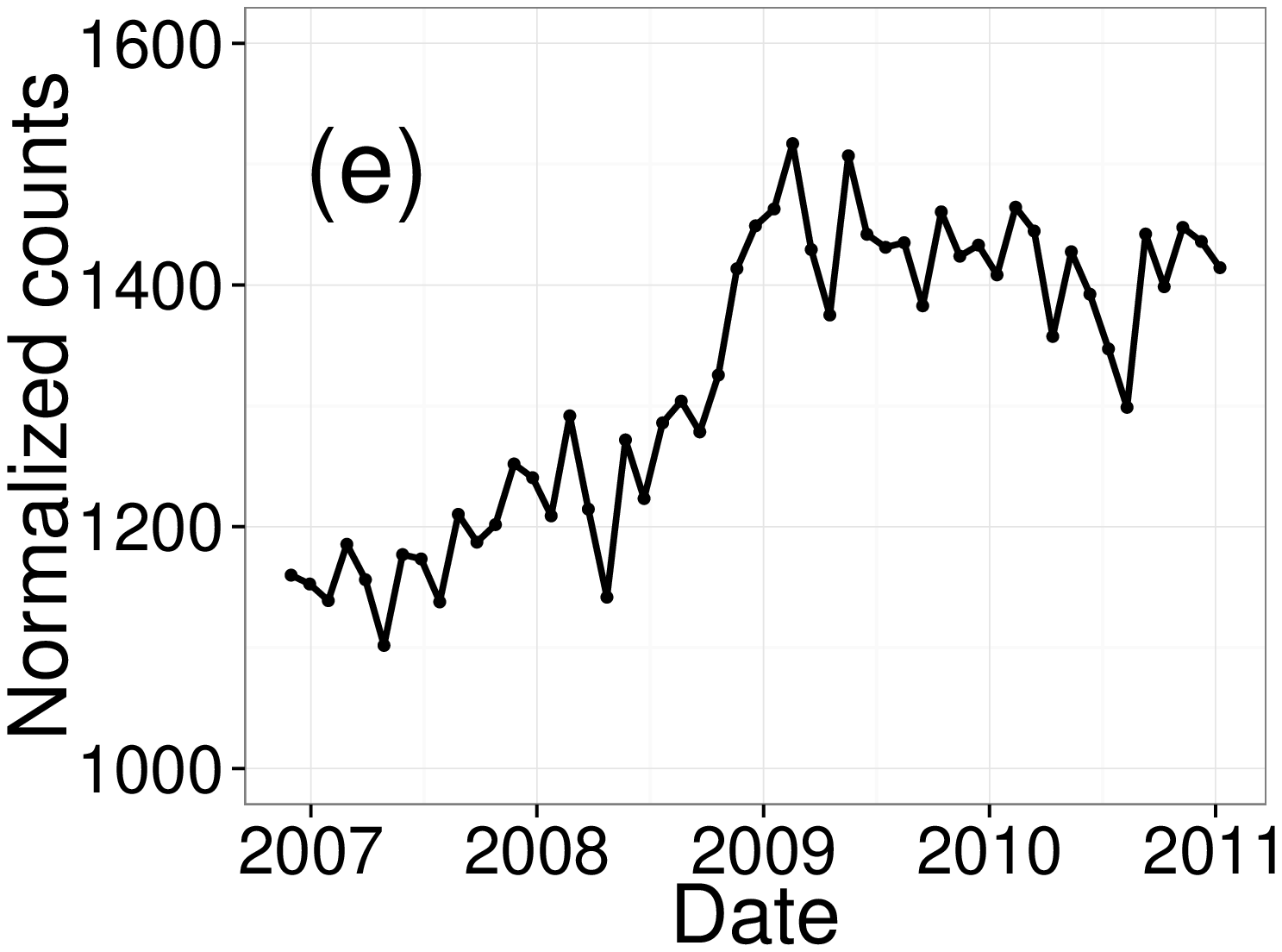}
\end{minipage}
\begin{minipage}{0.48\hsize}
\centering
\includegraphics[width=7cm,height=4cm]{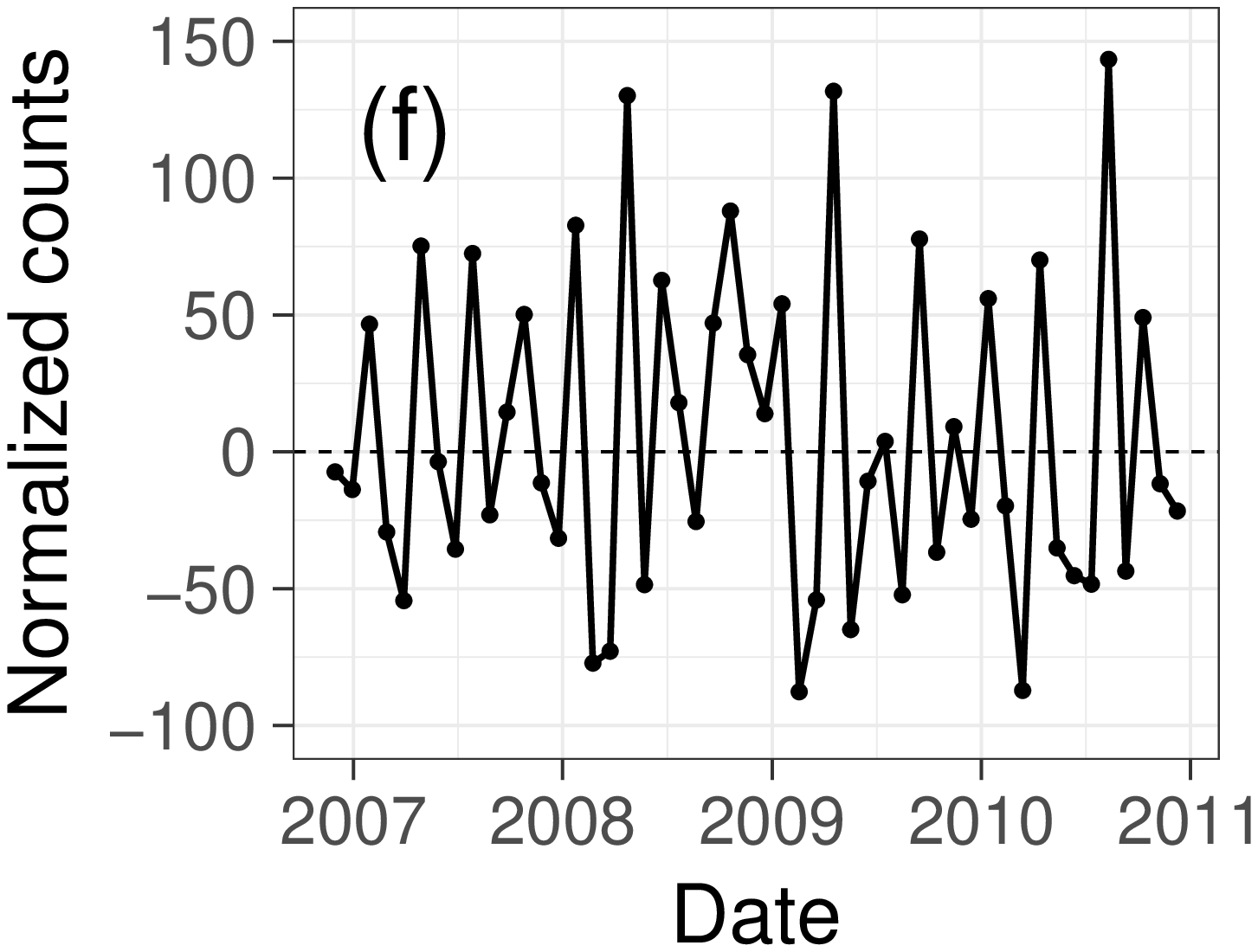}
\end{minipage}
\caption{
(a) An example of a daily time series of raw word appearances for ``oishii'' (delicious), $g_j(t)$. 
(b) The daily time series of the normalised total number of blogs, $m(t)$ (see Appendix \ref{app_m}).  
(c) The daily time series of word appearances scaled by the normalised total number of blogs for ``oishii'', $f_j(t)=g_j(t)/m(t)$. 
(d) The differential time series of the word appearances scaled by the normalised total number of blogs for ``oishii'', $\delta f_j(t) \equiv f_j(t)-f_j(t)$.
(e) The time series of the box means of the word appearances scaled by the normalised total number of blogs for ``oishii'', $F_j^{(L)}(\tau^{(L)}) \equiv \sum^{L-1}_{k=0}f_j (L \cdot \tau^{(L)}+k)/L. \quad (\tau^{(L)}=1,2,\cdots, T^{(L)})$;$L=30$
(f) The differential time series of the box means word appearances scaled by the normalised total number of blogs for “oishii”, $\delta F(\tau^{(L)})=F_j^{(L)}(\tau^{(L)})-F_j^{(L)}(\tau^{(L)}-1)$;$L=30$.
We can confirm that the time-variation of raw word appearances $g_j(t)$ shown in panel (a) is almost the same as that of the total number of blogs $m(t)$ shown in panel (b).
}
\label{tseries}
\end{figure*}

\section{Data set}
In the data analysis, we analysed the word frequency time series of blogs, that is, the time series of the numbers of word occurrences in Japanese blogs per day.
To obtain these time series, we used a large database of Japanese blogs ("Kuchikomi@kakaricho"), which was provided by Hottolink, Inc. This database contains 3 billion articles of Japanese blogs, which covers 90 percent parcent of Japanese blogs from 1 Nov., 2006 to 31 Dec., 2012. We used 1,771 basic adjectives as keywords \cite{RD_base}. \par
\subsection{The normaised time series of word appearances}
Here, we define the notation of the time series of the word appearances $g_j(t)$ and $f_j(t)$ as follows:
\begin{itemize}  
\item $g_j(t)$ $(t=1,2,\cdots T)$ $(j=1,2,3,W)$ is a raw daily count of the appearances of the j-th word within the dataset (see Fig.\ref{tseries}(a)). 
\item $f_j(t)=g_j(t)/m(t)$ is time series of daily count normalised by the total numbers of blogs $m(t)$.
\end{itemize}
Here, $m(t)$ is the normalised total number of blogs assuming that $\sum_{t=1}^{T}m(t)/T=1$ for normalisation (see Fig.\ref{tseries}(c)), 
where $m(t)$ is estimated by the ensemble median of the number of words at time t, as described in the Appendix \ref{app_m}.  
Note that $f_j(t)$ corresponds to the original time deviation of the $j$-th word separated from the effects of deviations in the total number of blogs $m(t)$ (see Figs. \ref{tseries}(b) and (c)).  
\begin{figure*}
\begin{minipage}{0.48\hsize}
\centering
\includegraphics[width=7cm]{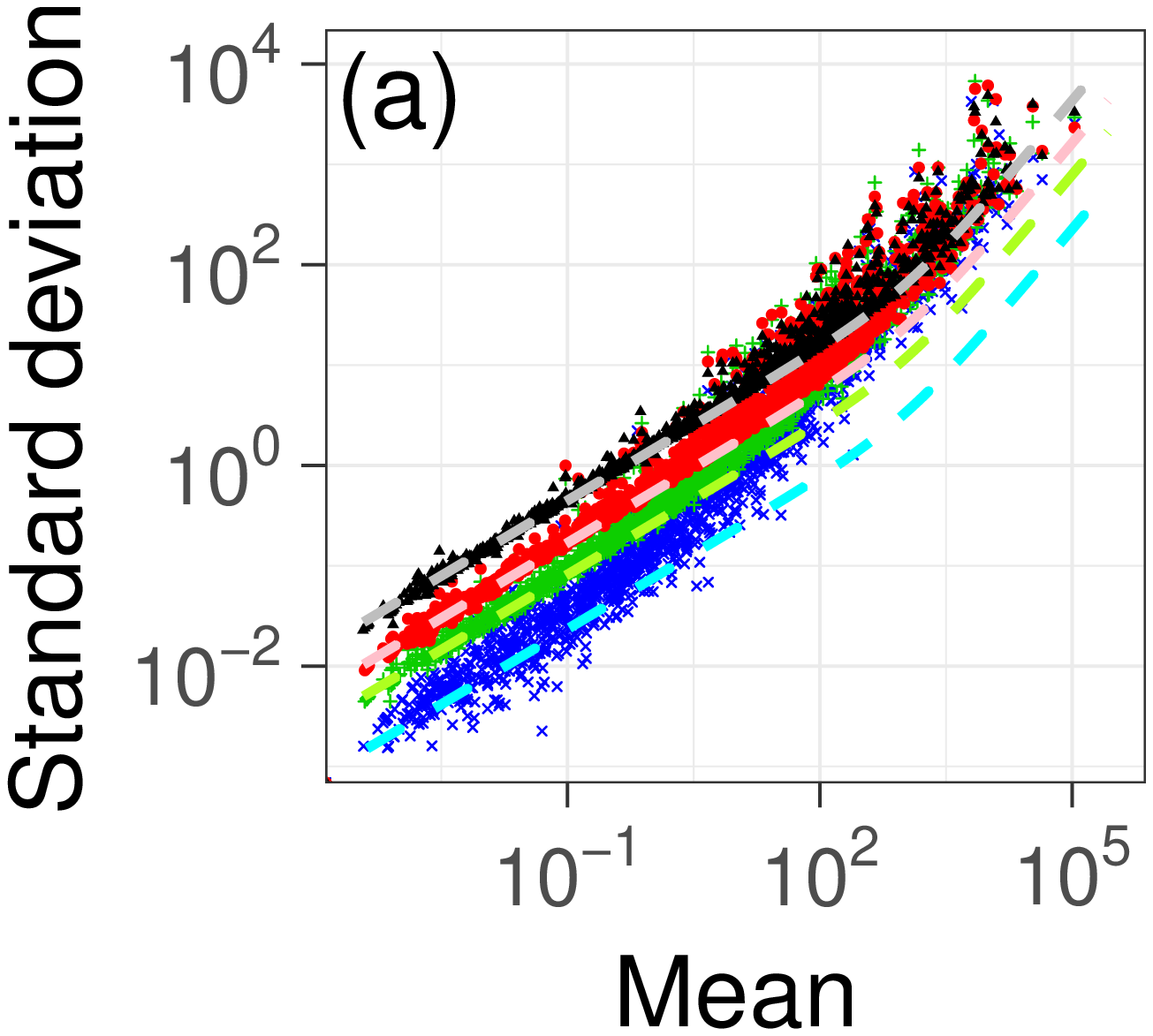}
\end{minipage}
\begin{minipage}{0.48\hsize}
\centering
\includegraphics[width=7cm]{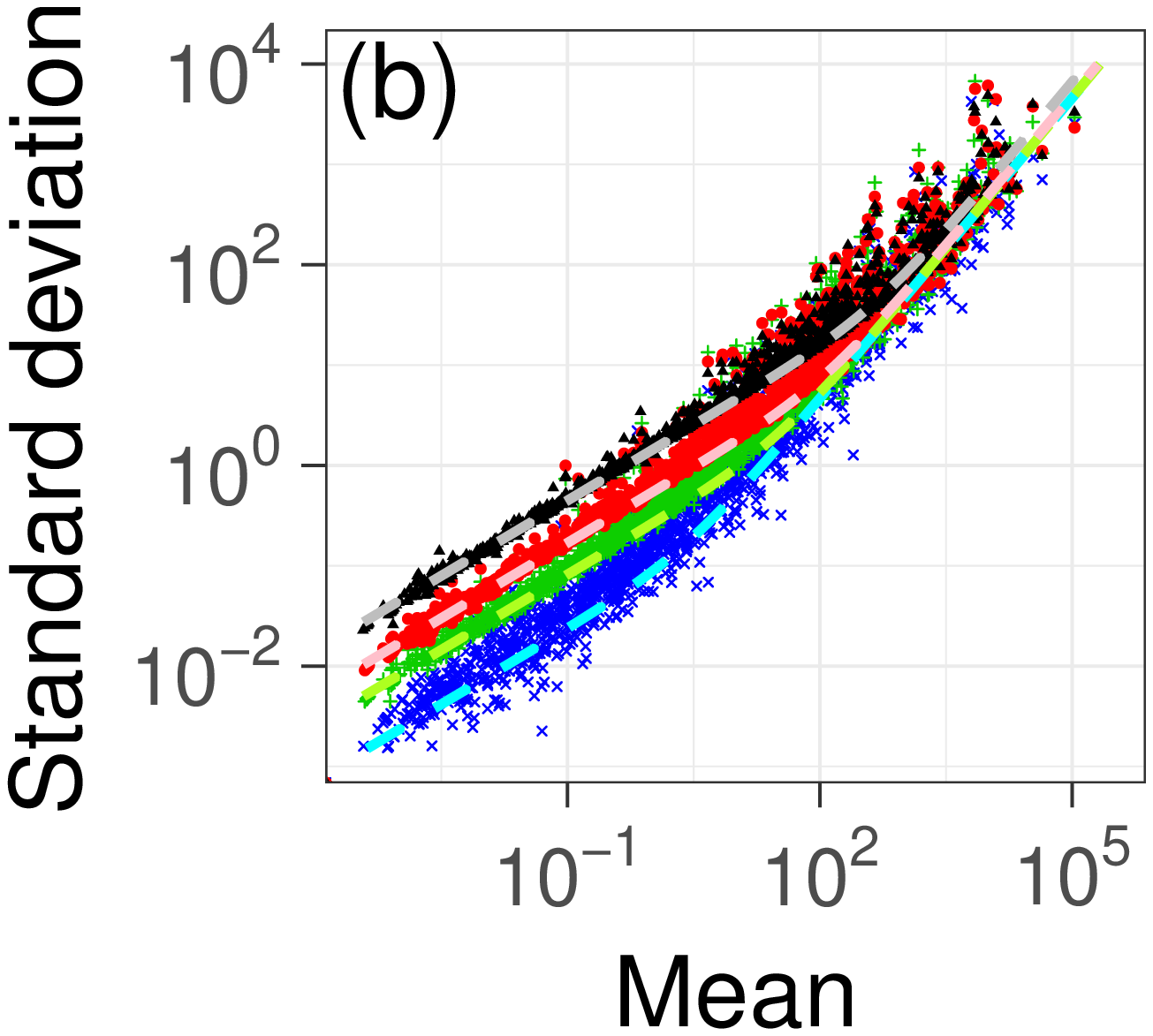}
\end{minipage}
\caption{(a) The TFS of the box mean time series, mean $E[F^{(L)}_j]$, versus standard deviation $V[\delta F^{(L)}_j]^{1/2}$. The data shown are empirical results of 1,771 adjectives for $L=1$ (black triangles),
$L=7$ (red circle), $L=30$(green plus), $L=365$(blue cross), and the corresponding theoretical curves given in Eq. \ref{V_df_emp} under the conditions that $a(L)=a_0/L$, $b(L)=b_0/L$, $a_0=E[1/m] \approx 1$ and $b_0=\check{\Delta}_0^2=0.030^2$ which are deduced from the assumption that $\{f_j(t)\}$ is sampled from independent random variables, for $L=1$ (grey dashed line), $L=7$ (pink dashed line), $L=30$ (mint green dashed line), and $L=365$ (cyan dashed line) from the top to bottom. We can see that the empirical data are in good agreement with the theoretical lower bound for the small mean $V[\delta E^{(L)}_j]$, and are not in agreement for the large mean $V[\delta E^{(L)}_j]$.  
(b) The corresponding figure shows the theoretical lower bound calculated by the random diffusion model based on the power-law forgetting process, in which $b(L)$ is given by Eq. \ref{u_ans} and Eq. \ref{b_LL}. From panel (b), we can confirm that the empirical lower bound are in agreement with the theoretical curve at all scales.}
\label{fig_TFS}
\end{figure*}

\begin{figure*}
\begin{minipage}{0.48\hsize}
\includegraphics[width=7cm]{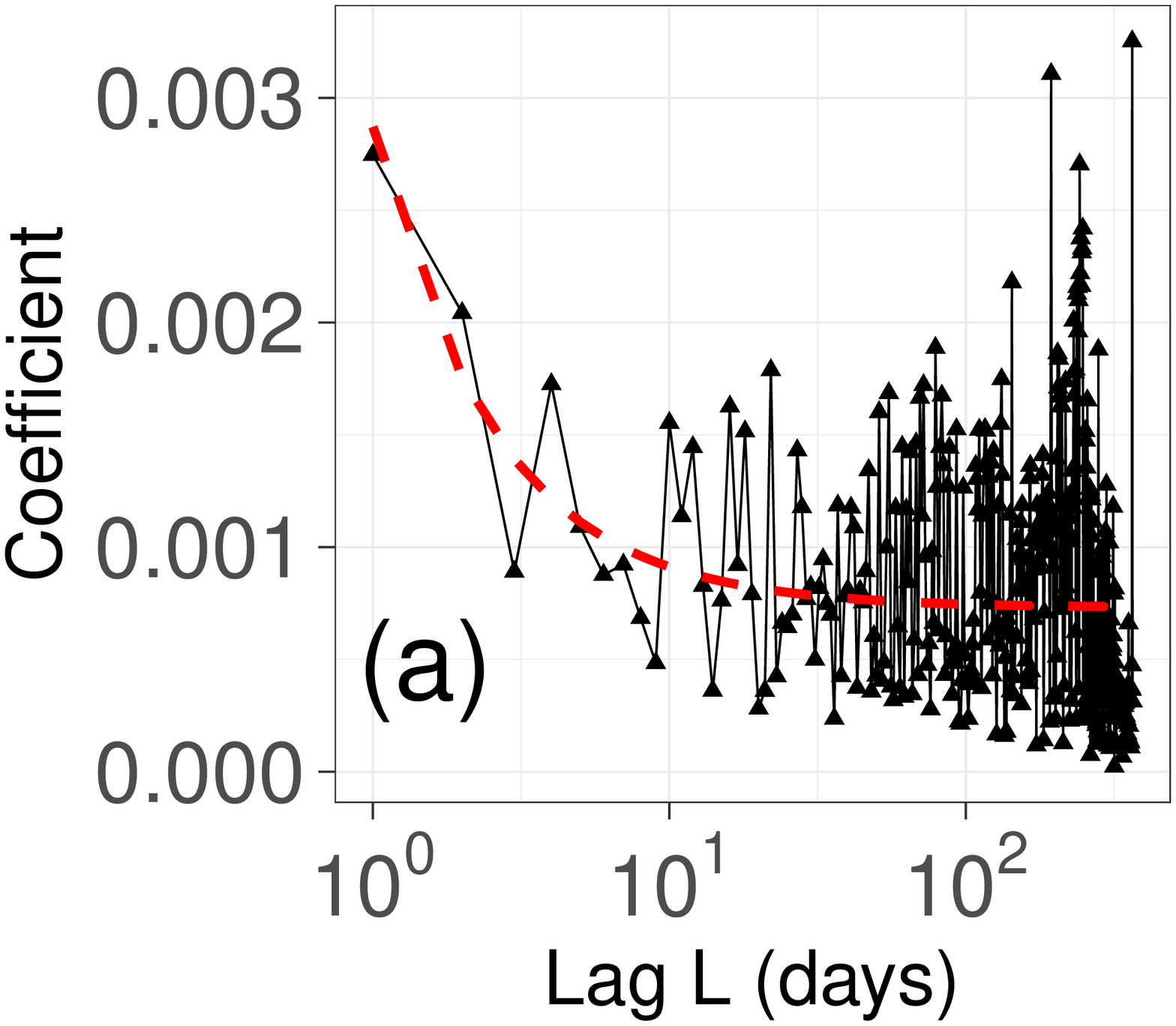}
\end{minipage}
\begin{minipage}{0.48\hsize}
\includegraphics[width=7cm]{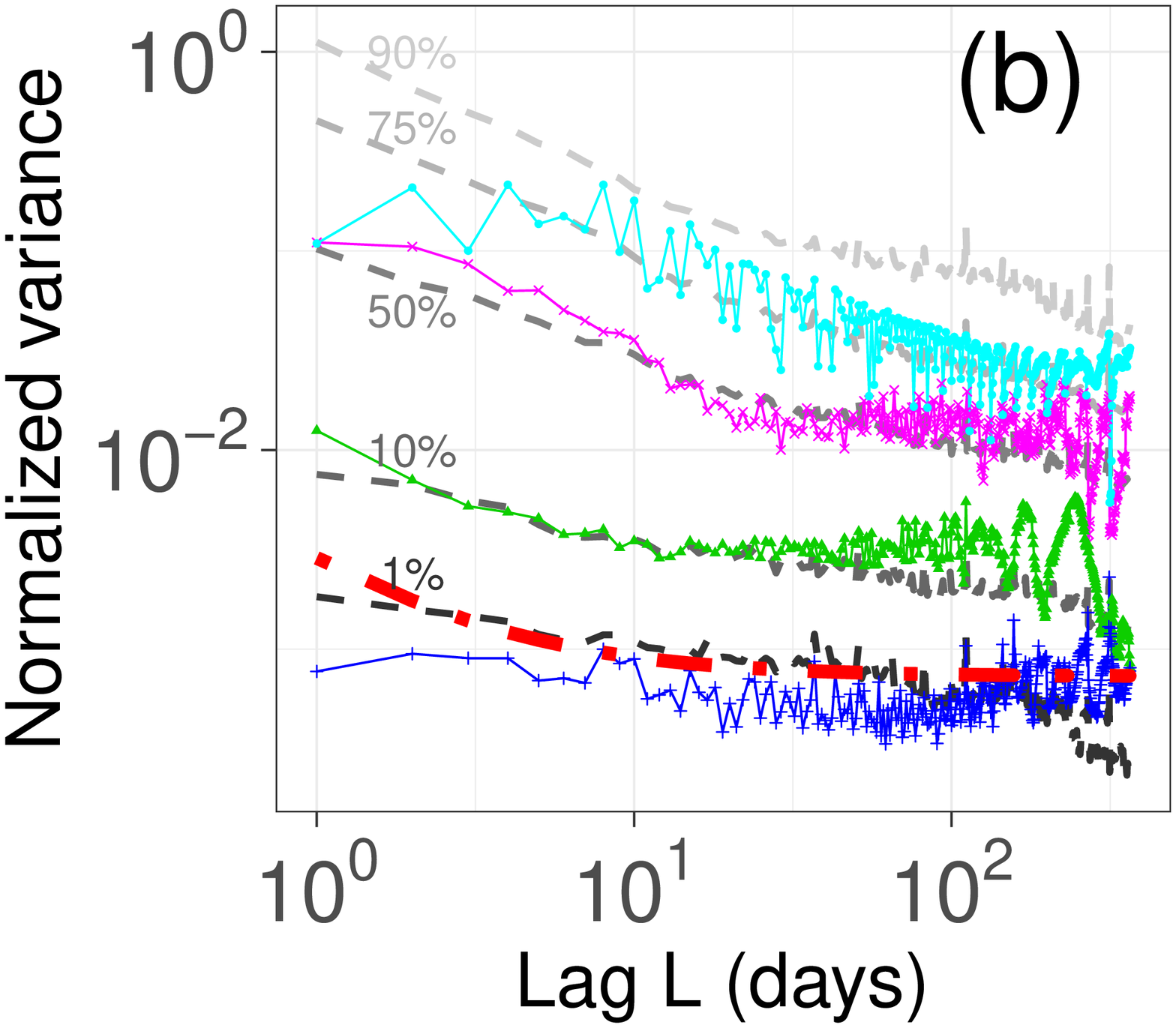}
\end{minipage}
\caption{(a)The coefficients of the fluctuation scaling, $b(L)$ given by Eq. \ref{V_df_emp}, which is directly estimated from the data using the method given in Appendix \ref{app_coef} (black triangle). In addition, the red dashed line is the theoretical curve calculated by the random diffusion model based on the power-law forgetting process given in Eq. \ref{u_ans} and Eq. \ref{b_LL}. 
From panel (a), we can see that $b(L)$ does not have a clear dependence on a lag $L$ for a large lag $L$, namely, $b(L) \approx const$ ($L>>1$). 
(b) The examples and the distribution of $\hat{b}_j(L)=V[\delta F_j^{(L)}]/E[F_j^{(L)}]^2$ nearly correspond to $b(L)$ for the particular words with a \textcolor{black}{large $E[F_j^{(L)}]$} (see Eq. \ref{b_LL}). The points show the empirical \textcolor{black}{data for ``ooi'' (many in English; blue plus) , ``otonashii'' (quiet personality in English; green triangles), ``tuyoi'' (strong in English; pink cross)  and ``kowai'' (scary in English; cyan circle). }
 The gray dashed lines indicate the percentiles of the distribution $\hat{b}_j(L)$ as a function of $L$. \textcolor{black}{The red dash-dotted line indicates the same theoretical curve with panel (a).}
Panel (b) implies that the particular words also obey a similar FS with the lower bound shown in panel (a) 
 because the points in the plot indicate an approximately $\hat{b}_j(L) \approx O(1)$ for a large $L$. 
In addition, 
this result suggests there is a common dynamics, which is independent of the types of words.
} 
\label{fig_coef}
\end{figure*}

\section{The fluctuation scaling in empirical data}
First, we discuss the temporal fluctuation scaling (TFS) to obtain a clue to describe the basic dynamics of the time series.
 The TFS of the difference in the time series $\{A(t)\}$ ($t=1,2,\cdots T$) 
is defined by the scaling between a temporal mean $E[A]$ and a temporal variance of the difference in time series $V[\delta A]$, 
\begin{equation}
V[\delta A] \propto E[A]^{\alpha}. \label{TFS_of_A}
\end{equation}
Here, the temporal mean $E[A]$ and temporal variance $V[A]$ are defined by 
\begin{equation}
E[A] \equiv \sum^{T}_{t=1}A(t)/T,
\end{equation}
\begin{equation}
V[\delta A] \equiv \sum^{T-1}_{t=1}(\delta A(t)-E[\delta A])^2/(T-1), 
\end{equation}
where $\delta A(t)$ means the difference at time $t$, $\delta A(t)=A(t)-A(t-1)$.
 The reasons why we investigate the TFS in the first step are as follows:
 \begin{itemize}
 \item The fluctuation scaling for the daily time scale have been studied intensively in \cite{RD_base}, and its statistical properties are explained through the model consistently.
 \item According to the study \cite{RD_base}, the deviation from the lower bound of the plots of TFSs for various words (see Fig \ref{fig_TFS}) is caused by the word-dependent special individual effects, such as news or seasonality.
Thus, it is expected that by focusing on the lower bounds of the TFS plot, we can obtain information of relatively ``normal'', words which only slightly affected by special effects.  
 \end{itemize}
 Note that the above definition of the TFS in Eq. \ref{TFS_of_A} is expressed in terms of the variance, although the standard deviation is usually used in observations.
 Under this condition, the TFS expressed by the standard deviation can be written as $V[\delta A]^{1/2} \propto E[A]^{\alpha/2}$. 
\textcolor{black}{ In addition, we assume in this section that $T>>1$ for simplicity.}
\subsection{Temporal fluctuation scaling for daily time scale}
Herein, we investigate the TFS of the daily differential of the number of word appearances $\delta f_j(t)=f_j(t)-f_j(t-1)$ (see Fig. \ref{tseries}(d)), 
which has been already studied intensively in Ref. \cite{RD_base}. 
From the black triangles in Fig. \ref{fig_TFS}, we can confirm the scaling with two exponents: 
\begin{equation}
V[\delta f] \propto
\begin{cases}
 E[ f_j]^{1} & \text{$E[f_j] < \mu^{*}$} \\
 E[ f_j]^{2} & \text{$E[f_j] > \mu^{*}$}, 
\end{cases} 
\end{equation}
where $\mu^{*} \approx 100$. 
In addition,
we can also calculate the theoretical lower bound of this scaling by using the random diffusion model, which is mentioned in section \ref{sec_model} \cite{RD_base},  
\textcolor{black}{
\begin{eqnarray}
V[\delta f_j]&\geq& 2 \cdot E[f_j] \cdot E[\frac{1}{m}]+E[f_j]^2 \cdot \{ 2 \cdot  \check{\Delta}_0^2 \}, \nonumber \\ \label{v_delta_tilda_f2}
\end{eqnarray}
} where \textcolor{black}{ $\check{\Delta}_0^2=0.030^2$} and $\{m(t)\}$ estimated by the method described in Appendix \ref{app_m}.
This lower bound is shown in the grey dashed line in Fig. \ref{fig_TFS} (a). 
\par
\subsection{Analysis of the rescaling of the TFS of word appearences data}
In this section, we investigate the time-scale-dependence of the TFS (i.e., an analysis of the rescaling) to extract essential information of the dynamics of the time series.
In particular, we use the box means for the time-scale coarse-graining, 
\begin{eqnarray}
F_j^{(L)}(\tau^{(L)}) \equiv \sum^{L-1}_{i=0}f_j(L \cdot \tau^{(L)}+i)/L. \quad (t=1,2,\cdots, T^{(L)})
\end{eqnarray}
where $\tau^{(L)}=1,2,T^{(L)}$ is an index of time for the $L$-day scale, namely, $t=\tau^{(L)} \times L$. 
For example, $L \times F_j^{(L)}(\tau^{(L)})$ corresponds closely to a (normalised) weekly word-appearance time series for $L=7$，monthly time series for $L=30$, and yearly time-series for $L=365$. \par
The mean and variance of the difference of this value are defined in the same way as shown in Eq. \ref{TFS_of_A} 
\begin{equation}
E[F_j^{(L)}]=\sum^{T^{(L)}}_{\tau^{(L)}=1}F_j(\tau^{(L)})/T
\end{equation}
and
\begin{equation}
V[\delta F_j^{(L)}] \equiv \sum^{T^{(L)}-1}_{\tau^{(L)}=1} (\delta F_j(\tau^{(L)})-E[\delta F_j(\tau^{(L)})])^2/(T^{(L)}-1).  
\end{equation}
The TFS of the coarse-grained time series $\{F_j^{(L)}(\tau^{(L)})\}$ for $L=1$, $L=7$, $L=30$, and $L=365$ is plotted in Fig \ref{fig_TFS}(a).
From this figure, we can observe that the scaling with two exponents (i.e., kinked lower bounds) is similar to the time scale of a day ($L=1$).   
The lines in Fig \ref{fig_TFS} indicate the theoretical curve 
\begin{equation}
V[\delta F_j^{(L)}] = a(L) \cdot E[F_j^{(L)}]+ b(L) \cdot E[F_j^{(L)}]^2.   \label{V_df_emp}
\end{equation}
In Fig \ref{fig_TFS}(a), we set $a(L)=2 \times a_0/L$ and $b(L)=2 \times b_0/L$, which are obtained using the central limit theorem under the assumption that $\{f_j(t)\}$ is independent 
(We use \textcolor{black}{$a_0=E[1/m] \approx 1$ and $b_0=\check{\Delta}_0=0.030$}). 
These lines are in good agreement with empirical lower bound for a small mean $E[F^{(L)}]$. However, they are in disagreement for a large mean $E[F^{(L)}]$.  These results imply that the assumption that  $\{f_j(t)\}$ is independent does not holds. \par
Fig. \ref{fig_coef}(a) shows a result in which $b(L)$ is directly estimated from data using the method described in Appendix \ref{app_coef}.
From this figure, we can see that $b(L)$ does not have a clear dependence on $L$ for a large $L$, namely, 
\begin{equation}
b(L) \approx const \quad (L>>1).
\end{equation} 
%
%
%
  \begin{table*}
 \begin{tabular}{cl cc}
 \hline
 & Parameter & Meaning & Estimation or Definition \\
 \hline
  \multicolumn{4}{l}{$g_j \sim Poi(c_j(t) \cdot \Lambda_j(t)); \quad f_j(t)=g_j(t)/m(t)$} \\
  (i)& \multicolumn{3}{l}{Dynamic noise: $c_j(t)=\check{c}_j \dot r_j(t); \quad r_j(t)= \sum^{-\infty}_{s=0}\theta_j(s) \eta_j(t-s); \quad \theta(s)=(s+\Gamma(1-\beta)^{-1/\beta})^{-\beta}/\Gamma(1-\beta)$$^{(*1)}$; \quad $\beta=0.5$} \\
  &\textbullet$\check{c}_j$ & scale factor of the $j$-th word & $\sum^{T}_{t=1}f_j(t)/T$ \\
  &\textbullet$\check{\eta}_j $ & The standard deviation of the dynamic noise of $j$-th word $\eta_j$ & from graphs$^{(*2)}$  \\
  &\textbullet$\eta_j(t)/\check{\eta}_j$ & The scaled distribution of the dynamic noise $\eta_j$ &  t-distrubition ($df=2.57$) \\
   (ii)  & \multicolumn{3}{l}{Ensemble noise : $\Lambda_j(t) \sim \{$distribution with mean $m(t)$ and standard deviation $\Delta^{(m)}_j(t)$\}, $\Delta^{(m)}_j(t)=\Delta_j^{(0)} \cdot m(t)$} \\
  &\textbullet $m(t)$ & mean of $\Lambda_j(t)$ (scaled number of blogs) & Appendix \ref{app_m}\\ 
   &\textbullet $\Delta^{(0)}_j$ & The standard deviation of the ensemble noise $\Lambda_j(t)$ &  from graphs$^{(*2)}$ \\
  &\textbullet $(\Lambda_j(t)-m(t))/\Delta^{(m)}_j(t)$ &  The scaled distribution of the ensemble noise $\Lambda_j(t)$ &  t-distrubition ($df=2.57$) \\
   \hline
     \multicolumn{4}{l}{${}^{*1}$We set $Z(\beta)=\Gamma(1-\beta)$} \\
  \multicolumn{4}{l}{${}^{*1}$ $\check{\eta}_j$ and $\Delta^{(0)}_j(t)$ are tuned depending on a word to reproduce empirical results consistently (see Figs. \ref{fig_spect}, \ref{fig_pdf}).  } \\
 \end{tabular}
 \caption{Summary of the model and parameters for comparison with actual data}
 \label{table_parameter}
 \end{table*}
\section{Model}
\label{sec_model}
We now introduce the random diffusion model (RD model) to explain the TFSs, as mentioned in the previous section.
The RD model can be used to explain the TFS and other statistical properties of a daily word appearance time series \cite{sano2009, RD_base}.  \par
 The RD model, which is a non-stationary Poisson process formed through a stochastic process, whose Poisson parameter (mean value) varies randomly, is defined for $t=1,2,3,\cdots,T$, $j=1,2,\cdots,J(T)$ as follows :
\begin{eqnarray}
g_j(t) &\sim& Poi(c_j(t) \Lambda_j(t)). \label{rd_model} \\
\Lambda_j(t) &\sim& Distribution  \\
&& s. t.  \nonumber \\
&& <\Lambda_j(t)>
=m(t), \nonumber \\ 
&& <(\Lambda_j(t)-<\Lambda_j(t)>)^2>= {\Delta^{(m)}_j}^2=m(t)^2 {\Delta^{(0)}_j}^2.  \nonumber \\
\end{eqnarray}
\textcolor{black}{
\textcolor{black}{The first equation means that the random variable $g_j(t)$ is sampled from the Poisson distribution whose Poisson parameter takes the value $c_j(t) \cdot \Lambda_j(t)$.}  $c_j(t) \geq 0$ is a scale factor of the Poisson parameter of the model, $\Lambda_j(t)$ is a random factor and \textcolor{black}{$J(t)$ is the total number of word types at time $t$.} In the case of a time series of blogs, the observable $g_j(t)$ corresponds to the frequency with which the $j$-th word occurs on the $t$-th day, and larger values of $c_j(t)$ indicate that the j-th word appears more frequently at time $t$ on average.} 
Note that \textcolor{black}{this model is a kind of doubly stochastic Poisson process \cite{lowen2005fractal}}. 
 \par
 
%
\textcolor{black}{$\Lambda_j(t)$ is a non-negative random variable with mean $m(t) \geq 0$ and standard deviation $\Delta^{(m)}_j(t)=m(t) \Delta^{(0)}_j \geq 0$, $\Delta_j^{(0)}>0$. }  
  Here, $m(t)$ is a shared time-variation factor for the entire system, and 
we assume that $(\sum_{t=1}^{T}m(t)/T=1)$, for normalization.
In the case of a time series of blogs, $m(t)$ closely corresponds to the normalised number of blogs. \par
For convenience of analysis, we assume that $c_j(t)$ can be decomposed into 
a scale component $\check{c}_j$, which corresponds to the temporal mean of the count of the j-th word during the observation period, and a time variance component $r_j(t)$, such that
\begin{equation}
c_j(t)=\check{c}_j  r_j(t). \label{cj}
\end{equation}
Here, we also assume \textcolor{black}{for normalization} that $(\sum_{t=1}^{T}r_j(t)/T \approx 1)$. \par
\subsection{Estimation of parameters of the model}
In comparison with the data, we estimate the parameters as follows: (i)$\check{c}_j=\sum^{t=1}_{T}f_j(t)$, (ii)$m(t)$ is estimated by the ensemble median of the number of words at time t, as described in Appendix \ref{app_m}, (iii)$\{r_j(t)\}$ is determined based on the model of the time evolution, as described in section \ref{sec_fs_model} and (iv)\textcolor{black}{$\Delta_j^{(0)}$ is determined depending on a word to reproduce empirical results consistently.}  A summary of the parameter estimations of the model is presented in table \ref{table_parameter}. \par
\section{Fluctuation scaling of the RD model}
\label{sec_fs_model}
Here, we calculate the fluctuation scaling of the RD model.
First, we introduce random variables $w_j(t)$, $\check{C}_j^{(L)}$, $R_j^{(L)}(t^{(L)})$ and $W_j^{(L)}(t^{(L)})$ for simplicity. \par
$w_j(t)$ is defined as
\begin{equation}
w_j(t) \equiv g_j(t)/m(t)-c_j(t)=f_j(t)-c_j(t), \label{base_u}
\end{equation}
Using this variable, we can write $f_j(t)$,
\begin{equation}
f_j(t)=\check{c}_j r_j(t)+w_j(t). \label{base_u2}
\end{equation}
From the definition，the mean of $w(t)$ is  
\begin{equation}
<w_j(t)>_w=0
\end{equation}
and from Ref. \cite{RD_base}, the variance of $w_j(t)$ is 
\begin{eqnarray}
&&<(w_j(t)-<w_j(t)>)^2>_w=<w_j(t)^2>_w \nonumber \\
&& = 1/m(t) \cdot c_j(t) + {\Delta^{(0)}_j}^2 \cdot c_j(t)^2,  \label{ww} 
\end{eqnarray}
where，$<A>_w$ is the mean respect to $w$, $<A>_w=\int A(w) p_w(w)dw$.
 \par
We also define the box mean $\check{C}_j^{(L)}$, $R_j^{(L)}(t^{(L)})$ and $W_j^{(L)}(t^{(L)})$, corresponds to $\check{c}_j$, $r_j(t)$ and $w_j(t)$, 
\begin{eqnarray}
&&\check{C}_j = \check{c}_j \\
&&R_j^{(L)}(t^{(L)}) \equiv \frac{1}{L}\sum^{L-1}_{t=0}r_j(L  t^{(L)}+t) \label{sd_L_0}
\end{eqnarray}
\begin{equation}
W_j^{(L)}(t^{(L)}) \equiv \frac{1}{L}  \sum_{t=0}^{L-1}w_j(L  t^{(L)}+t)  \label{W0}
\end{equation}
Using these values, we can write the time-scale coarse-grained equation corresponding to Eq. \ref{base_u}, 
\begin{equation}
F_j^{(L)}(t^{(L)})=\check{C}_j R_j^{(L)}(t^{(L)})+W_j^{(L)}(t^{(L)}).  \label{large_f}
\end{equation}
\par
Second, calculating the variance $V[\delta F^{(L)}]$, we can obtain 
\begin{eqnarray}
&&V[\delta F^{(L)}] \approx a(L) E[ F^{(L)}] +b(L) E[ F^{(L)}]^2,  \label{df_ans0}
\end{eqnarray}
where 
\begin{equation}
a(L)=\frac{2}{L} a^{(0)}
\end{equation}
\begin{eqnarray}
b_j(L)&=&V[\delta R_j^{(L)}]+\frac{2b^{(0)}_j(1+V[r_j])}{L}, \label{b_LL} \\
     &\geq&V[\delta R_j^{(L)}]+\frac{2b^{(0)}_j}{L}
\end{eqnarray}
$a^{(0)}=E[1/m]$ and $b^{(0)}_j={\Delta^{(0)}_{j}}^2$. 
The details of the derivation of the variance are provided in Appendix \ref{app_long_rd}. 
From Eq. \ref{df_ans0}, we can confirm that the RD model reproduces the empirical properties for a small mean $E[F^{(L)}]$, that is, $a(L) \propto 1/L$ (Fig. \ref{fig_TFS}(a)). In addition, we can also confirm that $b_j(L)$ (i.e., the properties for a large mean $E[F^{(L)}]$ in Fig. \ref{fig_TFS}(a)) is determined by $V[\delta R_j^{(L)}]$, that is, it is determined based on the  property of the dynamics of $\{R_j(t)\}$.
\par
\subsection{Relation between TFS and dynamics of $\{r_j(t)\}$}
What are the dynamics that make $b_j(L) \approx const.$ for $L>>1$, which is an empirical finding shown in Fig. \ref{fig_coef}(a)? 
To clarify this question, we studied the relation between $b_j(L)$ or $V[R^{(L)}_j]$ and the dynamics of $\{r_j(t)\}$.
\paragraph{{\bf Random walk}}
Firstly, we consider the case in which $\{r_j(t)\}$ is generated by the following simple random walk model, 
\begin{equation}
r_j(t+1)=r_j(t)+\eta_j(t), \label{rw}
\end{equation}
where $\eta_j(t)$ is an i.i.d random variable whose mean is zero, and we assume that the variance is $\check{\eta}_j^2<<1$, and that $r_j(t)>0$ takes nearly $1$. \par
From Appendix \ref{app_random_walk} (in the case of $\kappa=1$ and $u(t)=0$), we obtain
\begin{equation}
V[\delta R_j^{(L)}] \approx \frac{1}{3} \check{\eta}_j^2 \cdot [2L+\frac{1}{L}]  \label{result}.
\end{equation}
This result, namely, $b_j(L) \propto L$ for $L>>1$ is in disagreement with the empirical result $b_j(L) \approx const. $ (see Eq. \ref{b_LL}).   \par
In the case of random walk with dissipation $\kappa \geq 0$ and an external force $u_j(t)$, 
\begin{equation}
r_j(t+1)= \kappa \cdot r_j(t)+u_j(t)+\eta_j(t), 
\end{equation}
we can also obtain the variance for $T>>1$, $L<<T$, 
\begin{equation}
V[\delta R_j^{(L)}] \approx
\begin{cases}
 \check{\eta}_j^2 \cdot (\kappa-1)^{-2} \cdot L^{-1} & (0<\kappa<1) \\
 \check{\eta}_j^2 \cdot 1/3 \cdot (2L+1/L) & (\kappa=1)\\
  O( \frac{\kappa^{2T}}{L(T-L)}) &  (\kappa>1)\\  
\end{cases}.  
\end{equation}
This result also disagrees with the empirical result. 
The details of the derivations and the results of the variance $V[\delta R_j^{(L)}]$ for the case of a random walk are provided in Appendix \ref{app_random_walk}. 

\subsection{Power-law forgetting process}
We confirmed that neither the random walk model (described in the previous section) nor the independent and identical steady noise (see Fig. \ref{fig_TFS}(a)) can reproduce the empirical results $b_j(L) \approx const.$
Thus, we introduce the following stochastic process with power-law forgetting (a long memory process), \textcolor{black}{which is an extension of the random walk model}: 
\begin{equation}
r_j(t)=\sum_{s=0}^{\infty} \theta(s) \cdot \eta_j(t-s). \label{r_forget}
\end{equation}
where 
\begin{equation} 
\theta(s)=\frac{(s+a)^{-\beta}}{Z(\beta)},  \label{theta}
\end{equation}
\begin{equation}
a \equiv a(\beta) \equiv Z(\beta)^{-1/\beta}, 
\end{equation}
and $Z(\beta)>0$ is an arbitrary coefficient.
We call this model the power-law forgetting process.  
For $\beta=0$, this model is in agreement with the random walk model ($\kappa=1, u(t)=0$).
\par
In addition, under the conditions $0<\beta<1$ and $Z(\beta)=1-\beta$, 
this model is approximated through $ARFIMA(0,1-\beta,0)$ \cite{baillie1996analysing}, which is given by
\begin{equation}
r_j(t)=\sum_{s=0}^{\infty} \frac{\Gamma(s+1-\beta)}{\Gamma(s+1) \Gamma(1-\beta)} \cdot \eta(t-s), 
\end{equation}
because of the approximation of the constant, 
\begin{equation}
 \frac{(s+a)^{-\beta}}{\Gamma(1-\beta)} \approx \frac{\Gamma(s+1-\beta)}{\Gamma(s+1) \Gamma(1-\beta)}.
\end{equation}
Note that $ARFIMA(0,1-\beta,0)$ can also be written using a lag operator as follows:
\begin{equation}
(1-\hat{d})^{(1-\beta)} X(t)=\eta(t), \label{arfima2}
\end{equation}
or
\begin{equation}
\hat{D}^{(1-\beta)} X(t)=\eta(t), \label{arfima3} 
\end{equation} 
where $\hat{d}$ is the lag operator satisfied with $\hat{d}A(s)=A(s-1)$, $\hat{D}=(1-\hat{d})$ is a difference operator, and 
$(1-\hat{d})^{\alpha}=\sum^{\infty}_{k=0} \frac{\Gamma(\alpha+1)}{\Gamma(k+1)\Gamma(\alpha-k+1)} (-\hat{d})^k$, namely, $((1-\hat{d})^{m/n})^n=1-\hat{d}$ (e.g, $(\hat{D}^{0.5})(\hat{D}^{0.5})x(t)=\hat{D}x(t)=x(t)-x(t-1)$). 
From this point of view, our model may be interpreted as a fractional order integral of white noise.  For instance, ARFIMA(0,0.5,0) may be able to be interpreted  as a half-order integral of white noise ($\beta=0.5$).
In addition, the first order corresponds to a random walk ($\beta=0$), and the zero-th order corresponds to i.i.d. noise ($\beta=1$).  
\par
An approximation formula of $V[\delta R^{(L)}_j]$ for this model is given by Eq. \ref{u_ans} in Appendix \ref{app_forget}.
For $L>>1$, the main terms of Eq. \ref{u_ans} are given by
\begin{eqnarray}
&&V[\delta R^{(L)}_j] \approx \frac{\check{\eta}_j^2}{Z(\beta)^2} \nonumber \\
&&\begin{cases}
u_1(\beta)L^{1-2\beta}+u_2(\beta)L^{-\beta}+u_3(\beta)L^{-1} &\text{($0<\beta<1$)} \\
u_a \log(L)^2 L^{-1} + u_b \log(L) L^{-1}+u_c L^{-1} &\text{($\beta=1$)} \\
u_3(\beta)L^{-1} & \text{($\beta>1$)}, 
\end{cases}
\end{eqnarray}
where $u_1(\beta)$, $u_2(\beta)$, $u_3(\beta)$, $u_a$, $u_b$, and $u_c$ are $L$-independent coefficients given by from Eqs. \ref{u_ans_l} to \ref{u_ans_l_end} in Appendix \ref{app_l_forget}. 
Thereby, the maximum term of a series is written as 
\begin{eqnarray}
V[\delta R^{(L)}_j] \propto 
\begin{cases}
L^{1-2\beta} &\text{($0<\beta<1$)} \\
\log(L)^2 L^{-1} &\text{($\beta=1$)} \\
L^{-1} & \text{($\beta>1$)} 
\end{cases}
\end{eqnarray}
\par
From this result, we can confirm that the empirical result,  
 $b_j(L) \approx const.$ , namely, $V(\{R_j\}) \approx const.$ 
is reproduced  under the condition 
\begin{eqnarray}
\beta \approx 0.5.
\end{eqnarray}
\par
The red dashed line in Fig. \ref{fig_coef}(a) indicates the theoretical curve in which we insert Eq. \ref{u_ans} into Eq. \ref{b_LL} for the parameter $\beta=0.5$ and $Z(\beta)=\Gamma(1-\beta)$.
From this figure, we can confirm that the theoretical curve is in accordance with the empirical lower bound. 
 In addition, the corresponding theoretical curve in Fig \ref{fig_TFS}(b) is in 
agreement with the empirical data (\textcolor{black}{for $\check{\eta}_{0}=0.029$, $\Delta^{(0)}_{0}=0.030$}). \par
\begin{figure}
\includegraphics[width=6cm,angle=270]{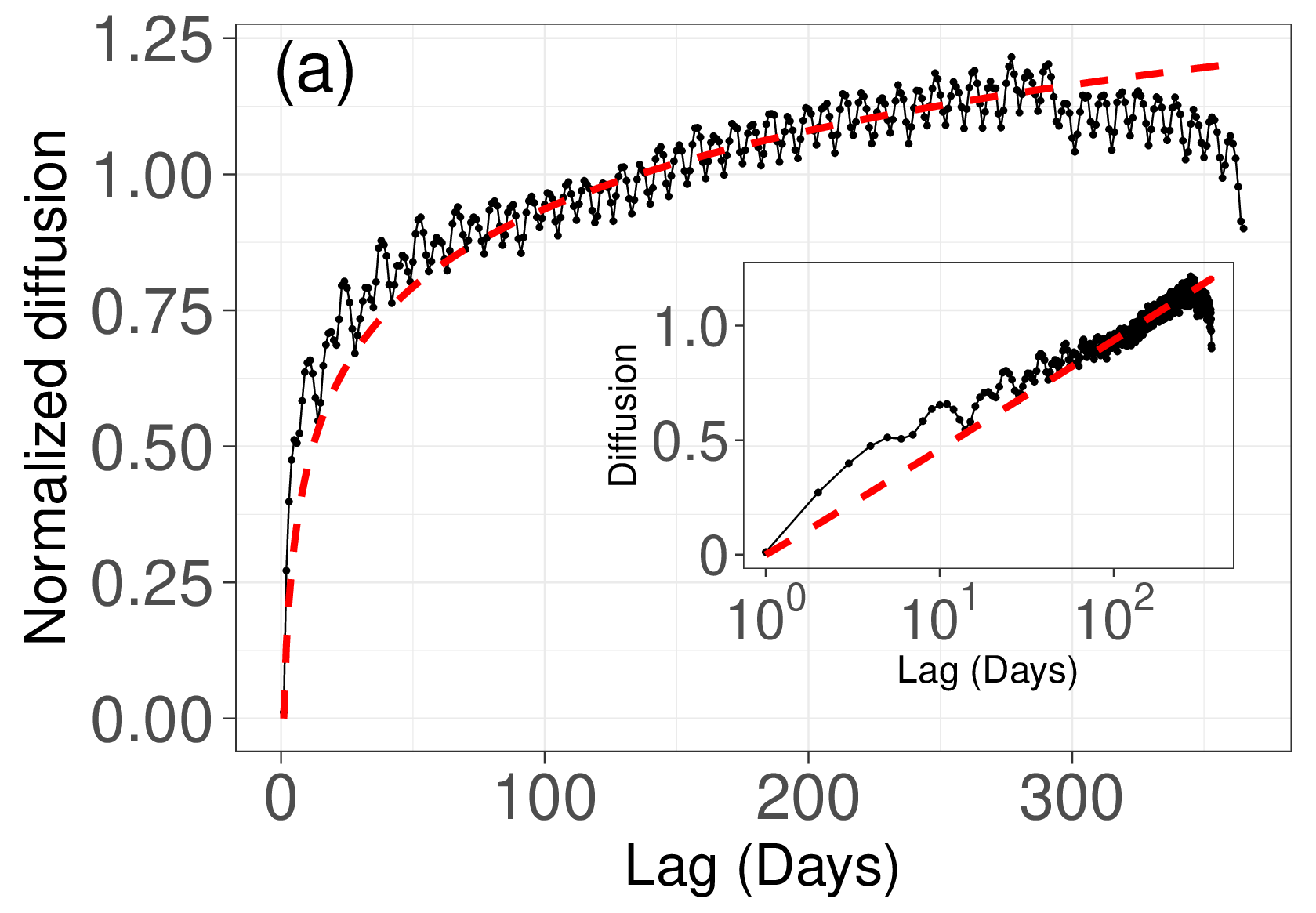}
\includegraphics[width=6cm,angle=270]{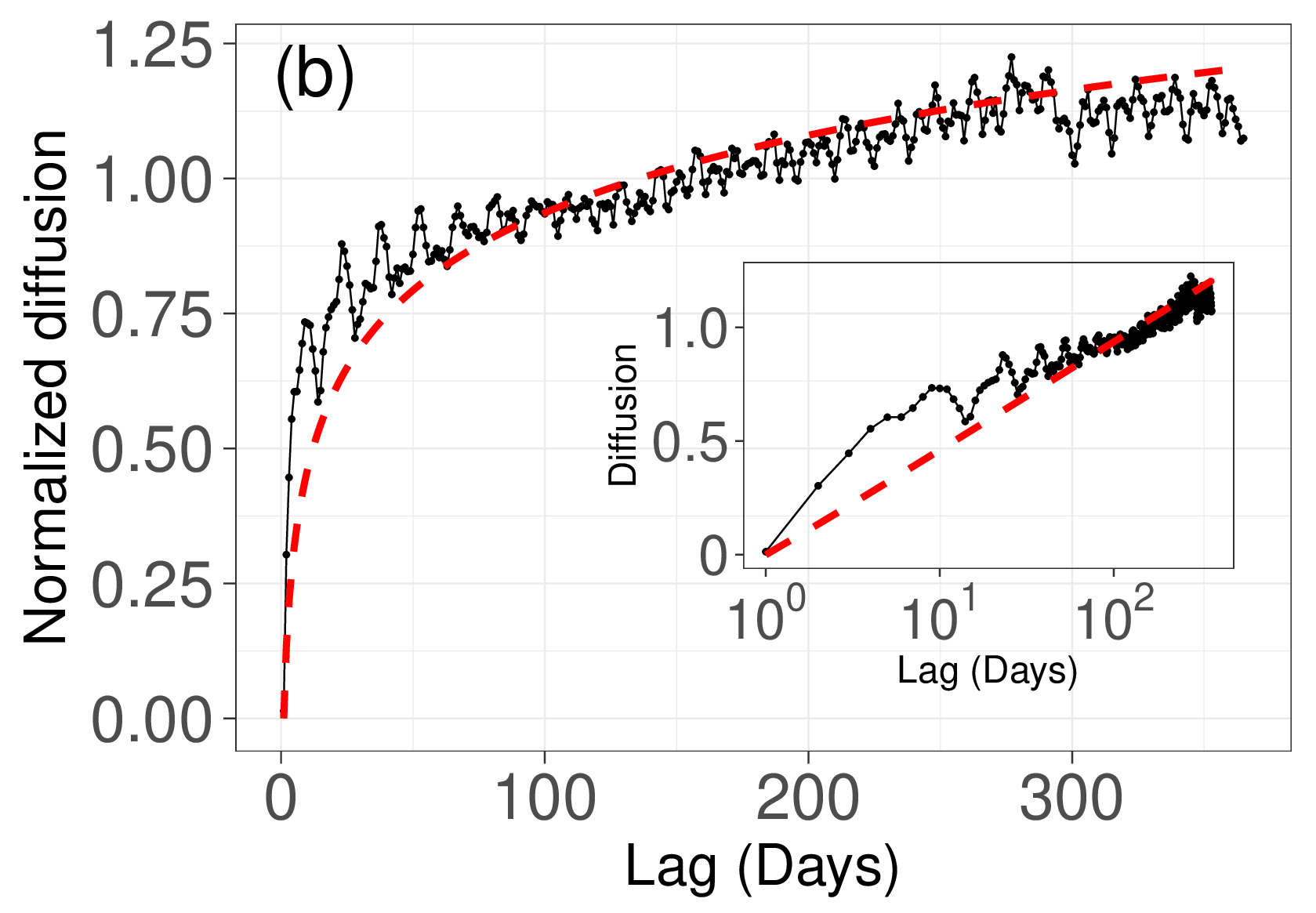}
\includegraphics[width=6cm,angle=270]{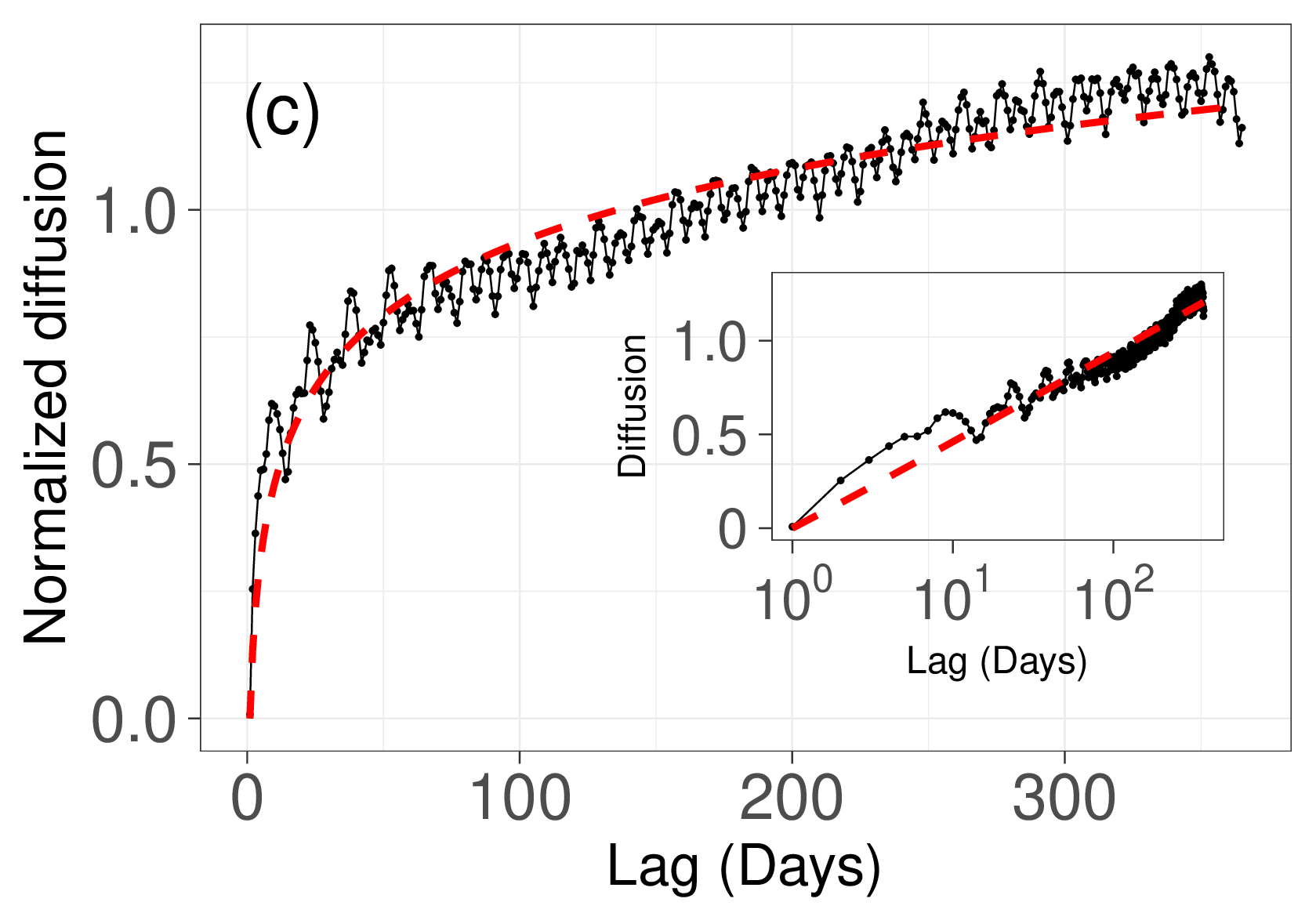}
\caption{
(a)The ensemble average of the scaled MSD $\hat{\Pi}_0(L)$ is given by Eq. \ref{normal_msd} for words with a mean $c_j$ above 20, 
and the red dashed line is the theoretical curve given by Eq. \ref{msd_norm}, which is obtained by the random diffusion model based on the power-law forgetting process.The inserted small graph is the corresponding data plotted in a semilog plot. 
 (We exclude words with a small $\check{c}_j$ because they have relatively large signal-to-noise ratios.)
 From this figure, we can confirm that the theoretical curve substantially agrees with the corresponding empirical data. 
(b) The corresponding graph in which the words with significant seasonality such as “atsui” (“hot” in English) and “suzushi” (“cool” in English) are removed from the word sets. (c) The graph in which the effects of outliers caused by special news events such as the great east Japan earthquake by using the trimmed mean given in Eq. \ref{trim_mean} (i.e., in panel (c), we removed both the effects of the seasonality and the outliers from the data analysis.).
From these figures, we can confirm that the curves approach the logarithmic function by removing the effects that are not considered in the model, such as the effects from special news events and seasonality.
}
\label{fig_msd}
\end{figure}

\section{Diffusion Properties}
In this section, we investigate the diffusion properties of the RD model based on the power-law forgetting process and compare it with the empirical observation.
Herein, we consider the case of $Z(\beta)=\Gamma(1-\beta)$ and $0 \leq \beta<1$. \par
The MSD of the forgetting process is given by (See Eq. \ref{diff_ans} in Appendix \ref{MSD_forget})．
\begin{eqnarray}
&&\Pi(L;\{r_j(t)\})=<(r_j(t+L)-r_j(t))^2>  \nonumber \\
&& \approx  \check{\eta}_j^2/\Gamma(1-\beta)^2 \Pi_0(L) 
\end{eqnarray}
Here, 
\begin{eqnarray}
&&\Pi_0(L) = \nonumber \\
&&- 2a^{-\beta}(a+L)^{-\beta} \nonumber \\
&&- (a+1)^{-\beta} (a+1+L)^{-\beta} \nonumber \\
&&-\frac{\beta}{6}((a+1)^{-\beta}(a+1+L)^{-\beta-1} \nonumber \\
&&+(a+1)^{-\beta-1}(a+1+L)^{-\beta}) \nonumber \\ 
&&+2L^{1-2\beta} G((a+1)/L,\beta)+ \nonumber \\
&&\begin{cases}
\text{($0<\beta < 1$, $\beta \neq 1/2$)} \\
-2L^{1-2\beta} \frac{\Gamma(2\beta-1)\Gamma(2-\beta)}{(1-\beta)\Gamma(\beta)}+2\zeta(2\beta,a)  \\
(\beta=1/2) \\
2\log(L)-2\log(4)-2\psi^{(0)}(a)  \\
\end{cases} \nonumber \\
&& \label{diff_ans0}
\end{eqnarray} 
where $\zeta(\alpha,x)$ is the Hurwitz zeta function $\zeta(s,q)=\sum^{\infty}_{n=0}1/(q+n)^s$, $\psi^{(0)}(x)$ is the digamma function, $\psi^{(0)}(x)=d/dx \log(\Gamma(x))$, and $G(x,\beta)$ for $0<\beta<1$ is given by 
\begin{eqnarray}
&&G(x,\beta)= \int x^{-\beta} (x+1)^{-\beta}dx=  \nonumber \\
&&\frac{x^{1-\beta}{}_{2}F_{1}(1-\beta,\beta,2-\beta,-x)}{1-\beta} \nonumber \\
\end{eqnarray}
\textcolor{black}{where ${}_{2}F_{1}(a,b,c;x)$ is the Gaussian hypergeometric function defined by Eq. \ref{hyper_geom}.} 
Thus, the asymptotic behaviour for a large $L$ is written as 
\begin{eqnarray}
<(r_j(t+L)-r_j(t))^2> \propto 
\begin{cases}
L^{1-2\beta}  & (0 \leq \beta <0.5) \\
\log(L) &(\beta=0.5)\\
O(1) &(1>\beta > 0.5). 
\end{cases}
\nonumber \\ 
\end{eqnarray}
\par
\begin{figure*}
\begin{minipage}{0.3\hsize}
\includegraphics[width=6cm]{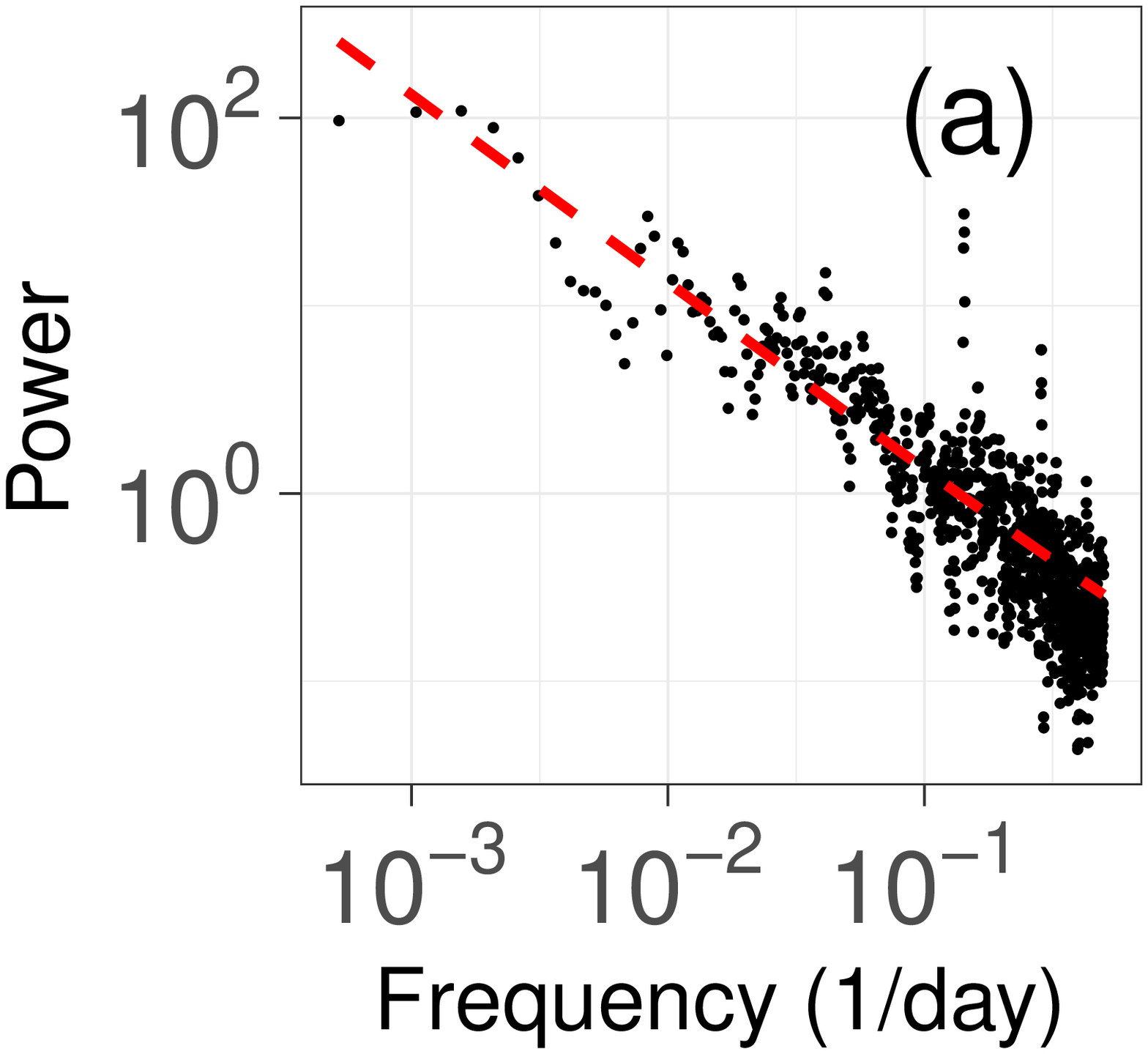}
\end{minipage}
\begin{minipage}{0.3\hsize}
\includegraphics[width=6cm]{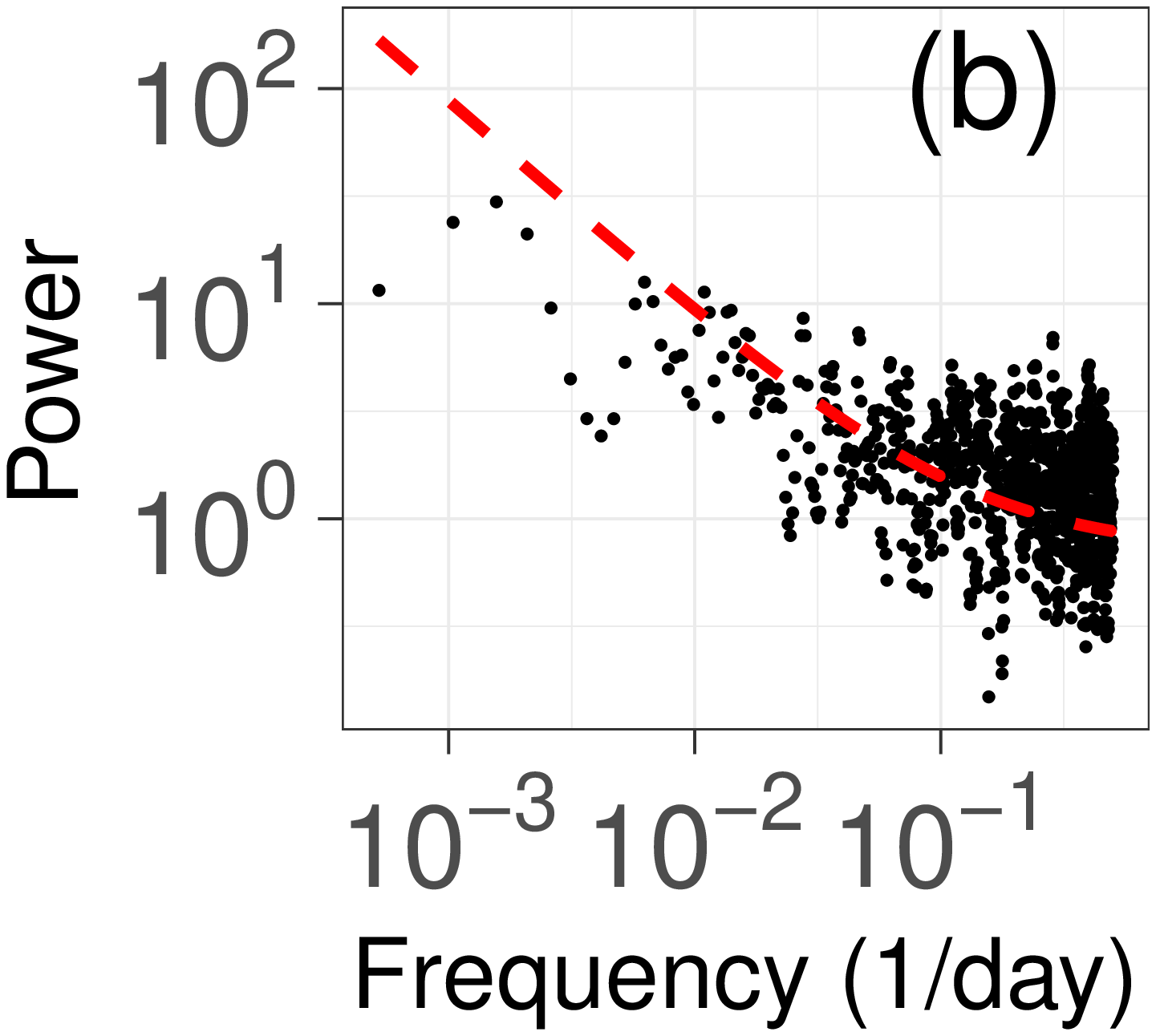}
\end{minipage}
\begin{minipage}{0.3\hsize}
\includegraphics[width=6cm]{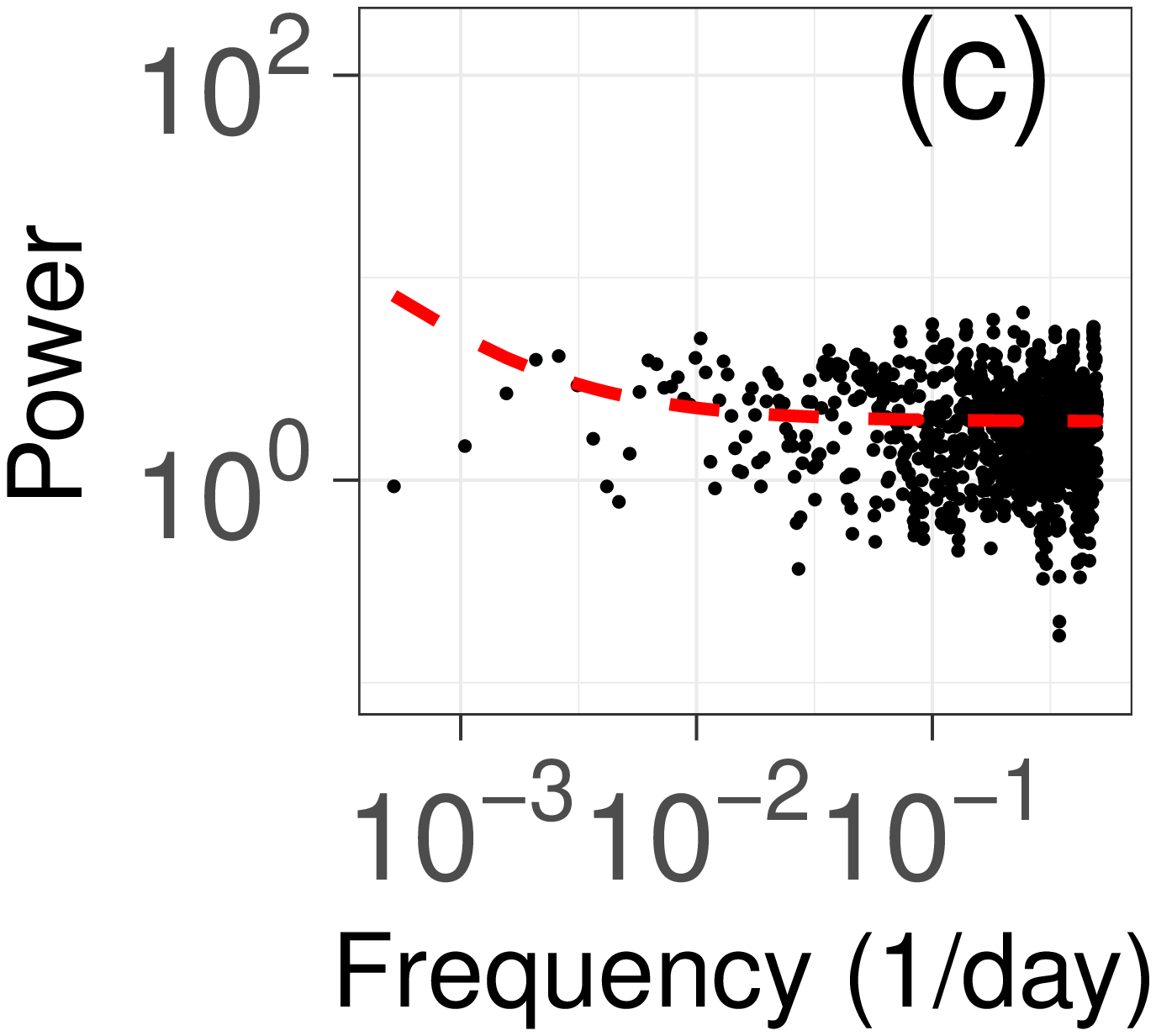}
\end{minipage}
\begin{minipage}{0.3\hsize}
\includegraphics[width=6cm]{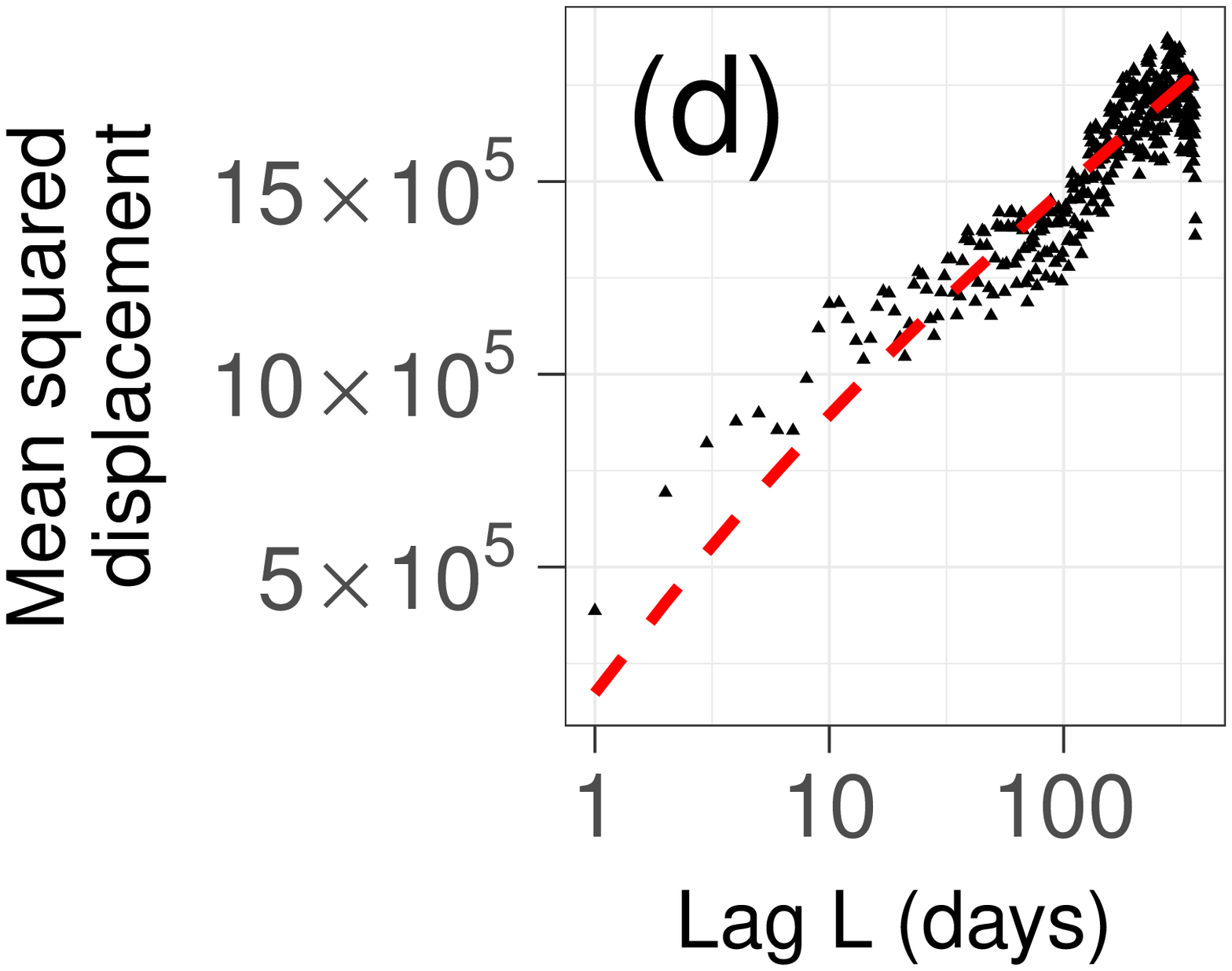}
\end{minipage}
\begin{minipage}{0.3\hsize}
\includegraphics[width=6cm]{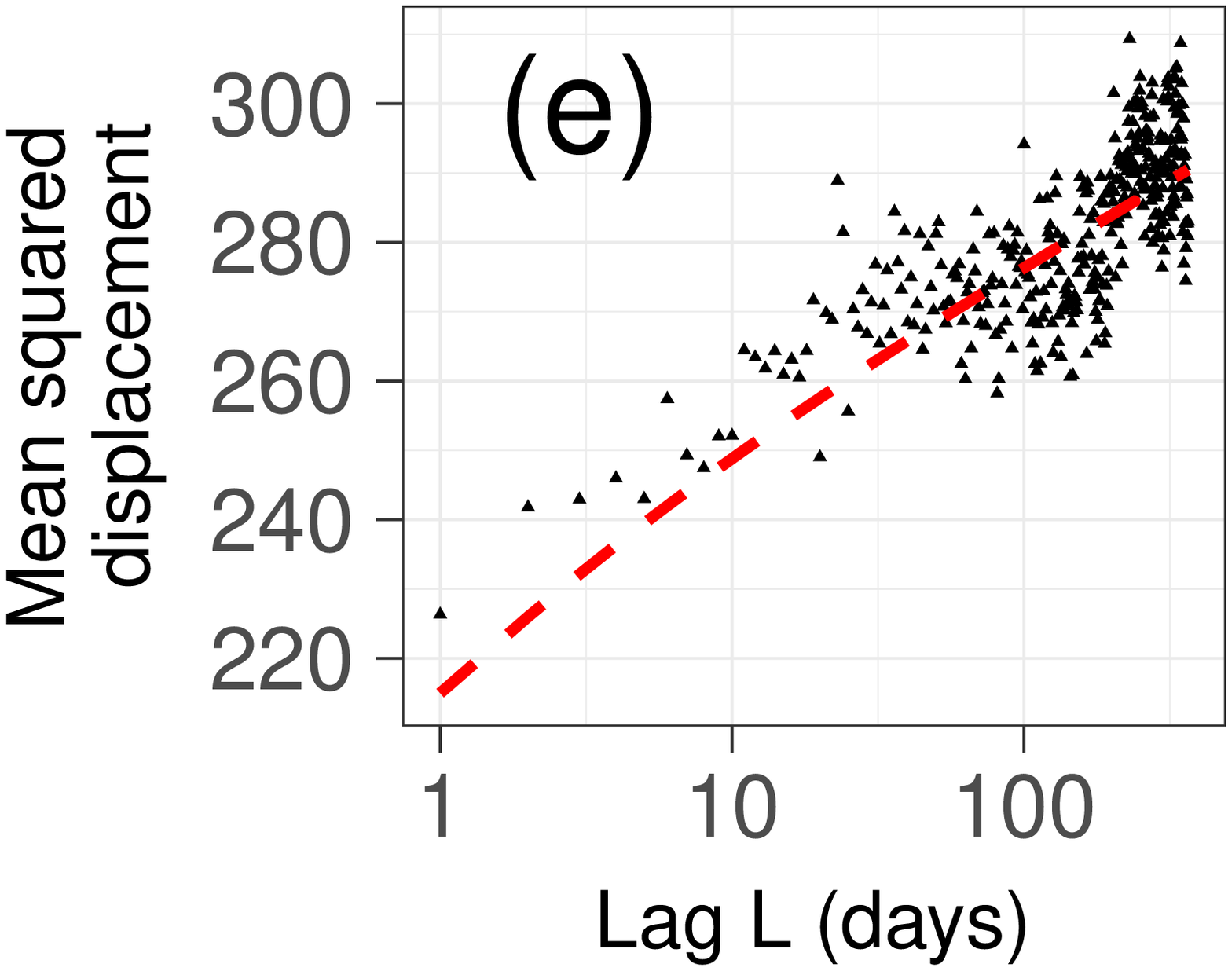}
\end{minipage}
\begin{minipage}{0.33\hsize}
\includegraphics[width=6cm]{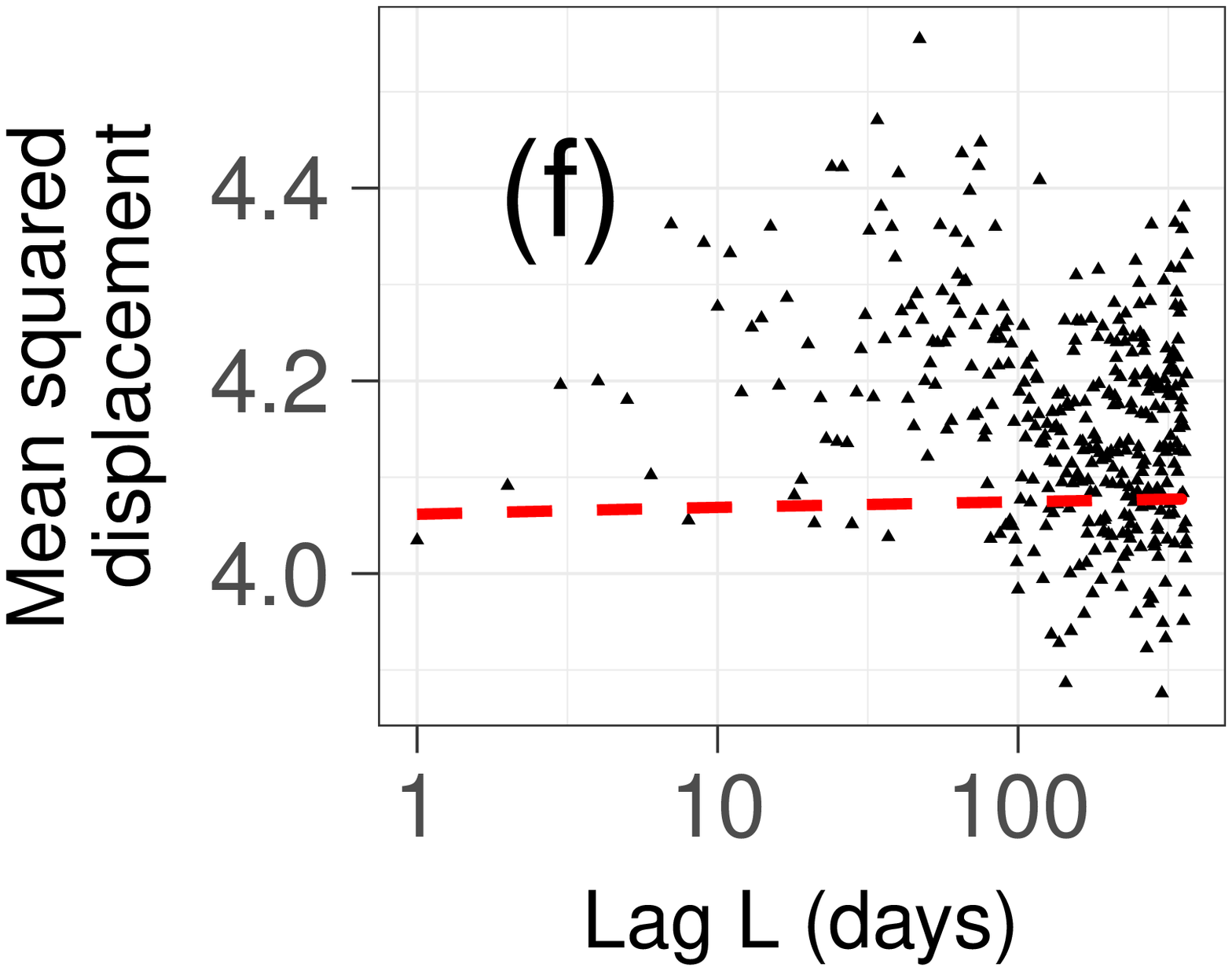}
\end{minipage}

\caption{(a)The empirical spectral density for the time series of ``ooi'' (``many'', $\check{c}_j=22369.8$; $\check{\eta}_j=0.026$, $\Delta_j^{(0)}= 0.012$). The dashed red line corresponds to the theoretical curve given by Eq. \ref{spect}. The corresponding figures for
(b)``zuzushii'' (``impudent'', $\check{c}_j=92.9$; $\check{\eta}_j=0.043$, $\Delta_j^{(0)}= 0.065$) and (c) ``komuzukasii'' (``troublesome'', ``difficult'', or ``tortuous'', $\check{c}_j=2.03$; $\check{\eta}_j=0.029$, $\Delta_j^{(0)}= 0.030$). 
From panels (a), (b), and (c), we can confirm that the theoretical curve is in good agreement with the empirical data. In addition, we can confirm that the data approach white noise with a decrease in $\check{c}_j$ (the second term of Eq. \ref{spect} is dominant for a small $\check{c}_j$). 
Note that in panels (a),(b) and (c), we plotted the scaled the spectral densities satisfied with $\int P_{f}(\nu) d\nu=1$ for comparisons. \\
(d)The empirical MSD for the time series of ``ooi'' (``many''). The dashed red line corresponds to the theoretical curve given by Eq. \ref{spect}. Corresponding figures for
(e) ``zuzushii'' (``impudent'') and (f) ``komuzukasii'' (``troublesome''). 
}
\label{fig_spect}
\end{figure*}

Next, we calculate the diffusion properties of $\{f_j(t)\}$, which is generated by the RD model based on the power-law forgetting process. 
Because $f_j(t)$ can be decomposed into independent random variables $r_j(t)$ and $w_j(t)$ by Eq. \ref{base_u2}, 
we can obtain
\begin{eqnarray}
&&\Pi[L; \{f_j(t)\}]=\Pi[L; \{r_j(t)+w_j(t)\}] \\
&&=\Pi[L; \{r_j(t)\}]+\Pi[L; \{w_j(t)\}]\nonumber  \\
&&\approx 2 a_0 \check{c}_j+(\check{\eta}_j^2/\Gamma(1-\beta)^2 \Pi_0(L)+2 {\Delta^{(0)}_j}^2 ) \check{c}_j^2. \label{MSD_f}
\end{eqnarray}
Under the condition $\Pi_0(L) >> 2 a_0 \Gamma(1-\beta)^2/(\check{c}_j \check{\eta}_j^2)$ and $\Pi_0(L) >> 2 {\Delta^{(0)}_j}^2 \Gamma(1-\beta)^2/\check{\eta}_j^2$ ($L>>1$), in the same way as $\{r_j(t)\}$,  
the asymptotic behaviour for a large $L$ is written as 
\begin{eqnarray}
<(f_j(t+L)-f_j(t))^2> \propto 
\begin{cases}
L^{1-2\beta}  & (0 \leq \beta <0.5) \\
\log(L) &(\beta=0.5)\\ 
O(1) &(\beta > 0.5) .
\end{cases} \label{MSD_f_l}
\end{eqnarray}
These results imply the following:
\begin{enumerate}
\item The first term is dominant for a small $\check{c}_j$ in Eq. \ref{MSD_f}. Thus, like a steady process, the MSD does not depend on $L$ for small $\check{c}_j$ 
superficially (see Fig. \ref{fig_spect} (f)).
\item  Eq. \ref{MSD_f_l} predicts that the MSD of the actual data, which corresponds to the model for $\beta=0.5$ obeys a logarithmic-like diffusion (ultraslow diffusion) for a large $\check{c_j}$ (see Fig. \ref{fig_msd}).
\item From Eq. \ref{MSD_f_l}, we can confirm that the parameter of the actual data, $\beta \approx 0.5$, corresponds to the boundary parameter between a stationary and non-stationary time series. 
 \end{enumerate}
 Figs .\ref{fig_spect} (d), (e), and (f) show a comparison of the MSD of the model for $\beta=0.5$ given by Eq. \ref{MSD_f} with the actual corresponding data.
 \par
To check the validity of the logarithmic diffusion, we introduce a value, $\hat{\Pi}_0(L)$, which has a universal curve as a function of $L$ (i.e., a word-independent curve), as follows:
\begin{eqnarray} 
&&\hat{\Pi}_0(L) \equiv \frac{\Pi[L; \{f_j(t)\}]-\Pi[1;\{f_j(t)\}]}{\sum^{L_{max}}_{L=1}\Pi[L; \{f_j(t)\}]-\Pi[1; \{f_j(t)\}]} \\
&&=\frac{\Pi_0(L)-\Pi_0(1)}{\sum^{L^{max}}_{L=1}\Pi_0(L)-\Pi_0(1)} \label{msd_norm} \\
&&\approx \log(L) \quad (L>>1, \beta \approx 0.5),  \label{normal_msd}
\end{eqnarray}
where $L^{max}$ is the maximum $L$ in the observation, i.e., $L^{max}=365$. \par
Fig .\ref{fig_msd}(a) shows an ensemble average of $\hat{\Pi}_0(L)$ for words in which the mean $c_j$ is above 20.
We exclude words with a small $\check{c}_j$ from the analysis because these words have a relatively large signal-to-noise ratio (see Fig.\ref{fig_spect} (f)). From this figure, we can confirm that the theoretical curve given by Eq. \ref{msd_norm} substantially agrees with the corresponding empirical data. \par
Fig. \ref{fig_msd}(b) shows a corresponding graph in which we removed the words with significant seasonality, such as ``atsui'' (``hot'' in English) and ``suzushi'' (``cool'' in English) from the word sets. In addition, Fig. \ref{fig_msd}(c) shows a corresponding figure where we additionally remove the effects of the outliers caused by a special news event, such as the great east Japan earthquake, using a trimmed mean, 
\begin{equation}
\Pi_{rob}[\{f_j(t)\}; L] \approx \frac{\sum_{\{t| \in |f_j(t+L)-f_j(t)| \leq \theta_0 \}}(f_j(t+L)-f_j(t))^2}{\sum_{\{t| \in |f_j(t+L)-f_j(t)| \leq \theta_0 \}}1}, \label{trim_mean}
\end{equation}
where we set $\theta_0=8 \times IQR(\{(f_j(t+L)-f_j)^2\})$ and $IQR(\{A\})$ as the interquartile range of sets $\{A\}$.
From these figures, we can confirm that the curves approach the logarithmic function by removing those effects that are not considered in the model, such as the effects of seasonality or special news event. \par 
%
%
%

\begin{figure*}
\begin{minipage}{0.48\hsize}
\includegraphics[width=8cm]{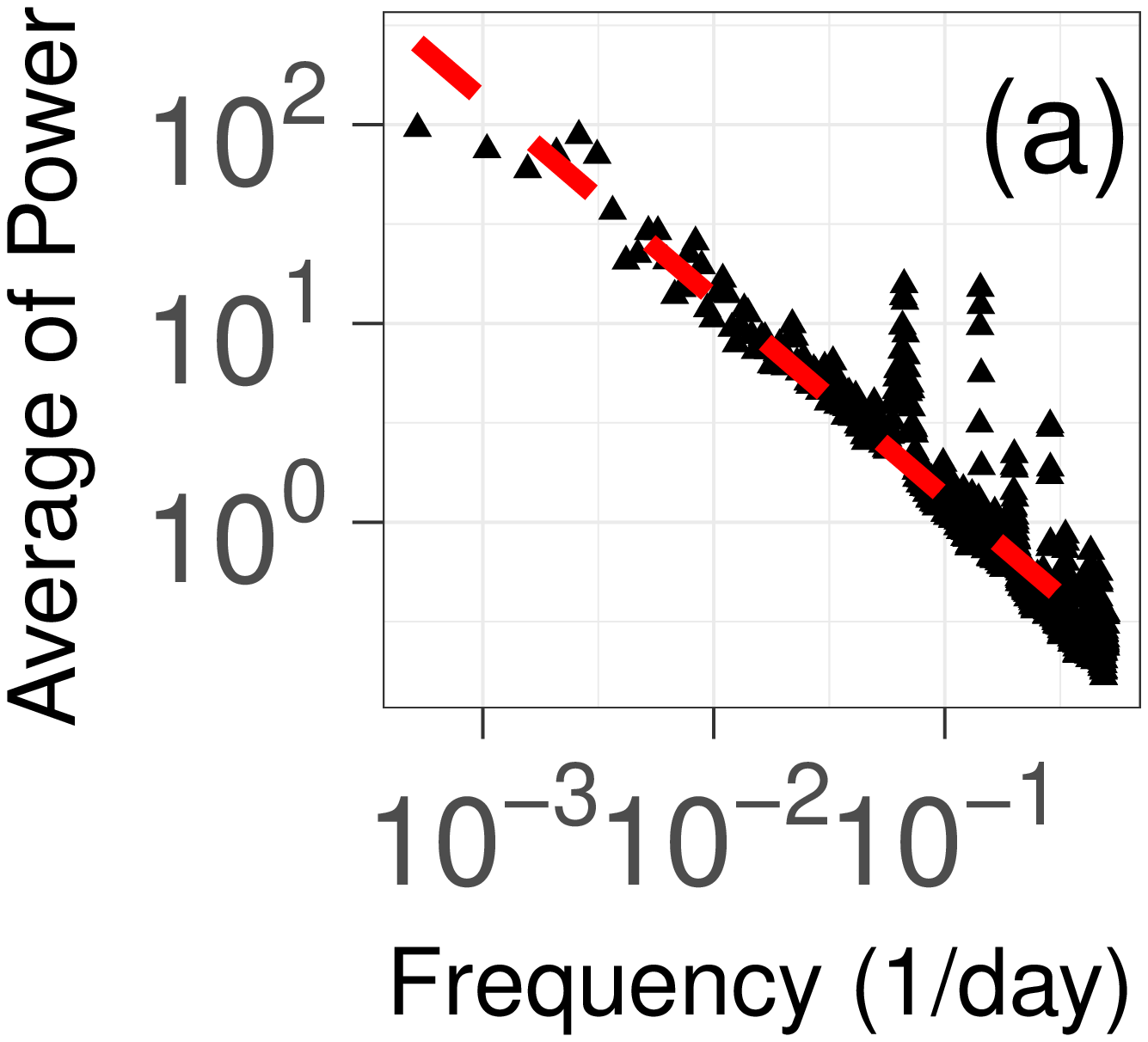}
\end{minipage}
\begin{minipage}{0.48\hsize}
\includegraphics[width=8cm]{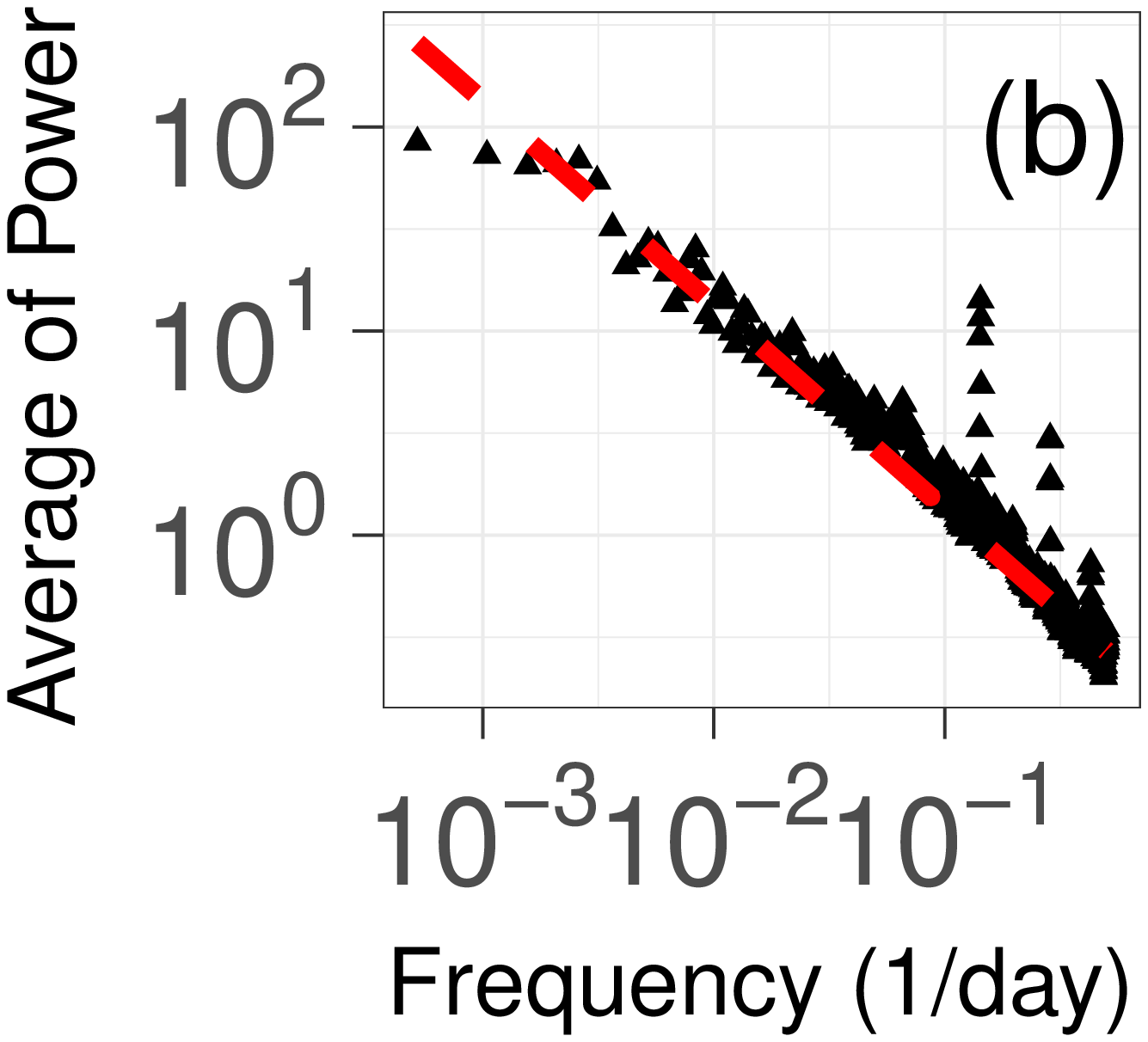}
\end{minipage}
\caption{
(a)The ensemble average of the word-independent normalised spectral density  $\hat{P}_f(\nu)$ given by Eq. \ref{normal_peri} for words with a mean $c_j$ above 20. 
\textcolor{black}{The black triangle is empirical data and the red dashed line is the theoretical curve.} 
(b)The corresponding figure of an ensemble with a 5 percent trimmed mean.
From these figures, we can confirm that the theoretical curve is almost in agreement with the empirical data.
In addition, we can also see peaks at $0.14$, $0.29$, and $0.57$ corresponding to a period of 7 days (1 week), which are not considered in the model. 
The peak at $0.067$ for the corresponding period of 15 days in panel (a) was caused by a few exceptional words that were probably posted by robots. 
}
\label{fig_norm_spect}
\end{figure*}

\begin{figure*}
\begin{minipage}{0.322\hsize}
\includegraphics[width=6cm]{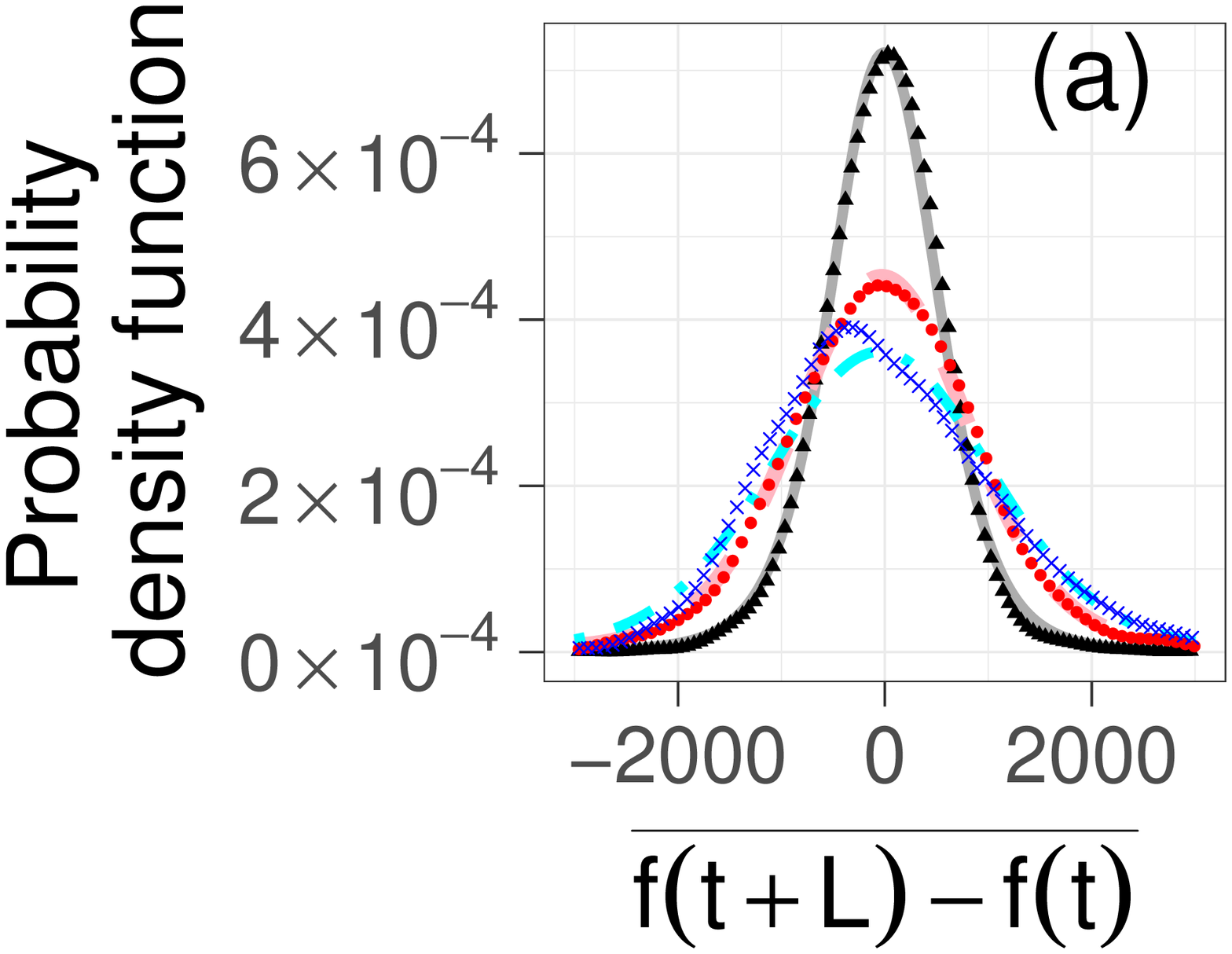}
\end{minipage}
\begin{minipage}{0.322\hsize}
\includegraphics[width=6cm]{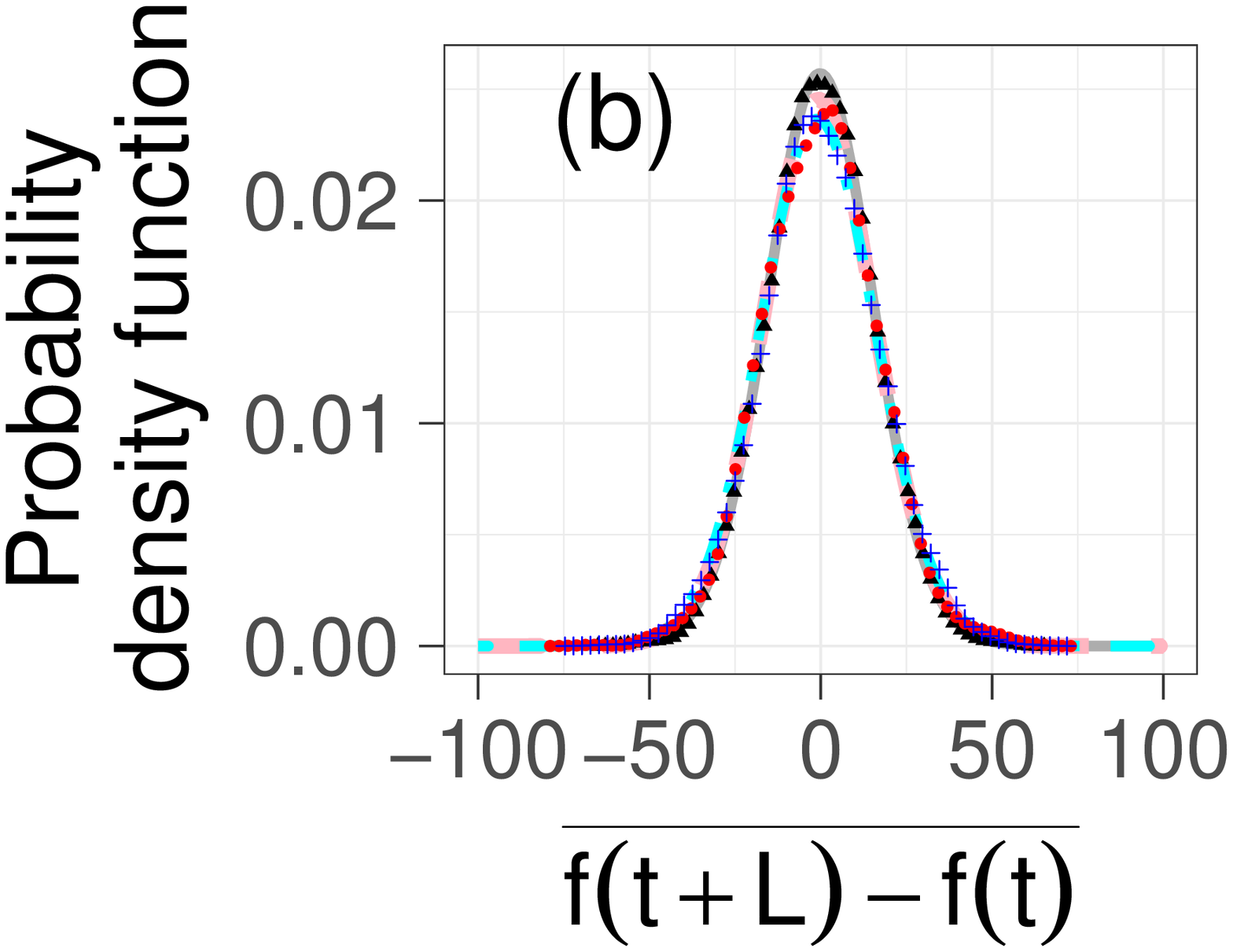}
\end{minipage}
\begin{minipage}{0.322\hsize}
\includegraphics[width=6cm]{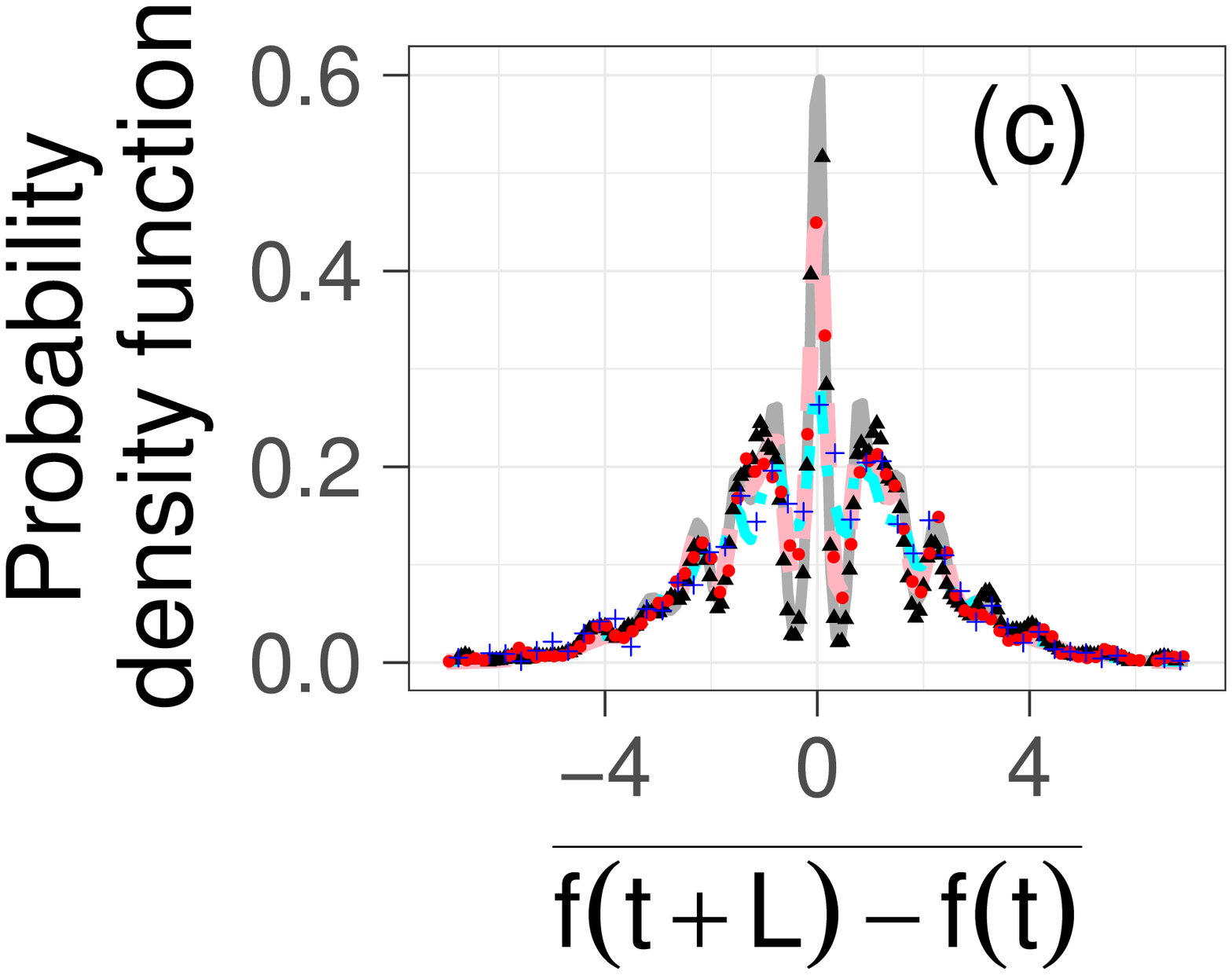}
\end{minipage}
\begin{minipage}{0.322\hsize}
\includegraphics[width=6cm]{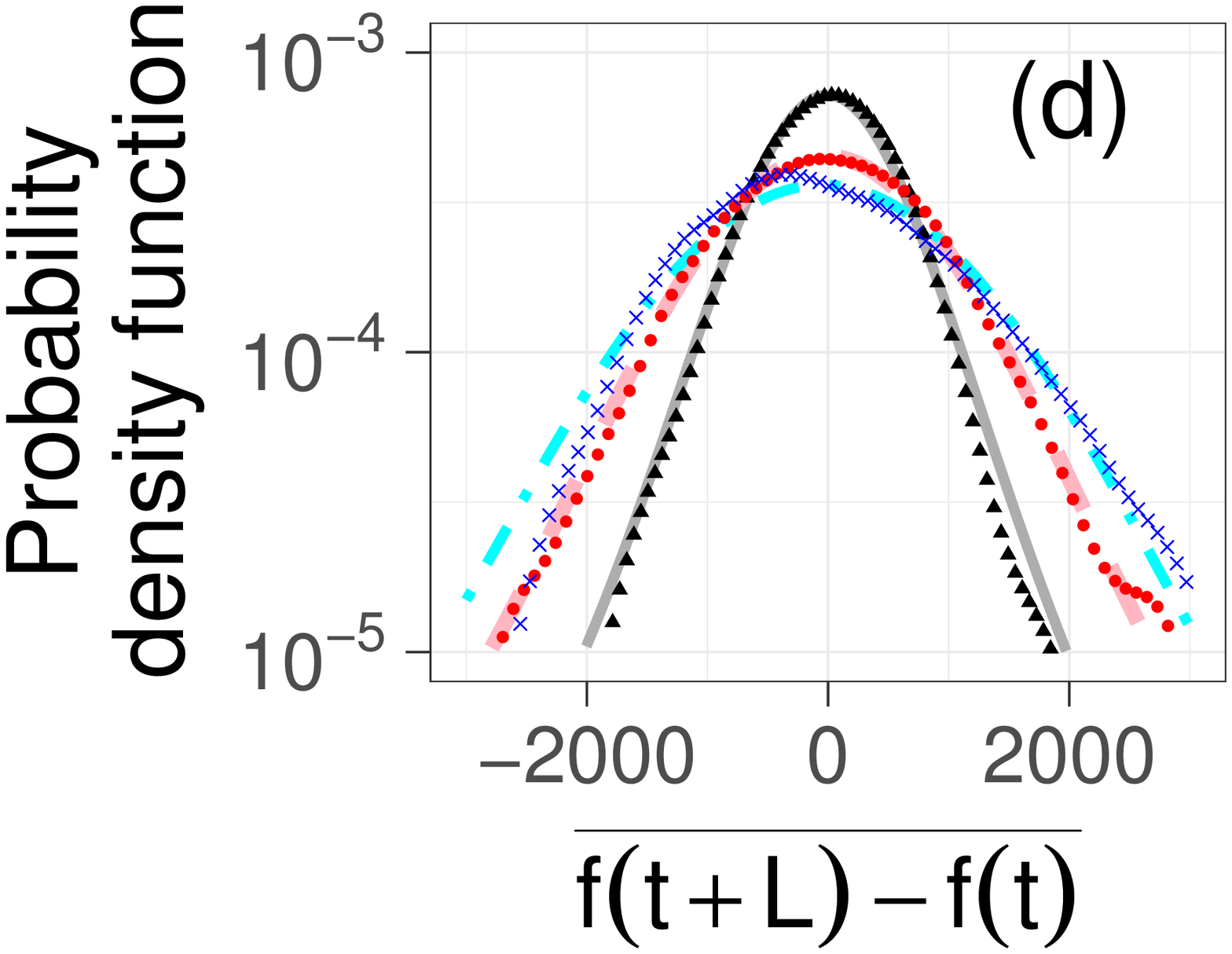}
\end{minipage}
\begin{minipage}{0.322\hsize}
\includegraphics[width=6cm]{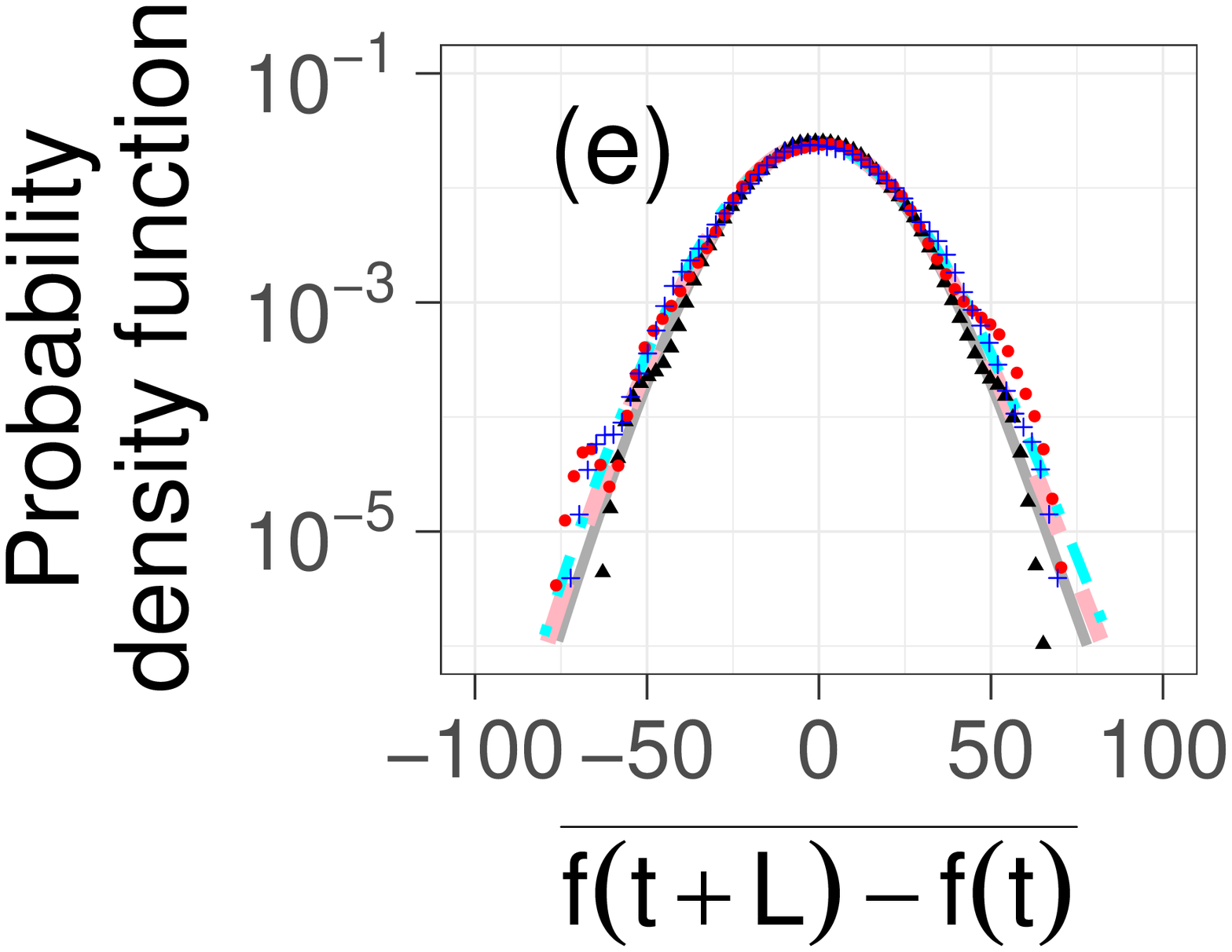}
\end{minipage}
\begin{minipage}{0.322\hsize}
\includegraphics[width=6cm]{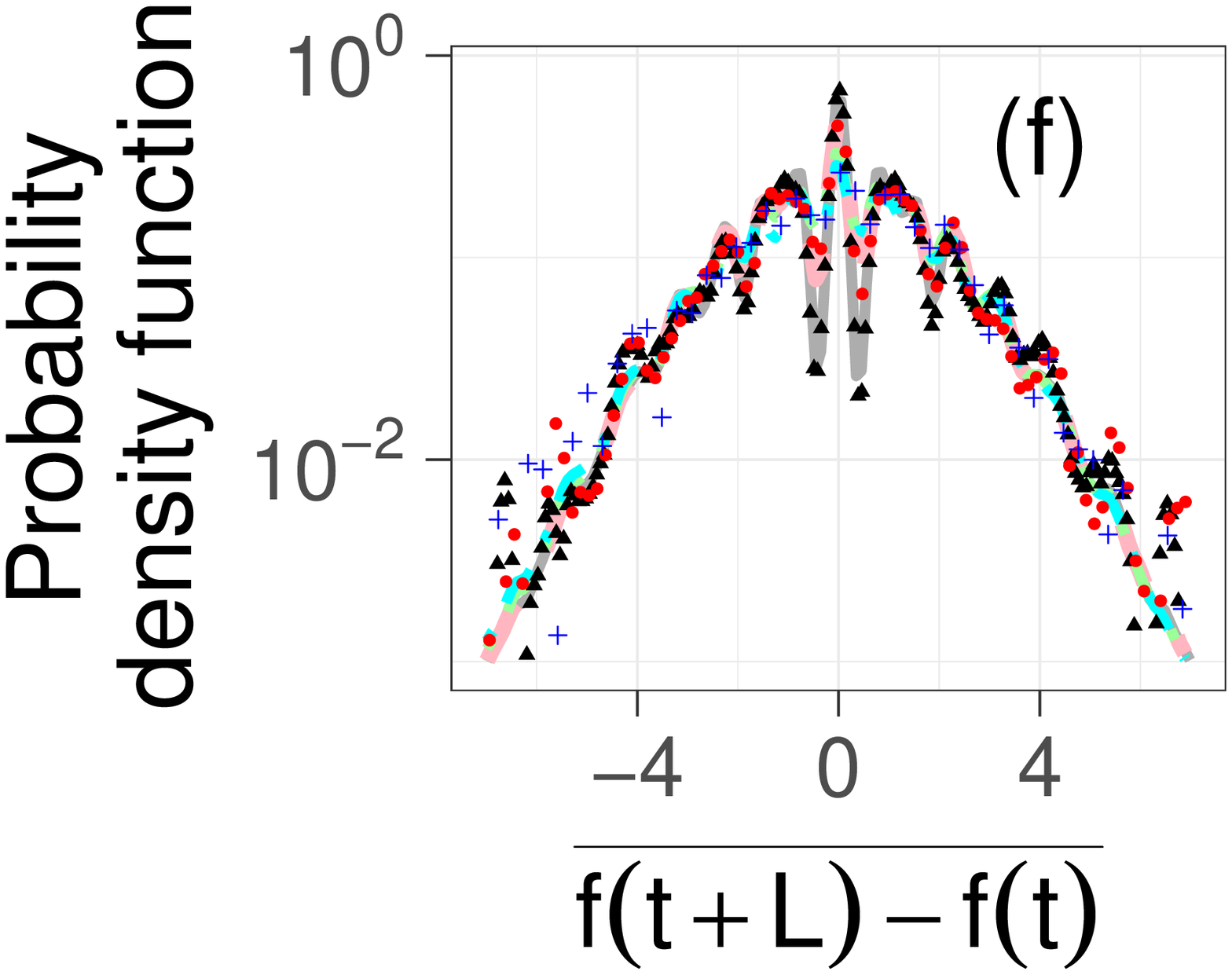}
\end{minipage}
\caption{The PDF of the centralized difference in the appearance of words $\overline{f_j(t+L)-f_j(t)}=f_j(t+L)-f_j(t)-\sum^{T-L}_{s=1}(f_j(s+L)-f_j(s))/(T-L)$ for $L=1$ (black triangles), $L=30$ (red circles) and $L=365$ (blue pluses). 
The points show for (a)``ooi'' (``many'', $\check{c}_j=22369.8$),(b)``zuzushii'' (``impudent'', $\check{c}_j=92.9$), and (c) ``komuzukasii'' (``troublesome'', ``difficult'', or ``tortuous'', $\check{c}_j=2.03$).  The lines are the corresponding theoretical curves, which are obtained by the random diffusion model based on the power-law forgetting process, given by Eq. \ref{r_forget} for $L=1$ (gray solid line) , $L=30$ (pink dashed line), and $L=365$ (cyan \textcolor{black}{dashed-dotted} line)  under the condition of $\Lambda_j(t)$ and $\eta(t)$ obeys the scaled t-distribution, whose degree of freedom is \textcolor{black}{2.64} using the same parameter sets $\check{\eta}_j$ and $\Delta_j^{(0)}$ as Fig. \ref{fig_spect}.
From panels (a), (b), and (c), we can confirm that the theoretical curves are in accordance with the empirical data.  
}
\label{fig_pdf}
\end{figure*}

\begin{figure*}
\begin{minipage}{0.32\hsize}
\end{minipage}
\begin{minipage}{0.32\hsize}
\includegraphics[width=6cm]{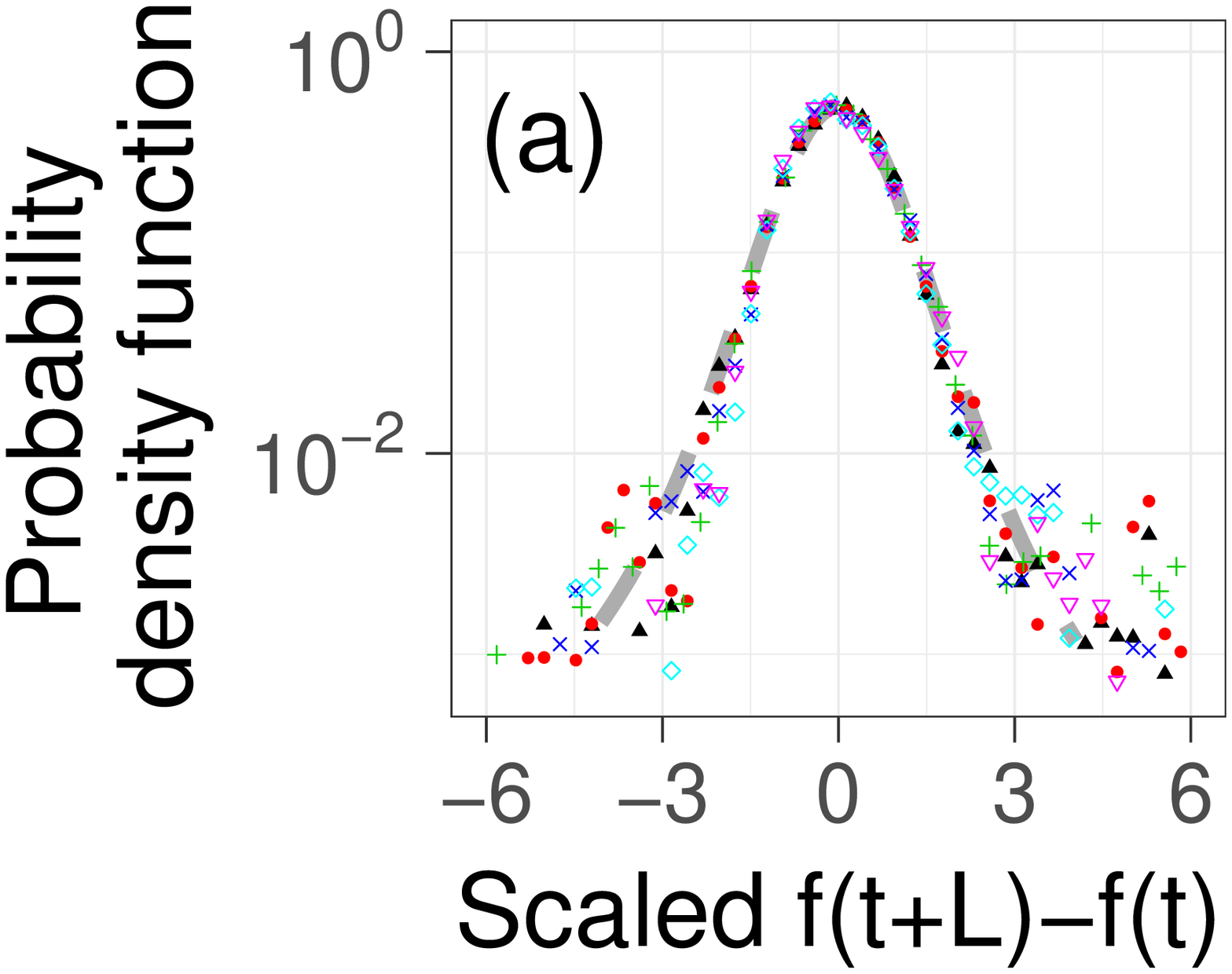}
\end{minipage}
\begin{minipage}{0.32\hsize}
\end{minipage}
\begin{minipage}{0.32\hsize}
\includegraphics[width=6cm]{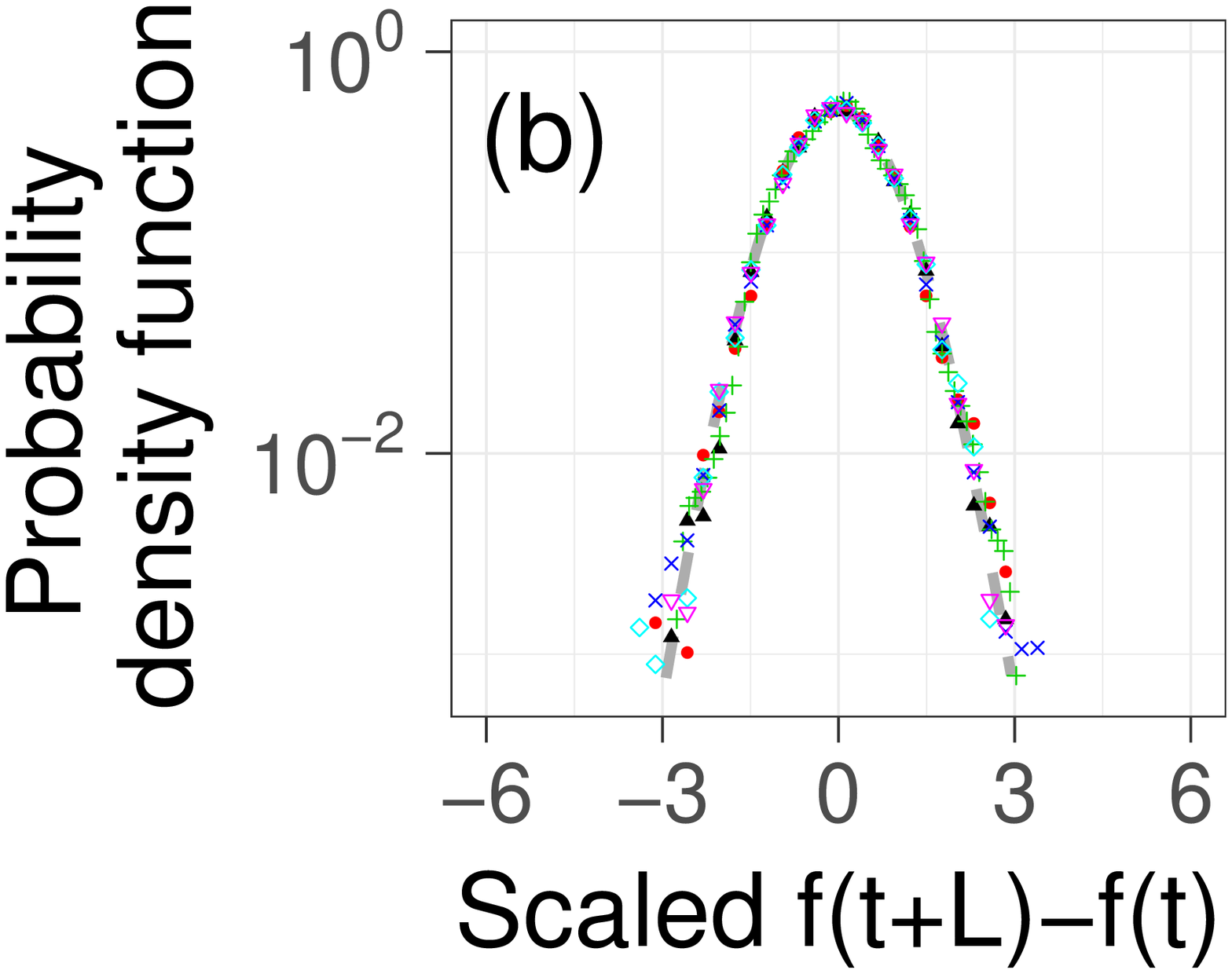}
\end{minipage}
\begin{minipage}{0.32\hsize}
\includegraphics[width=6cm]{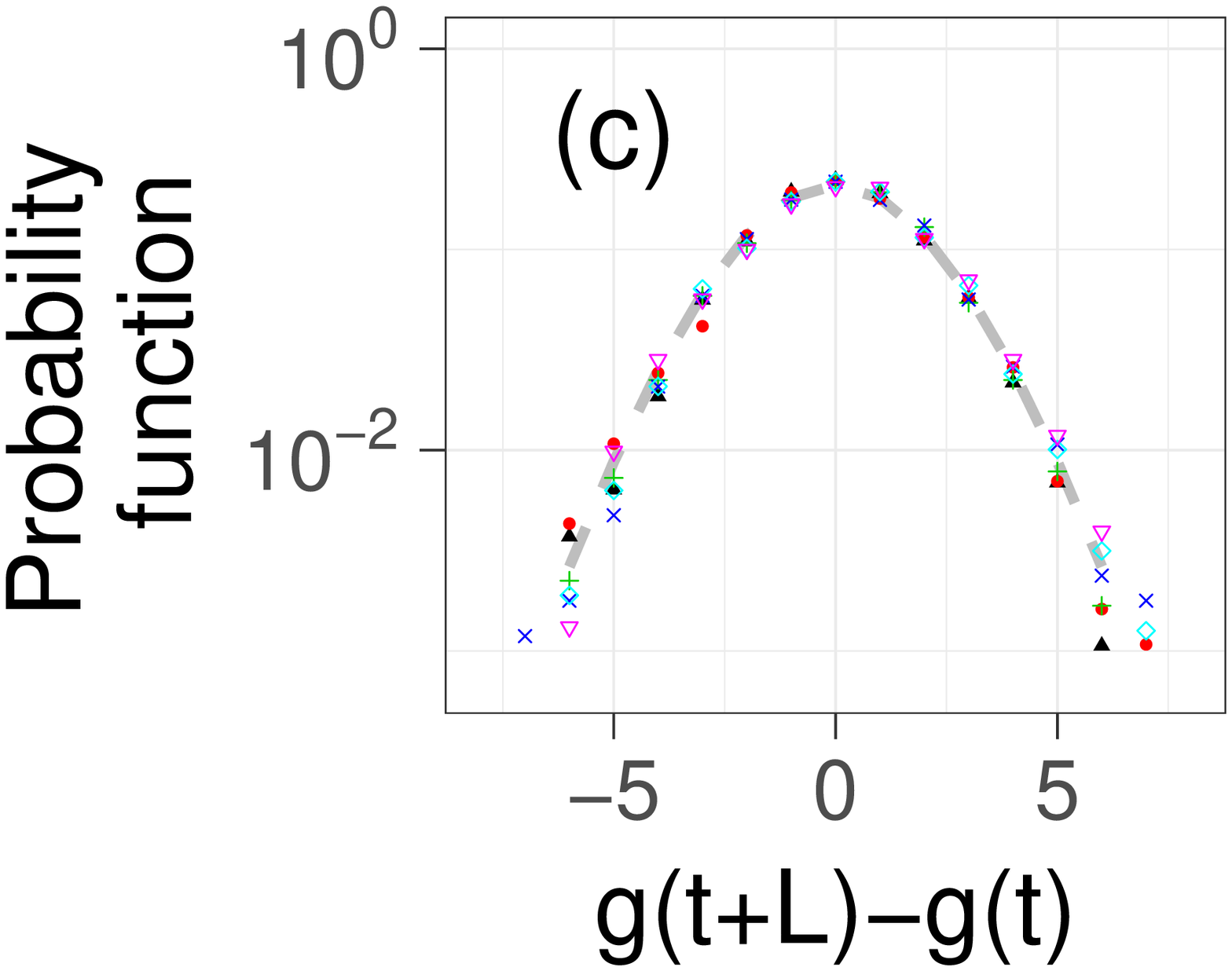}
\end{minipage}
\caption{(a)The PDF of standardized difference of appearance of words, $f_j(t+L)-f_j(t)$ (i.e, centralized by the mean and scaled by the interquartile range) in the case of ``ooi'' (``many'', $\check{c}_j=22369.8$)  for $L=1$ (black triangles), $L=30$ (red circles), $L=100$ (geen pluses), $L=200$ (blue crosses), $L=300$ (cyan diamonds) and $L=365$ (pink upside-down triangles).  The gray dashed line is the corresponding theoretical curve, which is obtained by the the random diffusion model based on the power-law forgetting process for $L=1$ given by Eq. \ref{r_forget}.  Here, we assume that $\Lambda_j(t)$ and $\eta(t)$ obey the scaled t-distribution whose degree of freedom is \textcolor{black}{whose degree of freedom is 2.64} using the same parameter sets $\check{\eta}_j$ and $\Delta_j^{(0)}$ as Fig \ref{fig_spect}. 
(b) The corresponding data are shown for ``zuzushii'' (``impudent'', $\check{c}_j=92.9$).  
From panels (a) and (b), we can confirm that all of the same shapes of PDFs are almost independent of $L$ in all of the figures.  
 (c) The corresponding probability function of appearance of words, $g_j(t+L)-g_j(t)$  in the case of ``komuzukasii''(``troublesome'', ``difficult'', or ``tortuous'', $\check{c}_j=2.03$). The gray dashed line is the Skellam distribution, which is the difference in two independent random Poisson variables, independently of $L$.
 \textcolor{black}{This distibution is  predicted by random diffusion model based on the power-law forgetting process for a small $\check{c}_j$.}
}
\label{fig_norm_pdf}
\end{figure*}
\section{Other properties}
Lastly, we compare the other commonly used time-series features of the model with the empirical data.
\subsection{Power spectrum}
The spectral density of the power-law forgetting process under the condition $\beta \to 0.5$ and $Z(\beta)=\Gamma(1-\beta)$ is given by 
\begin{equation}
P_r(\nu) \approx \frac{1}{2\pi} \check{\eta}_j^2 (2 \sin(\nu/2))^{-1},
\end{equation}
where we employed the spectral density of $ARFIMA(0,0.5,0)$.
Because $f_j(t)$ is independently decomposed into random variables $r_j(t)$ and $w_j(t)$ from Eq. \ref{base_u2}, 
the spectral density of the RD model based on the power-law forgetting process is given by
\begin{equation}
P_f(\nu) \approx \frac{1}{2\pi} \{ (\check{\eta}_j^2 (2 \sin(\nu/2))^{-1}+{\Delta^{(0)}_j}^2) \check{c}_j^2+ a_0 \check{c}_j \}. \label{spect}
\end{equation} 
Here, the spectral density of the random time series $\{x(t)\}$ is defined by
\begin{equation}
P_x(\nu)= \lim_{T \to \infty } \frac{1}{2T} \left<\left|\int_{T}^{-T} \exp(- i \nu t) x(t)  dt \right|^2 \right>, 
\end{equation}
where we denote $<A(\{x(t)\})>$ as the expectation value of $A(\{x(t)\})$ over $\{x(t)\}$.
From Figs. \ref{fig_spect}(a), (b), and (c), which show a comparison of Eq. \ref{spect} with the empirical spectral density of typical words, we can see that the theoretical curve is in agreement with the empirical data. In addition, we can confirm that the data approaches white noise with a decrease in $\check{c}_j$ (the second term of Eq. \ref{spect} is dominant for a small $\check{c}_j$). \par
Additionaly, we check the validity of Eq. \ref{spect} by the word-independent normalised value, 
\begin{eqnarray}
&&\textcolor{black}{\hat{P}_f(\nu)=\frac{(P_f(\nu)-Min[P_f(\nu)])}{\int^{\nu_{min}}_{\pi}[P_f(\nu)-Min[P_f(\nu)]]d \nu }} \\
&&=\frac{1}{2\pi} \frac{(2 \sin(\nu/2))^{-1}}{(\int^{\nu_{min}}_{\pi}(2 \sin(\nu/2))^{-1} d \nu )}. \label{normal_peri}
\end{eqnarray}
where $\nu_{min}$ is the minimum $\nu$ in the observation.
Fig. \ref{fig_norm_spect}(a) shows Eq. \ref{normal_peri} and the ensemble average of the $\hat{P}_f(\nu)$ of the actual data for words in which the mean $c_j$ is above 20, and
Fig. \ref{fig_norm_spect}(b) shows the corresponding ensemble for a 5 percent trimmed mean.
From these figures, we can confirm that the theoretical curve is almost in accordance with the empirical data.
In addition, we can also see peaks at $0.14$ ($\approx 1/7$), $0.29$ ($\approx 2/7$), and $0.57$ ($\approx 4/7$) corresponding to a period of 7 days (1 week), which are not considered in the model. 
Note that the peak at $0.067$ of the corresponding 15-day period in Fig. \ref{fig_norm_spect}(a) was caused by a few exceptional words that were probably posted by robots. 
\subsection{Probability density function}
In the above discussions, we discussed only the summary statistics.   
Herein, we investigate the probability distribution function (PDF) of the time series directly.
Fig. \ref{fig_pdf} shows a comparison of the PDFs of the time series for examples of actual data with those of the theoretical model given by Eq. \ref{pdf_ans} in Appendix \ref{app_pdf} for $L=1$, $L=7$, and $L=365$ \textcolor{black}{under the condition that $\Lambda_j(t)$ and $\eta(t)$ obey the scaled t-distribution whose degree of freedom is 2.64.} 
We can confirm that observations are also in agreement with the theoretical curves. 
 Note that the PDFs of words with unmodelised effects such as the seasonality or continued increase deviate from theory. \par
 Figs. \ref{fig_norm_pdf} provide the PDFs scaled by the width (interquartile range).  From this figure, we can see that the shapes of the PDF are almost unchanged regardless of the lag time $L$. From the viewpoint of our model, in the case of a large $\check{c}_j$, the shape of the distribution is unchanged because the speed of convergence of the distribution toward a normal distribution is decreased by the uneven distributed weights ($\beta=0.5$) of the sum in Eq \ref{r_forget}. \textcolor{black}{ Note that for $\beta<0.5$,  this discussion implies that the PDF of the difference in the time series of word appearance $r_j(t+L)-r_j(t)$ does not converge toward a universal distribution (i.e., Gaussian), which is independent of the detailed structure for even a very large $L$.} 
 \textcolor{black}{On the other hand, in the case of a very small $\check{c}_j$, $v_j(t+L)-v_j(t)$ approximately obeys the Skellam distribution, which is the difference in two independent random Poisson variables, independently of $L$ because $w_j(t)$, which obeys a Poisson distribution ($O(\sqrt{\check{c}_j})$) under this condition \cite{RD_base}, is dominant in Eq. \ref{base_u2}.}
\begin{table*}
\begin{tabular}{lccc}
\hline
& Random walk ($\beta=0$) & Blog time series ($\beta=0.5$) & IID noise ($\beta=1$)  \\
\hline 
 \multicolumn{4}{l}{(i) Time evolution} \\ 
Difference form; $Dx(t) \equiv x(t)-x(t-1)$; & $D^1 x(t)= \eta(t)$ &  $\sqrt{D} x(t) \approx \eta(t)$ & $D^0 x(t)=x(t)= \eta(t)$  \\
Summation form ; $x(t)=\eta(t)+\sum^{s=\infty}_{s=1} \theta(s) \eta(t-s)$; &  $\theta(s) \propto 1$ & $ \theta(s) \propto s^{-0.5}$ & $ \theta(s)=0$  \\
 \multicolumn{4}{l}{(ii)Dynamics statistics} \\ 
MSD  $<(r(t+L)-r(t))^2>$ for $L>>1$ & $L$ & $\log(L)$ & $L^0$ \\
Power Spectrum & $\nu^{-2}$ & $\nu^{-1}$ & $\nu^{0}$  \\  
 \multicolumn{4}{l}{(iii)Fulctuation scaling coefficients}  \\ 
$a(t;L)$ & \multicolumn{3}{c}{$L^{-1}$} \\  
 $b(t;L)$ & $L$  &  $L^{0}$ & $L^{-1}$  \\ 
 \multicolumn{4}{l}{(iv)Shape of distribution }  \\ 
$\{r(t-L)-r(t)\}$ & Normal & Normal? & $\eta$-depend \\
$g(t)$ for $c<<1$ & \multicolumn{3}{c}{Poisson}  \\ 
$f(t+L)-f(t)$ for $c>>1$ & $\Lambda$ and $\eta$-depend &  $\Lambda$ and $\eta$-depend & $\Lambda$ and $\eta$-depend \\
\hline
\end{tabular}
\caption{Summary of the model properties for $Z(\beta)=\Gamma(1-\beta)$} \label{table1}
\end{table*}
\section{Conclusions and discussion}
In this paper, we investigate what dynamics govern the appearances of already popularized words in nation-wide blogs.
In other words, we investigate the pure elementary dynamics, from which word-dependent special effects such as breaking news, increasing (or decreasing) concerns, or seasonality are segregated. \par
Through an analysis of nation-wide Japanese blog data, we found that a word appearance can be explained using the random diffusion model based on the power-law forgetting process, which is a type of long memory point processes related to ARFIMA (0,0.5,0), and we found that the diffusion can be approximated through ultraslow diffusion (i.e., the mean squared displacement grows logarithmically), which is given by Eq. \ref{rd_model} and Eq. \ref{r_forget} intrinsically as follows: 
%
%
%
\begin{enumerate}
\item For a small temporal mean $\check{c}_j$, the distribution of a row word appearance $g_j(t)$ approximately obeys the steady Poisson process in a superficial manner. Correspondingly, the MSD is nearly non-dependent on $L$ and the white spectral density (Figs. \ref{fig_spect}(c) (f), Figs. \ref{fig_pdf} (c) (f), and Table \ref{table1}).
\item  For a large temporal mean $\check{c}_j$, the width of the PDF of the distribution of the difference in words count $f_j(t+L)-f_j(t)$ is increasing depending on the lag $L$. This increase is related to the ``ultraslow diffusion'' (i.e., the mean squared displacement grows logarithmically), which is predicted by the power-law forgetting process for $\beta=0.5$. (Fig. \ref{fig_msd}, Figs. \ref{fig_spect} (a) (d), Figs. \ref{fig_pdf}(a) (d), and Table \ref{table1}).
\item For the any temporal mean  $\check{c}_j$, the model can consistently reproduce statistical properties of the blog time series: (i) the fluctuation scaling, (ii) MSD, (iii) spectrum density, and (iv) the shapes of the probability density functions. These properties are intermediate between the cases of a large  $\check{c}_j$ and small $\check{c}_j$ (Fig. \ref{fig_TFS}, Figs. \ref{fig_spect}(b) (e), and Figs. \ref{fig_pdf} (b) (e)). 
\item Based on the result of our model in Eq. \ref{MSD_f_l}, the actual time series, which corresponds to the model for the parameter $\beta \approx 0.5$, is within the parameter boundary between a stationary and non-stationary time series. In addition, because the model can be approximated by $ARFIMA(0,0.5,0)$ (see Eq. \ref{arfima2}), the blog time series may be able to be interpreted as a half order integral of white noise for $\beta \approx 0.5$.
 \end{enumerate} 
 \par
In this study, we only examined adjectives on blogs for $L \leq 364$ owing to a data limitation. 
Thus, it is necessary to examine other parts of speech and other corpuses in order to know the applicability of our theoretical framework for general language phenomena.
However, because our theoretical framework does not use the peculiarity of adjectives, a wide applicability is expected.
\par
 Although our model can explain the dynamical properties of the blog time series, our framework cannot explain the model parameter $\beta \approx 0.5$ in Eq. \ref{r_forget}. It is also necessary for the theme to clarify the origin of the parameter of the speed of forgetting $\beta \approx 0.5$, which may be related to not only complex system science but also to neuroscience. \par
As far as we know, the only example observation of ``ultraslow diffusion'' (i.e., in which the mean squared displacement grows logarithmically) in human or animals behaviour is the mobility of humans and monkeys.  \cite{song2010modelling,boyer2011non}. \textcolor{black}{In addition, we could hardly find empirical studies of ultraslow diffusion even in material science, in spite of huge amount of theoretical studies.
}  
We think one of the possible reasons for the few observations of “ultraslow diffusion” in non-material phenomena is the difficulty in distinguishing between logarithmic behaviour and steady behaviour based on insufficient accuracy. 
However, nowadays, it is expected that high-precision data accumulation of human behaviour will remedy this difficulty.
We hope that our study will contribute to quantitative studies of very slow changes or ``almost'' stationary phenomena regarding the methodology of precise observations and the empirical example.
\begin{acknowledgments}
The authors would like to thank Hottolink, Inc. for providing the data. This work was supported by JSPS KAKENHI, Grant Number \textcolor{black}{JP17K13815}.
\end{acknowledgments}
%
\bibliography{adj6c_pnas}
\appendix
\renewcommand{\theequation}{A\arabic{equation}}
\renewcommand{\thefigure}{A-\arabic{figure}}
\renewcommand{\thetable}{S-A\arabic{table}}
\setcounter{figure}{0}
\setcounter{equation}{0}
\section{Estimation of $b(L)$}
\label{app_coef}
We use the following procedure to estimate $b(L)$ given by Eq. \ref{V_df_emp} from the actual data, 
\begin{enumerate}
\item We fix $L$.
\item We calculate $E[F_j^{(L)}]$ and $V[\delta F_j^{(L)}]$ for all words $(j=1,2,\cdots,W)$. 
\item We miminize $Q^{(L)}$ with respect to $b(L)$ under the condition $b(L)>0$. 
\end{enumerate}
Here, $Q^{(L)}$ is defined by
\begin{eqnarray}
&&Q^{(L)} \equiv s \cdot \{ \sum_{j \in \{j|Q_j^{(L)}>0, C_j \geq 100\}}\frac{{Q_j^{(L)}}^2}{N_p}\} \nonumber \\
&& +(1-s) \cdot \{\sum_{j \in \{j|Q_j^{(L)}<0, C_j \geq 100\}} 
 \frac{{Q_j^{(L)}}^2}{N_m} \}
\end{eqnarray}
where  
\begin{equation}
Q_j^{(L)}=\log{(V[\delta F_j^{(L)}(\tau^{L})]^{1/2})}-\log{((\frac{2}{L} \cdot C_j+b(L) \cdot C_j^2)^{1/2})} \label{QjL},
\end{equation}.
$N_p=\sum_{j \in \{j|Q_j^{(L)}>0, C_j \geq 100\}} 1$ and $N_m=\sum_{j \in \{j|Q_j^{(L)}<0, C_j \geq 100\}}1$
Note that the minimization of the first term of Eq. \ref{QjL} corresponds to a reduction of the data beneath the theoretical lower bound in Eq. \ref{V_df_emp}.
However, when we use only the first term, the estimation of $b(L)$ is strongly affected by outliers.  
Thus, we use the second term in order to accept the data beneath the theoretical curve, and $s$ is the parameter used to control the ratio of acceptance.  
Here, we use $s=0.9$ in our analysis. In addition, the reason why we only use $c_j \geq 100$ is that we neglect words with a small $c_j$, which do 
not affect the estimation $b(L)$ (see Eq. \ref{V_df_emp} and Fig. \ref{fig_TFS}). 
\par
\renewcommand{\theequation}{B\arabic{equation}}
\renewcommand{\thefigure}{B-\arabic{figure}}
\renewcommand{\thetable}{S-B\arabic{table}}
\setcounter{figure}{0}
\setcounter{equation}{0}
\section{$V[\delta F^{(L)}_j]$ for given $\{r_j(t)\}$} 
\label{app_long_rd}
We calculate $V[ \delta F_j^{(L)}]$ for given $\{r_j(t)\}$. 
Using Eq. \ref{large_f}, we can decompose $V[ \delta F_j^{(L)}]$: 
\begin{eqnarray}
&&V[ \delta F_j^{(L)} (\tau^{(L)})]=\check{C_j}^2 V[\delta R_j^{(L)}]+ V[\delta W_j^{(L)}] \nonumber \\
&&+ 2 \cdot \check{C}_j E[ \delta R_j^{(L)}  \delta W_j^{(L)}]. \label{deltaF_d}
\end{eqnarray}
\par
First, we calculate the second term in Eq. \ref{deltaF_d}.  
The second term is written as $V[\delta W_j^{(L)}]=E[\delta {W_j^{(L)}}^2]-E[\delta {W_j^{(L)}}]^2$．\par
Here, $E[\delta {W_j^{(L)}}^2]$ is given by 
\begin{eqnarray}
&&E[\delta {W_j^{(L)}}^2]= \frac{1}{T^{(L)}-1} \sum^{T^{(L)}-1}_{t^{(L)}=1} \{W_j^{(L)}(t^{(L)}+1) \nonumber \\
&&-W^{(L)}(t^{(L)})\}^2  \\
&\approx& \frac{1}{L} \{ \sum^{T}_{t=L+1}\frac{w_j(t)^2}{T-L} + \sum^{T-L}_{t=1} \frac{w_j(t)^2}{T-L} \},  \label{E_W3}
\end{eqnarray}
where we use the assumption that $T>>1$, $<w_j(t)>=0$, and $\{w_j(t)\}$ are independently distributed random variables.
Approximating the sums in Eq. \ref{E_W3} by Eq. \ref{ww}  (using the assumption $T>>1$ and $L \leq T/2$), we write 
\begin{equation}
E[\delta {W_j^{(L)}}^2] \approx \frac{2}{L}\{ E[1/m] \cdot \check{c}_j + {\Delta^{(0)}_j}^2  (1+V[r_j]) \check{c}_j^2 \}.   \label{EWW}
\end{equation}
In the calculation, we also use the approximations,  
\begin{eqnarray}
&& \sum^{T}_{t=L+1} \frac{r_j(t)}{T-L} \approx 1 \\
&& \sum^{T-L}_{t=1} \frac{r_j(t)}{T-L} \approx 1 
\end{eqnarray}
and
\begin{eqnarray}
\bar{V}[r_j]&=&  \frac{1}{2}\{\sum^{T}_{t=L+!}\frac{\{r_j(t)-\sum^{T}_{t=L+1}\frac{r_j(t)}{T-L}\}^2}{T-L} \nonumber \\
&&+ \sum^{T-L}_{t=1}\frac{\{r_j(t)-\sum^{T-L}_{t=1}\frac{r_j(t)}{T-L}\}^2}{T-L}\} \nonumber \\
&\approx & V[r_j].
\end{eqnarray}
These approximations are based on the assumption that $T>>L$ and $\{r_j(t)\}$ do not have a particular trend.
 \par
Next，we calculate $E[\delta {W_j^{(L)}}]^2$. Using Eq. \ref{W0}, we can estimate $E[\delta {W_j^{(L)}}]^2$ as follows:
\begin{eqnarray}
E[\delta {W_j^{(L)}}]^2&=&\{\frac{W_j^{(L)}(T^{(L)})-W_j^{(L)}(1)}{T^{(L)}-1} \}^2 \nonumber  \\
&& \approx \{\frac{O(1/\sqrt{L})}{T^{(L)}-1} \}^2 \approx O(\frac{L}{(T-L)^2}) \\
&&\approx
\begin{cases}
 O(1/T^2) &(L<<T) \\
 O(1/T)  & (L \approx T).
\end{cases} 
\label{EWEW}
\end{eqnarray}
Thus, we can neglect this term for $T>>1$. \par
Consequently, inserting Eq. \ref{EWW} and Eq. \ref{EWEW} into Eq. \ref{deltaF_d}, and using $E[\delta C_j \delta W_j] \approx O(\sqrt{1/T}) \approx 0$ $(T>>1)$, 
we can obtain 
\begin{eqnarray}
&&V[\delta F_j^{(L)}] \approx a(L) \check{C}_j +b(L) \check{C}_j^2  \label{df_ans0}, 
\end{eqnarray}
where 
\begin{equation}
a(L)=\frac{2}{L} a_0
\end{equation}
and 
\begin{equation}
b(L)= V[\delta R_j^{(L)}]+\frac{2b_0(1+V[r_j])}{L}.
\end{equation}
Here, \textcolor{black}{$a_0=E[1/m]$ and $b_0={\Delta^{(0)}_j}^2$}.
%
\renewcommand{\theequation}{C\arabic{equation}}
\renewcommand{\thefigure}{C-\arabic{figure}}
\renewcommand{\thetable}{S-C\arabic{table}}
\setcounter{figure}{0}
\setcounter{equation}{0}
\section{$V[\delta R^{(L)}_j]$ for a random walk}
\label{app_random_walk}
We caluclate $V[\delta R^{(L)}_j]$ for the following random walk with dissipation $\kappa \geq 0$ and external force $u(t)$, 
\begin{equation}
r(t+1)=\kappa  \cdot r(t)+u(t)+\eta(t),  \label{s_ran_a}
\end{equation}
where $<\eta(t)>_{\eta}=0$, $<(\eta(t)-<\eta(t)>_{\eta})^2>_{\eta}=\check{\eta}^2<<1$ $u(t)>0$ and $\kappa \geq 0$ and we omit the subscript $j$. 
Using Eq. \ref{s_ran_a}, $R(I)$, defined by \ref{sd_L_0}, is written as 
\begin{eqnarray}
&& R(I) = \frac{1}{L} \sum^{(LI-1)}_{t'=(I-1)L}[\sum^{t'-1}_{k=0} \kappa^k u(1+t'-k)  \nonumber \\
&& + \kappa^{t'} r(1) + \sum^{t'-1}_{k=0}\kappa^k \eta(1+t'-k)] \nonumber \\
&& = P_1(I)+P_2(I)+P_3(I).
\end{eqnarray}
Here, we define $P_1(I)$, $P_2(I)$, and $P_3(I)$ as follows:
\begin{eqnarray}
P_1(I) \equiv \frac{1}{L}  \sum^{(LI-1)}_{t'=(I-1)L} \sum^{t-1}_{k=0} \kappa^k  \cdot u(1+t'-k) \label{p1i}
\end{eqnarray}
\begin{eqnarray}
P_2(I) \equiv \frac{1}{L} \sum^{(LI-1)}_{t'=(I-1)L} \kappa^{t'} \cdot r(1) \label{p2i}
\end{eqnarray}
\begin{eqnarray}
P_3(I) \equiv \frac{1}{L}  \sum^{(LI-1)}_{t'=(I-1)L} \sum^{t-1}_{k=0} \kappa^k \cdot \eta(1+t'-k).  
\end{eqnarray}
Because $V[\delta R^{(L)}]$ can be decomposed
\begin{equation}
V[\delta R^{(L)}_j] =V[\delta R]=E[\delta R^2]-E[\delta R]^2, \label{VR_rand}
\end{equation}
we calculate $E[\delta R^2]$ and $E[\delta R]^2$, respectively. \par
{\bf Calculation of $E[\delta R^2]$.}
Here, we calculate the first term of Eq. \ref{VR_rand}, $E[\delta R^2]$.
$\delta R(I)^2$ is denoted by 
\begin{eqnarray}
 \delta R_j(I)^2 &=& (\delta P_1(I)+ \delta P_2(I)+ \delta P_3(I))^2 \\
&\approx& \delta P_1(I)^2+ \delta P_2(I)^2+\delta P_3(I)^2+ 2 \cdot \delta P_1(I) \cdot \delta P_2(I) \nonumber. \\ \label{RJ2}
\end{eqnarray}
\par
We estimate the effects of the first term in Eq. \ref{RJ2}, $\delta P_1(I)^2$.
$\delta P_1(I)$ can be written as
\begin{equation}
\delta P_1(I)=\frac{1}{L^2} \cdot \sum^{L(I+1)}_{t=2}g_0(t,I) u(t), 
\end{equation}
where 
\begin{eqnarray}
g_0(t,T) = 
\begin{cases}
\frac{ (\kappa^L-1)^2 \cdot \kappa^{-t+1} \cdot \kappa^{(I-1)L} }{\kappa-1}  & 2 \leq t \leq (I-1)L+1 \\ 
\frac{(\kappa^L-2) \cdot \kappa^{(-t+1)} \cdot \kappa^{I \cdot L}+1}{\kappa-1} & (I-1)L+2 \leq t \leq LI+1  \\
\frac{\kappa^{(-t+1)} \cdot \kappa^{(I+1)L}-1}{\kappa-1} & IL+2 \leq t \leq L(I+1). \\
\end{cases} 
\nonumber \\
\end{eqnarray}  
In addition, using these variables
\begin{equation}
u_0=\sum^{T}_{t=1}u(t)/T
\end{equation}
\begin{equation}
\delta u'(t)=u(t)-u_0, 
\end{equation}
we can write
\begin{eqnarray}
\delta P_1(I)^2 &=& \frac{1}{L^2} \{ \sum^{L(I+1)}_{t=2}g_0(t,I) u(t) \}^2 \\
&=& \frac{1}{L^2} \{u_0^2 \cdot [\sum^{L(I+1)}_{t=2}g_0(t,I)]^2+ [\sum^{L(I+1)}_{t=2}g_0(t,I) u'(t)]^2 \}^2 \nonumber \\
&\approx& \frac{1}{L^2} u_0^2 \cdot \{\sum^{L(I+1)}_{t=2}g_0(t,I) \}^2 \\
&=& \frac{1}{L^2} u_0^2 \cdot [\frac{\kappa^{(I-1)L} \cdot (\kappa^L-1)^2}{(\kappa-1)^2}]^2.
\end{eqnarray}
Hence, the temporal average of $\delta P_1(I)^2$ is obtained by 
\begin{eqnarray}
&&\bar{P}_1 \equiv \sum^{T^{(L)}-1}_{I=1} \frac{\delta P_1(I)^2}{T^{(L)}-1} \\
 &\approx& \frac{1}{T^{(L)}-1} \cdot \sum^{T^{(L)}-1}_{I=1}  \frac{1}{L^2} u_0^2 \cdot [\frac{\kappa^{(I-1)L} \cdot (\kappa^L-1)^2}{(\kappa-1)^2}]^2 \nonumber \\
&=& \frac{u_0^2}{L^2 \cdot(T^{(L)}-1)} \frac{(\kappa^L-1)^3 \cdot (\kappa^{2L(T^{(L)}-1)}-1)}{(\kappa-1)^4 \cdot (\kappa^L+1)}.  \label{P1_m} 
\end{eqnarray}
\par
Similarly, we estimate the effects of $\delta P_2(I)^2$,  
\begin{eqnarray}
\delta P_2(I)^2 &=&  \frac{r(1)}{L^2} \cdot \frac{\kappa^{2IL}(\kappa^L+\kappa^{-L}-2)^2}{(\kappa-1)^2}.
\end{eqnarray}
Thus, the temporal average of $\delta P_2(I)^2$ is obtained by 
\begin{eqnarray}
&&\bar{P}_2 \equiv \sum^{T^{(L)}-1}_{I=1}\frac{\delta P_2(I)^2}{T^{(L)}-1}  \\
&=& \frac{r(1)^2}{L^2 \cdot (T^{(L)}-1)} \frac{(\kappa^L-1)^3 \cdot (\kappa^{2L(T^{(L)}-1)}-1)}{(\kappa-1)^2 \cdot (\kappa^L+1)}.  \label{P2_m}
\end{eqnarray}
 \par 
Lastly, we investigate the effects of $\delta P_3(I)^2$, i.e., 
\begin{eqnarray}
\delta P_3(I)^2 &=&  \frac{1}{L^2} \{ \sum^{L(I+1)}_{t=2}g_0(t,I) \eta(t) \}^2 \\
&\approx& \frac{1}{L^2} \eta_0^2 \cdot \sum^{L(I+1)}_{t=2}g_0(t,I)^2,   
\end{eqnarray}
where, from the definition, 
\begin{equation}
\eta_0^2 \approx \sum^{T}_{t=1}\eta(t)^2/T.
\end{equation}
%
We can calculate the sum of $g(t,I)$ with respect to $t$, 
\begin{equation}
 \sum^{L(I+1)}_{t=2}g_0(t,I)^2=Q_1(I)+Q_2+Q_3, 
\end{equation}
where 
\begin{eqnarray}
Q_1(I)\equiv \sum^{(I-1)L+1}_{t=2}g_0(t,I)^2  =\frac{(\kappa^L-1)^4 \cdot(\kappa^{2(I-1)L}-1)}{(\kappa-1)^3 \cdot (\kappa+1)}
\end{eqnarray}
\begin{eqnarray}
&&Q_2 \equiv \sum^{LI+1}_{(I-1)L+2}g_0(t,I)^2  \nonumber \\
&&=\frac{L(\kappa^2-1)+(\kappa^{2L}-3\kappa^L+2)(\kappa^{2L}-\kappa^{L}+2 \kappa)}{(\kappa-1)^3 \cdot (\kappa+1)}
\end{eqnarray}
\begin{eqnarray}
&&Q_3 \equiv \sum^{(I+1)L}_{LI+2}g_0(t,I)^2   \nonumber \\
&&=\frac{(\kappa^L-1)(\kappa^L-2\kappa-1)+L(\kappa^2-1)}{(\kappa-1)^3 \cdot (\kappa+1)}. 
\end{eqnarray}
From these results, we can obtain the temporal average of $\delta P_3(I)^2$, 
\begin{eqnarray}
&&\bar{P}_3 \equiv \sum^{T^{(L)}-1}_{I=1}\frac{\delta P_3(I)^2}{T^{(L)}-1}  \\
&=&\frac{\check{\eta}^2}{L^2(T^{(L)}-1)}\cdot(\bar{P}_{3a}+\bar{P}_{3b}+\bar{P}_{3c}).  \label{P3_m}
\end{eqnarray}
Here, 
\begin{eqnarray}
\bar{P}_{3a}=\frac{(\kappa^L-1)^3 \cdot (\kappa^{2L(T^{(L)}-1)}-(T^{(L)}-1)\kappa^{2L}+T^{(L)}-2)}{(\kappa-1)^3 (\kappa+1) (\kappa^L+1)} \nonumber \\
\end{eqnarray}
\begin{eqnarray}
&&\bar{P}_{3b}=(T^{(L)}-1) \nonumber \\
&&\times \frac{L \cdot (\kappa^2-1) +(\kappa^{2L}-3\kappa^L+2)(\kappa^{2L}-\kappa^L+2\kappa )}{(\kappa-1)^3 (\kappa+1)} \nonumber \\
\end{eqnarray}
\begin{eqnarray}
\bar{P}_{3c}=(T^{(L)}-1) \cdot \frac{L \cdot (\kappa^2-1) + (\kappa^L-1)(\kappa^L-2\kappa-1)}{(\kappa-1)^3 (\kappa+1)}. \nonumber \\
\end{eqnarray}
\par
{\bf Calculation of $E[\delta R]^2$.}
Next, we calculate $E[\delta R]^2$.
 $E[\delta R]^2$ can be decomposed as follows:
\begin{eqnarray}
&&E[\delta R]^2=(E[\delta P_1]+E[\delta P_2]+E[\delta P_3])^2 \\
&=& E[\delta P_1]^2+E[\delta P_2]^2+2 \cdot E[\delta P_1] \cdot E[\delta P_2] \nonumber \\
&+&2E[\delta P_3](E[\delta P_1]+E[\delta P_2]+E[\delta P_3]) \nonumber \\
&\approx& E[\delta P_1]^2+E[\delta P_2]^2+2 \cdot E[\delta P_1] \cdot E[\delta P_2],  \label{EQ2}
\end{eqnarray} 
where we use $E[\delta P_3] \approx 0$． \par
$E[\delta P_1]$ and $E[\delta P_2]$ are obtained as 
\begin{eqnarray}
&&E[\delta P_{1}]=P_{1}[T^{(L)}]-P_{1}[1] \\
&=&\frac{u_0}{T^{(L)}-1} \cdot \frac{1}{L}[\frac{-\kappa^{(T^{(L)}-1)L}+\kappa^{T^{(L)}L}-\kappa^L+1}{(\kappa-1)^2}].  \label{EQ1} \nonumber \\
\end{eqnarray}
\begin{eqnarray}
&&E[\delta P_{2}]=Q_{2}[T^{(L)}]-P_{2}[1] \\
&=&\frac{r(1)}{T^{(L)}-1} \cdot \frac{1}{L}[\frac{-\kappa^{(T^{(L)}-1)L}+\kappa^{T^{(L)}L}-\kappa^L+1}{\kappa-1}].  \label{EQ2} \nonumber \\
\end{eqnarray}
{\bf Calculation of $V[R^{(L)}]$.} 
Lastly, we calculate $V[R^{(L)}]$.
Substituting Eq. \ref{RJ2} and Eq. \ref{EQ2} for Eq. \ref{VR_rand}, we can obtain
\begin{eqnarray}
V[R^{(L)}] \approx R^{(1)} \cdot u_0^2 +R^{(2)} \cdot \check{\eta}^2 +R^{(3)} \cdot r(1)^2+R^{(4)} \cdot u_0 \cdot r(1), \nonumber \\
\end{eqnarray}
where from Eq. \ref{P1_m} and Eq. \ref{EQ1}
\begin{eqnarray}
R^{(1)} =R_1^{(1)}+R_2^{(1)}, 
\end{eqnarray}
\begin{eqnarray}
R_1^{(1)}=  \frac{1}{L^2 \dot (T^{(L)}-1)}[ \frac{(\kappa^L-1)^3 \cdot (\kappa^{2L(T^{(L)}-1)}-1)}{(\kappa-1)^4 \cdot (\kappa^L+1)}], 
\end{eqnarray}
\begin{eqnarray}
R_2^{(1)}=-\frac{1}{L^2 \cdot (T^{(L)}-1)^2} \cdot W_1^2, 
\end{eqnarray}
\begin{eqnarray}
W_1=\frac{\kappa^{T^{(L)}L}-\kappa^{(T^{(L)}-1)L}+1-\kappa^L}{(\kappa-1)^2},
\end{eqnarray}
from Eq. \ref{P3_m},  
\begin{eqnarray}
R^{(2)} =R_1^{(2)}+R_2^{(2)}+R_3^{(2)}, 
\end{eqnarray}
\begin{eqnarray}
&&R_{1}^{(2)}= \frac{1}{L^2 \cdot (T^{(L)}-1)}  \nonumber  \\
&& \times \frac{(\kappa^L-1)^3 \cdot (\kappa^{2L(T^{(L)}-1)}-(T^{(L)}-1)\kappa^{2L}+T^{(L)}-2)}{(\kappa-1)^3 (\kappa+1) (\kappa^L+1)}, \nonumber \\
\end{eqnarray}
\begin{eqnarray}
&&R_{2}^{(2)}=\nonumber  \\
&&\frac{1}{L^2} \cdot \frac{L \cdot (\kappa^2-1) +(\kappa^{2L}-3\kappa^L+2)(\kappa^{2L}-\kappa^L+2\kappa)}{(\kappa-1)^3 (\kappa+1)}, \nonumber \\
\end{eqnarray}
\begin{eqnarray}
{R}_{3}^{(2)}=\frac{1}{L^2} \cdot \frac{L \cdot (\kappa^2-1) + (\kappa^L-1)(\kappa^L-2\kappa-1)}{(\kappa-1)^3 (\kappa+1)},
\end{eqnarray}
from Eq. \ref{P2_m} and Eq. \ref{EQ2}, 
\begin{eqnarray}
R^{(3)} =R_1^{(3)}+R_2^{(3)}, 
\end{eqnarray}
\begin{eqnarray}
R_{1}^{(3)} = \frac{1}{L^2 \cdot (T^{(L)}-1)} \frac{(\kappa^L-1)^3 \cdot (\kappa^{2L(T^{(L)}-1)}-1)}{(\kappa-1)^2 \cdot (\kappa^L+1)}, 
\end{eqnarray}
\begin{eqnarray}
R_{2}^{(3)}=-\frac{1}{L^2 \cdot (T^{(L)}-1)^2} \cdot W_2^2, 
\end{eqnarray}
\begin{eqnarray}
W_2=\frac{\kappa^{T^{(L)}L}-\kappa^{(T^{(L)}-1)L}+1-\kappa^L}{\kappa-1}, 
\end{eqnarray}
and from Eq. \ref{EQ1} and Eq. \ref{EQ2} 
\begin{eqnarray}
R^{(4)}=-\frac{2}{L^2 \cdot (T^{(L)}-1)^2} \cdot W_1 \cdot W_2.
\end{eqnarray}
\par
\subsection{Calculation of $V[{R^{(L)}}^2]$ for large $L$}
We calulate $V[{R^{(L)}}^2]$ for $L>>1$. Here, we assume $L/2 \geq T$. \\
{\bf Case of $\kappa<1$:} \\
Using $\kappa^L \approx 0$ and $\kappa^T \approx 0$, we can obtain 
\begin{equation}
R_1^{(1)} \approx \frac{1}{L(T-L)} \frac{1}{(\kappa-1)^4}
\end{equation}
\begin{equation}
R_2^{(1)} \approx \frac{1}{(T-L)^2} \frac{1}{(\kappa-1)^4}
\end{equation}
\begin{equation}
R_1^{(2)} \approx \frac{2L-T}{L^2 (T-L)} \cdot  \frac{1}{(\kappa-1)^3 \cdot (\kappa+1)} 
\end{equation}
\begin{equation}
R_2^{(2)} \approx \frac{1}{L} \cdot \frac{1}{(\kappa-1)^2 }+\frac{1}{L^2} \cdot \frac{4k}{(\kappa-1)^3 \cdot (\kappa+1)}
\end{equation}
\begin{equation}
R_3^{(2)} \approx \frac{1}{L} \cdot \frac{1}{(\kappa-1)^2 }+\frac{1}{L^2} \cdot \frac{2\kappa+1}{(\kappa-1)^3 \cdot (\kappa+1)}
\end{equation}
\begin{equation}
R_1^{(3)} \approx \frac{1}{L(T-L)} \frac{1}{(\kappa-1)^2}
\end{equation}
\begin{equation}
R_2^{(3)} \approx \frac{1}{(T-L)^2} \frac{1}{(\kappa-1)^4}
\end{equation}
\begin{equation}
R^{(4)} \approx \frac{1}{(T-L)^2} \frac{1}{(\kappa-1)^3}. 
\end{equation}
Considering only dominant terms, we can obtain
\begin{equation}
V[\delta R^{(L)}]\approx    \frac{2}{L}  \frac{ \check{\eta}^2 }{(\kappa-1)^2 } + \frac{1}{L(T-L)}( \frac{u_0^2}{(\kappa-1)^4}+ \frac{r(1)^2}{(\kappa-1)^2}). \nonumber \\
\end{equation}
In the case of $T-L>>1$, we can obtain a simpler form 
\begin{equation}
V[\delta R^{(L)}] \approx \check{\eta}^2  \frac{2}{L} \cdot \frac{1}{(\kappa-1)^2 } \propto \frac{1}{L}. 
\end{equation}
\par
{\bf Case of $\kappa=1$:} 
Next, we calculate the case of $r=1$. 
Taking the limit of $r\to1$, 
\begin{equation}
R_1^{(1)} \approx  L^2
\end{equation}
\begin{equation}
R_2^{(1)} \approx -L^2
\end{equation}
\begin{equation}
R_1^{(2)} \approx 0
\end{equation}
\begin{equation}
R_2^{(2)} \approx \frac{2L^3+3L^2+L}{6} \cdot \frac{1}{L^2}
\end{equation}
\begin{equation}
R_3^{(2)} \approx \frac{2L^3-3L^2+L}{6} \cdot \frac{1}{L^2}
\end{equation}
\begin{equation}
R_1^{(3)} \approx 0
\end{equation}
\begin{equation}
R_2^{(3)} \approx 0
\end{equation}
\begin{equation}
R^{(4)} \approx 0. 
\end{equation}
From these results, we can obtain
\begin{equation}
V[R^{(L)}] \approx \frac{1}{3} \cdot \check{\eta}^2 \cdot (2L+1/L) \propto  L. 
\end{equation}
\par 
{\bf Case of $ \kappa >1$:} 
Lastly, we calculate the case of $ \kappa >1$ for $L>>1$．\par
Calculation of $R^{(1)}$. 
For $L>>1$, in the case of $ \kappa >1$, a dominant term of $R_1^{(1)}$ is given by 
\begin{eqnarray}
R_1^{(1)} \approx \frac{1}{L(T-L)} \frac{k^{2T}}{(k-1)^4}
\end{eqnarray}
In a similar way, 
\begin{equation}
R_2^{(1)} \approx  \frac{-1}{(T-L)^2}\frac{k^{2T}}{(k-1)^4}. 
\end{equation}
\par
We calculate $R^{(2)}$. \\
For $L>>1$,  
\begin{eqnarray}
R_1^{(2)} &\approx&  \frac{\kappa^{2T}}{L(T-L)} \frac{1}{(\kappa-1)^3 \cdot (\kappa+1)}. \\
\end{eqnarray}
Similarly,
\begin{equation}
R_2^{(2)} \approx  \frac{\kappa^{4L}}{L^2} \frac{1}{(\kappa-1)^3 \cdot (\kappa+1)},  
\end{equation}
\begin{equation}
R_3^{(2)} \approx  \frac{\kappa^{2L}}{L^2} \frac{1}{(\kappa-1)^3 \cdot (\kappa+1)}.
\end{equation}
Thus,
\begin{equation}
R^{(2)} \approx \frac{\kappa^{2T}}{L(T-L)} \frac{1}{(\kappa-1)^3 \cdot (\kappa+1)}.
\end{equation}
\par
Calculation of  $R^{(3)}$ and  $R^{(4)}$. 
For a large $L$, 
\begin{equation}
R_1^{(3)} \approx \frac{\kappa^{2T}}{L(T-L)} \frac{1}{(\kappa-1)^2} 
\end{equation}
\begin{equation}
R_2^{(3)} \approx-\frac{\kappa^{2T}}{(T-L)^2} \frac{1}{(\kappa-1)^2}. 
\end{equation}
Thus, 
\begin{equation}
R^{(3)} \approx \frac{\kappa^{2T}}{(T-L) \cdot (\kappa-1)^2} [\frac{1}{L}-\frac{1}{T-L}]. 
\end{equation}
$R^{(4)}$ is also approximated as 
\begin{equation}
R^{(4)} \approx -\frac{2 \cdot \kappa^{2T}}{(T-L)^2} [\frac{1}{(\kappa-1)^3}]. 
\end{equation}
Consequently, for $T>>L>>1$, we can obtain
\begin{equation}
V[\delta R^{(L)}] \propto \frac{\kappa^{2T}}{L(T-L)}.
\end{equation}
\renewcommand{\theequation}{D\arabic{equation}}
\renewcommand{\thefigure}{D-\arabic{figure}}
\renewcommand{\thetable}{S-D\arabic{table}}
\setcounter{figure}{0}
\setcounter{equation}{0}
\section{$V[\delta R_j^{(L)}]$ for the power-law forgetting process}
\label{app_forget}
We calculate $V[\delta R_j^{(L)}]$ for the power-law forgetting process given by  Eq. \ref{r_forget}.  
Here, we consider $L \geq 2$ and omit the suffix $j$ for simplification. 
$R(I)$ is defined by 
\begin{equation}
R(I)=\sum_{t=LI+1}^{L(I+1)}\frac{r(t)}{L}. 
\end{equation}
From the definiton of $r(t)$, 
\begin{equation}
r(t)=\sum_{s=0}^{\infty} \theta(s) \cdot \eta(t-s), 
\end{equation}
we can calculate 
\begin{equation}
R(I)=\frac{1}{L}\sum^{LI+1}_{t=-\infty}\theta_1'(t) \eta(t) + \sum^{L(I+1)}_{t=LI+2}\theta_2'(t)\eta(t), 
\end{equation}
where 
\begin{eqnarray}
\theta_1'(t)&=&\sum^{L-1}_{k=0} \theta(k+LI+1-t) \label{theta_1d} \\
\theta_2'(t)&=&\sum^{L-(t-LI)}_{k=0} \theta(k). \label{theta_2d} 
\end{eqnarray}
\par
In a similar way, we can also calculate $R(I+1)$, 
\begin{equation}
R(I+1)=\frac{1}{L}\sum^{L(I+1)+1}_{t=-\infty}\theta_3'(t) \eta(t) + \sum^{L(I+2)}_{t=L(I+1)+2}\theta_4'(t)\eta(t), 
\end{equation}
where
\begin{eqnarray}
\theta_3'(t)&=&\sum^{L-1}_{k=0} \theta(k+L(I+1)+1-t) \label{theta_3d} \\
\theta_4'(t)&=&\sum^{L-(t-L(I+1))}_{k=0} \theta(k). \label{theta_4d} \\
\end{eqnarray}
From these results, $\delta R(I)=R(I+1)-R(I)$ is calculated by 
\begin{eqnarray}
&&\delta R(I)= R(I+1)-R(I) \\
&=&\frac{1}{L} \sum^{LI+1}_{t=-\infty}(\theta_3'(t)-\theta_1'(t)) \eta(t) \nonumber  
+\sum^{L(I+1)}_{t=LI+2}(\theta_3'(t)-\theta_2'(t)) \eta(t) \nonumber  \\
&+&\sum^{L(I+2)}_{t=L(I+1)+1}\theta_4'(t) \eta (t).  
\end{eqnarray}
\par
Taking the average of $\delta R(I)$ with respect to $\eta$,   
\begin{eqnarray}
&&E_{\eta}[(R(I+1)-R(I))^2] = 
\frac{\check{\eta}^2}{L^2} \{\sum^{LI+1}_{t=-\infty}(\theta_3'(t)-\theta_1'(t))^2 \nonumber  \\
&&+\sum^{L(I+1)}_{t=LI+2}(\theta_3'(t)-\theta_2'(t))^2 
+\sum^{L(I+2)}_{t=L(I+1)+1}\theta_4'(t)^2 \}  \\
&&=\frac{\check{\eta}^2}{L^2 Z(\beta)^2} (U_1(\beta,L)+U_2(\beta,L)+U_3(\beta,L)) \label{u_ans}
\end{eqnarray}
Here, $U_1(\beta,L)$, $U_2(\beta,L)$ and $U_3(\beta,L)$ is defined as
\begin{equation}
U_1(\beta,L)=\sum^{LI+1}_{t=-\infty}\frac{ Z(\beta)^2}{\check{\eta}^2} (\theta_3'(t)-\theta_1'(t))^2 \label{U_1}
\end{equation}
\begin{equation}
U_2(\beta,L)=\sum^{L(I+1)}_{t=LI+2} \frac{ Z(\beta)^2}{\check{\eta}^2}(\theta_3'(t)-\theta_2'(t))^2, \label{U_2}
\end{equation}
and
\begin{equation}
U_3(\beta,L)=\sum^{L(I+2)}_{t=L(I+1)+1} \frac{ Z(\beta)^2}{\check{\eta}^2} (\theta_4'(t))^2. \label{U_3}
\end{equation}
We factor out $Z(\beta)^2$ for later calculations. 
$U_1(\beta,L)$ is given by Eq. \ref{U_1_ans}, $U_2(\beta,L)$ is given by Eq. \ref{u2_ans}, and $U_3(\beta,L)$ is given by Eq. \ref{u3_ans}.
The details of the derivations of these equations are mentioned in the following section and beyond.
 \par
\subsection{Calculation of $U_1(\beta,L)$}
Here, we calculate $U_1(\beta,L)$ defined by Eq. \ref{U_1}. 
\begin{eqnarray}
&&U_1(\beta,L)= \nonumber \\
&&\frac{ Z(\beta)^2}{\check{\eta}^2} \sum^{LI+1}_{t=-\infty}(\theta_3'(t)-\theta_1'(t))^2 \\
&&=\frac{ Z(\beta)^2}{\check{\eta}^2} \sum^{LI+1}_{t=-\infty}(\sum^{L}_{i=1}\theta(k+L(I+1)+1-t) \nonumber \\
&&-\sum^{L}_{i=1}\theta(k+LI+1-t))^2. \nonumber \\
\end{eqnarray}
Replacing the index $t$ with a new index $t'=t+LI+1$, $U_1(\beta,L)$ can be written as
\begin{eqnarray}
U_1(\beta,L)= \sum^{\infty}_{t'=0}\frac{ Z(\beta)^2}{\check{\eta}^2}(\sum^{L-1}_{k=0}\theta(t'+k+L)-\sum^{L-1}_{k=0}\theta(t'+k))^2. \nonumber \\
\end{eqnarray}
\par
Using the Euler-Maclaurin formula \cite{abramowitz1964handbook},  
\begin{equation}
\sum^{b}_{k=a}g(k) \approx \int^{b}_{a}g(x)dx+\frac{1}{2}(g(b)+g(a))+\frac{1}{12}(\frac{d}{dx}g(x)|_{b}-\frac{d}{dx}g(x)|_{a}) \label{euler}
\end{equation}
we can obtain the apploximation, 
\begin{eqnarray}
&&U_1 \approx  \nonumber \\
&&\frac{ Z(\beta)^2}{\check{\eta}^2} \sum^{\infty}_{t'=0} \left\{ \theta(t'+L)+\frac{\theta(t'+2L-1)+\theta(t'+L+1)}{2} \right. \nonumber  \\
&+&\frac{\theta'(t'+2L-1)-\theta'(t'+L+1)}{12} +\int^{L-1}_{1}\theta(t'+k+L)dk \nonumber \\
&-&  \theta(t')-\frac{\theta(t'+L-1)+\theta(t'+1)}{2} \nonumber \\ 
&-&\left. \frac{\theta'(t'+L-1)-\theta'(t'+1)}{12}-\int^{L-1}_{1} \theta(t'+k)  dk \right\}^2. \nonumber \\
\end{eqnarray}
Substituting Eq. \ref{theta} into $\theta(t')$ and conducting some calculations, 
for $\beta>0$ and $\beta \neq 1$, 
\begin{eqnarray}
&&U_1 \approx \nonumber \\
&&\sum^{\infty}_{t'=0}\left\{  \sum^{3}_{k=1}J^{(-1)}_{A,k} (t+A^{(-1)}_k)^{-(\beta-1)} \right. \nonumber \\
&&+\left. \sum^{6}_{k=1}J^{(0)}_{A,k} ((t+A^{(0)}_k)^{-\beta}+ \sum^{3}_{k=1}J^{(+1)}_{A,k}(t+A^{(+1)}_k)^{-(\beta+1)}\right\}^2. \nonumber \\
\end{eqnarray}
For $\beta=1$, 
\begin{eqnarray}
&&U_1 \approx \nonumber \\
&&\sum^{\infty}_{t'=0}\left\{  \sum^{3}_{k=1}J^{(-1)}_{A,k} \log(t+A^{(-1)}_k) \right. \nonumber \\
&&\left. + \sum^{6}_{k=1}J^{(0)}_{A,k} ((t+A^{(0)}_k)^{-\beta}+ \sum^{3}_{k=1}J^{+1}_{A,k}(t+A^{(+1)}_k)^{-(\beta+1)}\right\}^2. \nonumber \\
\end{eqnarray}
We combine the two equations into one,  
\begin{eqnarray}
&&U_1 \approx \nonumber \\
&&\sum^{\infty}_{t'=0}\left\{\sum^{+1}_{m=-1} \sum_{k=1}^{p_m}J^{(m)}_{A,k} f_A(t,\beta+m,A^{(m)}_k) \right\}^2,  \nonumber \\
\end{eqnarray}
where 
\begin{eqnarray}
f_A(x,\alpha,U)= 
\begin{cases}
(x+U)^{-\alpha} & (\alpha \neq 0) \\
\log(x+U) &  (\alpha=0)
\end{cases}
\end{eqnarray}
\begin{eqnarray}
A_1^{(-1)}&=&=A_1^{(+1)}=2L-1-a \\
A_2^{(-1)}&=&A_2^{(+1)}=L+1+a \\
A_3^{(-1)}&=&A_3^{(+1)}=L-1+a \\
A_4^{(-1)}&=&A_4^{(+1)}=1+a \\
A_1^{(0)}&=&2L-1+a \\
A_2^{(0)}&=&1+L+a \\
A_3^{(0)}&=&L+a \\
A_4^{(0)}&=&L-1+a \\
A_5^{(0)}&=&1+a \\
A_6^{(0)}&=&a, 
\end{eqnarray}
$p_{-1}=p_{1}=4$, $p_{0}=6$, 
\begin{eqnarray}
J^{(-1)}_{A,1}=J^{(-1)}_{A,4}=
\begin{cases}
1/(1-\beta) & \text{($\beta \neq 1$)} \\
1 & \text{($\beta = 1$)} \\
\end{cases}
\end{eqnarray}
\begin{eqnarray}
J^{(-1)}_{A,2}=J^{(-1)}_{A,3}=
\begin{cases}
-1/(1-\beta) & \text{($\beta \neq 1$)} \\
-1 & \text{($\beta = 1$)} \\
\end{cases}, 
\end{eqnarray}
\par
\begin{eqnarray}
J^{(0)}_{A,1}&=&J^{(0)}_{A,2}=1/2 \\
J^{(0)}_{A,3}&=&1  \\
J^{(0)}_{A,4}&=&J^{(0)}_{A,5}=-1/2 \\
J^{(0)}_{A,6}&=&-1 
\end{eqnarray}
and 
\begin{eqnarray}
J^{(+1)}_{A,1}&=&J^{(+1)}_{A,3}=1/12 \cdot \beta \\
J^{(+1)}_{A,2}&=&J^{(+1)}_{A,4}=-1/12 \cdot \beta.  
\end{eqnarray}
We expand $U_1$, 
\begin{eqnarray}
&&U_1 \approx \nonumber \\
&&\sum^{\infty}_{t'=0}\left\{\sum^{+1}_{m=-1} \sum_{k=1}^{p_m}J^{(m)}_{A,k} f_A(t,\alpha+m,A^{(m)}_k) \right\}^2= \nonumber \\
&&\sum^{+1}_{m_1=-1}U_1^{(2)}(t;\beta,m_1,m_1)+2 \sum_{m_1>m_2}U_1^{(2)}(t;\beta,m_1,m_2),  \nonumber \\
\end{eqnarray}
where 
\begin{eqnarray}
&&U^{(2)}_1(t;\beta,m_1,m_2)= \nonumber \\
&& \sum_{i=1}^{p_{m_1}} \sum_{j=1}^{p_{m_2}} J_{Ai}^{(m_1)} J_{Aj}^{(m_2)} \nonumber \\
&& \lim_{Q \to \infty} \sum^{Q}_{t=0}f_A(t,\beta+m_i,A^{(m)}_i)f_A(t,\beta+m_j,A^{(m)}_j) \nonumber \\
\end{eqnarray}
%
We calculate $U^{(2)}_1(t;\beta,m_1,m_2)$. 
Using the Euler-Maclaurin formula in Eq. \ref{euler}, we can obtain 
\begin{eqnarray}
&&\sum^{Q}_{t=0}f_A(t,\alpha_1,V_1)f_A(t,\alpha_2,V_2) \nonumber \\
&&\approx G_2^{(R)}(1,Q;\alpha_1,\alpha_2,V_1,V_2)+\int^Q_1 (t+V_1)^{-\alpha_1}(t+V_2)^{-\alpha_2} dt \nonumber \\
&&\approx G_2^{(R)}(1,Q;\alpha_1,\alpha_2,V_1,V_2)+G_2^{(I)}(Q;\alpha_1,\alpha_2,V_1,V_2) \nonumber \\
&&-G_2^{(I)}(1;\alpha_1,\alpha_2,V_1,V_2). \nonumber \\ \label{fafa}
\end{eqnarray}
(i) For $\alpha_1 \neq 1$, $\alpha_2 \neq 1$ $x_1=0$ and $x_2=0$, $G_2^{(R)}(x_1,x_2;\alpha_1,\alpha_2,V_1,V_2)$ is given by 
\begin{eqnarray}
&&G_2^{(R)}(x_1,x_2;\alpha_1,\alpha_2,V_1,V_2)=(V_1)^{-\alpha_1}(V_2)^{-\alpha_2}+ \nonumber \\
&&\frac{1}{2}\{(V_1+x_1)^{-\alpha_1}(V_2+x_1)^{-\alpha_2}+(V_1+x_2)^{-\alpha_1}(V_2+x_2)^{-\alpha_2}\}  \nonumber \\
&&-\frac{1}{12}\{(\alpha_2(V_2+x_2)+\alpha_1(V_1+x_2)) \nonumber \\
&&\times(V_1+x_2)^{-\alpha_1-1}(V_2+x_2)^{-\alpha_2-1} \nonumber \\
&&-(\alpha_2(V_1+x_1)+\alpha_1(V_2+x_1)) \nonumber \\
&&\times(V_1+x_1)^{-\alpha_1-1}(V_2+x_1)^{-\alpha_2-1}\}. \nonumber \\ 
\end{eqnarray}
Here, because $Q$ approachs zero for $Q>>1$, $G_2^{(R)}(x,Q;0,\alpha_2,V_1,V_2)$ is not dependent on $Q$.
Accordingly, we can denote $G_2^{(R)}(x,Q;0,\alpha_2,V_1,V_2)$ as $G_2^{(R \to Q)}(x;\alpha_1,\alpha_2,V_1,V_2)$. \\
$G_2^{(R \to Q)}(x;\alpha_1,\alpha_2,V_1,V_2)$ is written as 
\begin{eqnarray}
&&G_2^{(R \to Q)}(x;\alpha_1,\alpha_2,V_1,V_2)=(V_1)^{-\alpha_1}(V_2)^{-\alpha_2} \nonumber \\
&&+ \frac{1}{2}\{(V_1+x_1)^{-\alpha_1}(V_2+x_1)^{-\alpha_2}\}  \nonumber \\
&&-\frac{1}{12}\{ 
-(\alpha_2(V_1+x_1)+\alpha_1(V_2+x_1)) \nonumber \\
&&\times(V_1+x_1)^{-\alpha_1-1}(V_2+x_1)^{-\alpha_2-1}\}. \nonumber \\ 
\end{eqnarray}
As a special point, for $x_1=0$ and $x_2=0$, we determine
\begin{eqnarray}
&&G_2^{(R)}(0,0;\alpha_1,\alpha_2,V_1,V_2)=G_2^{(R \to Q)}(x;\alpha_1,\alpha_2,V_1,V_2) \nonumber \\
&&=(V_1)^{-\alpha_1}(V_2)^{-\alpha_2}.
\end{eqnarray}
\\
(ii) In the case of $\alpha_1 = 1$ and $\alpha_2 = 1$, we can also calculate 
\begin{eqnarray}
&&G_2^{(R)}(x_1,x_2;0,0,V_1,V_2)=\log(V_1) \log(V_2) \nonumber \\
&+&\frac{1}{2}\{\log(V_1+x_1)\log(V_2+x_1)+\log(V_1+x_2)\log(V_2+x_2)\}  \nonumber \\
&+&\frac{1}{12}\{\frac{\log(V_1+x_2)}{V_2+x_2}+\frac{\log(V_2+x_2)}{V_1+x_2} \nonumber \\
&-&(\frac{\log(V_1+x_1)}{V_2+x_1}+\frac{\log(V_2+x_1)}{V_1+x_1}),  \nonumber \\ 
\end{eqnarray}
and for $Q>>1$, we can obtain
\begin{eqnarray}
&&G_2^{(R)}(x_1,x_2;0,\alpha_2,V_1,V_2)=\log(V_1) V_2^{-\alpha_2}+ \nonumber \\
&&\frac{1}{2}\{\log(V_1+x_1)V_2^{-\alpha_2}+\log(V_1+x_2)(V_2+x_2)^{-\alpha_2}\}+\frac{1}{12} \times  \nonumber \\
&&\{(\frac{(V_2+x_2)^{-\alpha_2-1}(-\beta(V_1+x_2)\log(V_1+x_2)+V_2+x_2)}{V_1+x_2})  \nonumber \\
&&-(\frac{(V_2+x_1)^{-\alpha_2-1}(-\beta(V_1+x_1)\log(V_1+x_1)+V_1+x_1)}{V_1+x_1})\}. \nonumber \\ 
\end{eqnarray}
As a special point, for $x_1=0$ and $x_2=0$, we determine
\begin{eqnarray}
&&G_2^{(R)}(0,0;0,0,V_1,V_2)=G_2^{(R \to Q)}(x;\alpha_1,\alpha_2,V_1,V_2) \nonumber \\
&&=\log(V_1) \log(V_2).
\end{eqnarray}
 \\
(iii)In the case of $\alpha_1 = 1$ and $\alpha_2 \neq 0$, we can also obtain
\begin{eqnarray}
&&G_2^{(R)}(x_1,x_2;0,\alpha_2,V_1,V_2)=\log(V_1) V_2^{-\alpha_2} \nonumber \\
&&+\frac{1}{2}\{\log(V_1+x_1)V_2^{-\alpha_2}+\log(V_1+x_2)(V_2+x_2)^{-\alpha_2}\}  \nonumber \\
&&+\frac{1}{12}\{(\frac{(V_2+x_2)^{-\alpha_2-1}(-\beta(V_1+x_2)\log(V_1+x_2)+V_2+x_2)}{V_1+x_2})  \nonumber \\
&&-(\frac{(V_2+x_1)^{-\alpha_2-1}(-\beta(V_1+x_1)\log(V_1+x_1)+V_1+x_1)}{V_1+x_1})\}. \nonumber \\ 
\end{eqnarray}
As a special point, for $x_1=0$ and $x_2=0$, we determine
\begin{equation}
G_2^{(R)}(0,0;0,\alpha_2,V_1,V_2)= \log(V_1) V_2^{-\alpha_2}.
\end{equation}
\par
Next, we calculate the integration term of Eq. \ref{fafa}, $G_2^{(I)}(x;\alpha_1,\alpha_2,V_1,V_2)$. 
$G_2^{(I)}(x;\alpha_1,\alpha_2,V_1,V_2)$ is defined by 
\begin{eqnarray}
&&G_2^{(I)}(x;\alpha_1,\alpha_2,V_1,V_2)= \nonumber \\
&&
\begin{cases}
\int (x+V_1)^{-\alpha_1}(x+V_2)^{-\alpha_2} dx & (\alpha_1,\alpha_2>0) \\
\int \log(x+V_1) \log(x+V_2) dx & (\alpha_1,\alpha_2=0) \\
\int \log(x+V_1) (x+V_2)^{-\alpha_2} dx & (\alpha_1=0, \alpha_2>0) \\
\end{cases}, 
\end{eqnarray}
where we neglecte an integral constant. 
\par
We calculate $G_2^{(I)}(x;\alpha_1,\alpha_2,V_1,V_2)$. \\
(i) In the case of $V_1=V_2$ and $\alpha_1 \neq 0$, $\alpha_2 \neq 0$, we can write 
\begin{eqnarray}
&&G_2^{(I)}(x;\alpha_1,\alpha_2,V_1,V_2) =\frac{1}{1-\alpha_1-\alpha_2}(x+V_1)^{-\alpha_1-\alpha_2+1}. \nonumber \\
\label{GI_AA}
\end{eqnarray}
(ii) When $\alpha_1$ and $\alpha_2$ are non-integers, and $V_1 \neq V_1$, we obtain
\begin{eqnarray}
&&G_2^{(I)}(x;\alpha_1,\alpha_2,V_1,V_2)= \\
&&\frac{(V_j+x)^{(1-\alpha_j)}(V_i-V_j)^{(-\alpha_i)}{}_2F_1(\alpha_i,1-\alpha_j,2-\alpha_j,\frac{-(V_j+x)}{(V_i-V_j)})}{(1-\alpha_j)}  \nonumber \\
\label{GI_real}
\end{eqnarray}
where $i={\arg \max}_{k \in \{1,2\}}\{V_k\}$ and $j={\arg \min}_{k \in \{1,2\}}\{V_k\}$ (under this condition, $G_2^{(I)}$ takes a real number). \\
(iii) For $\alpha_1=1,2,3,\cdots$, $\alpha_2=1,2,3,\cdots$ and $V_1 \neq V_2$,  
%
%
%
\par
using a partial fraction decomposition, 
\begin{eqnarray}
\frac{1}{(z+q_1)^{n_1}(z+q_2)^{n_2}}=\sum_{k=1}^{n_1}\frac{h^{(1)}_k}{(z+q_1)^k}+\sum_{k=1}^{n_2}\frac{h^{(2)}_k}{(z+q_2)^k} \nonumber \\
\end{eqnarray}
\begin{eqnarray}
&&h^{(1)}_k=\frac{1}{(n_1-k)!}(-1)^{n_2}\frac{(n_1+n_2-k-1)!}{(n_2-1)!} \nonumber \\
&&\times (-q_2+q_1)^{-n_1-n_2+k} \nonumber \\
&&h^{(2)}_k=\frac{1}{(n_2-k)!}(-1)^{n_1}\frac{(n_1+n_2-k-1)!}{(n_1-1)!}\nonumber \\
&&\times (-q_1+q_2)^{-n_1-n_2+k}, \nonumber \\
\end{eqnarray}
we can write
\begin{eqnarray}
&&G_2^{(I)}(x;\alpha_1,\alpha_2,V_1,V_2)=h_1^{(1)}\log(x+V_1)+h_1^{(2)}\log(x+V_2)+ \nonumber \\ 
&&\sum_{k=2}^{\alpha_1}\frac{1}{1-k}h^{(1)}_k (x+V_1)^{1-k}+\sum_{k=2}^{\alpha_2} \frac{1}{1-k} h^{(2)}_k (x+V_2)^{1-k}.
\label{GI_int}
\end{eqnarray}
(iv) For $\alpha_1=0$, $\alpha_2=0$, and $V_1 \neq V_2$, we can write
\begin{eqnarray}
&&G_2^{(I)}(x;0,0,V_1,V_2)= \nonumber \\
&&Li_{2}(\frac{V_1+x}{V_1-V_2}) (V_2-V_1)+ \nonumber \\
&&\log(V_1+x)\{(V_1+x)\log(V_2+x) \nonumber \\
&&+(V_2-V_1)\log(-\frac{V_2+x}{V_1-V_2})-V_1-x\} \nonumber \\
&&+V_1-V_2\log(V_2+x)-x\log(V_2+x)+2x. \label{GI_00AB} 
\end{eqnarray}
(v) For $\alpha_1=0$, $\alpha_2=0$ and $V_1=V_2$, we can write
\begin{eqnarray}
&&G_2^{(I)}(x;0,0,V_1,V_2)=  \nonumber \\
&&\log(V_1+x)^2 (V_1+x)+2\log(V_1+x)(-V_1-x)+2x+V_1.  \nonumber \\
\label{GI_00AA} 
\end{eqnarray}
(vi) For $\alpha_1=0$, $\alpha_2=1$ and $V_1 \neq V_2$, we can write
\begin{eqnarray}
&&G_2^{(I)}(x;0,1,V_1,V_2)= \nonumber \\
&&Li_{2}(\frac{V_1+x}{V_1-V_2}) + \log(V_1+x)\{\log(-\frac{V_2+x}{V_1-V_2})\}. \nonumber \\
\label{GI_01AB} 
 \end{eqnarray}
(vii) For $\alpha_1=0$, $\alpha_2=1$ and $V_1=V_2$, 
\begin{eqnarray}
&&G_2^{(I)}(x;0,1,V_1,V_2)=\frac{1}{2}\log^2(V_1+x).
\label{GI_01AA} 
\end{eqnarray}
(viii) For $\alpha_1=0$, $\alpha_2=2$ and $V_1 \neq V_2$, we can write
\begin{eqnarray}
&&G_2^{(I)}(x;0,2,V_1,V_2)= \nonumber \\ 
&&\frac{(V_2+x)\log(V_2+x)-(V_1+x)\log(V_1+x)}{(V_1-V_2)(V_2+x)}. \nonumber \\
\label{GI_02AB}
 \end{eqnarray}
(ix) For $\alpha_1=0$, $\alpha_2=2$ and $V_1= V_2$, we can write
\begin{eqnarray}
&&G_2^{(I)}(x;0,2,V_1,V_2)=-\frac{\log(V_1+x)+1}{V_1+x}.
\label{GI_02AA}
\end{eqnarray}
\par
Consequently, the summary of $G_2^{(I)}$ is given by  
\begin{eqnarray}
&&G_2^{(I)}(x;\alpha_1,\alpha_2,V_1,V_2)= \nonumber \\
&&\begin{cases}
\text{($\alpha_1 \neq 0$, $\alpha_2 \neq 0$, $V_1=V_2$): }\\
\frac{1}{1-\alpha_1-\alpha_2}(x+V_1)^{-\alpha_1-\alpha_2+1} \\
\text{($\alpha_2$ and $\alpha_2$ are non-integer, $V_1 \neq V_2$): }\\
\frac{(V_j+x)^{(1-\alpha_j)}(V_i-V_j)^{(-\alpha_i)}{}_2F_1(\alpha_i,1-\alpha_j,2-\alpha_j,\frac{-(V_j+x)}{(V_i-V_j)})}{(1-\alpha_j)} \\
\text{($\alpha_1 \geq 2$, $\alpha_2 \geq 2$, $\alpha_1$ and $\alpha_2$ are integer, $V_1 \neq V_2$): }\\
h_1^{(1)}\log(x+V_1)+h_1^{(2)}\log(x+V_2)+\\
\sum_{k=2}^{\alpha_1}\frac{1}{1-k}h^{(1)}_k (x+V_1)^{1-k}+\sum_{k=2}^{\alpha_2} \frac{1}{1-k} h^{(2)}_k (x+V_2)^{1-k} \\
\text{($\alpha_1=0$, $\alpha_2=0$, $V_1= V_2$): }\\
\log(V_1+x)^2 (V_1+x)+2\log(V_1+x)(-V_1-x)+2x+V_1  \\
\text{($\alpha_1=0$, $\alpha_2=0$, $V_1 \neq V_2$): }\\
Li_{2}(\frac{V_1+x}{V_1-V_2}) (V_2-V_1)+  \\
\log(V_1+x)\{(V_1+x)\log(V_2+x)  \\
+(V_2-V_1)\log(-\frac{V_2+x}{V_1-V_2})-V_1-x\}  \\
+V_1-V_2\log(V_2+x)-x\log(V_2+x)+2x \\
\text{($\alpha_1=0$, $\alpha_2=1$, $V_1 = V_2$): }\\
\frac{1}{2}\log^2(V_1+x) \\
\text{($\alpha_1=0$, $\alpha_2=1$, $V_1 \neq V_2$): }\\
Li_{2}(\frac{V_1+x}{V_1-V_2}) + \log(V_1+x)\{\log(-\frac{V_2+x}{V_1-V_2})\} \\
\text{($\alpha_1=0$, $\alpha_2=2$, $V_1=V_2$): }\\
-\frac{\log(V_1+x)+1}{V_1+x} \\
\text{($\alpha_1=0$, $\alpha_2=2$, $V_1 \neq V_2$): }\\
\frac{(V_2+x)\log(V_2+x)-(V_1+x)\log(V_1+x)}{(V_1-V_2)(V_2+x)}.  \\
\end{cases} \nonumber \\
\end{eqnarray}
\subsection{Calculation of $G_2^{(Q)}(\beta,m_1,m_2)$.} 
$U_1(t;\beta,m_1,m_2)$ is decomposed into 
\begin{eqnarray}
&&U_1(t;\beta,m_1,m_2)= \nonumber \\
&& \sum_{i=1}^{p_{m_1}} \sum_{j=1}^{p_{m_2}} J_{Ai}^{(m_1)} J_{Aj}^{(m_2)}  G_2^{(R)}(1,Q;\beta+m_1,\beta+m_2,A^{(m_1)}_i,A^{(m_2)}_j) \nonumber \\&&- \sum_{i=1}^{p_{m_1}} \sum_{j=1}^{p_{m_2}} J_{Ai}^{(m_1)} J_{Aj}^{(m_2)}  G_2^{(I)}(0;\beta+m_1,\beta+m_2,A^{(m_1)}_i,A^{(m_2)}_j) \nonumber \\
&&+ \sum_{i=1}^{p_{m_1}} \sum_{j=1}^{p_{m_2}} J_{Ai}^{(m_1)} J_{Aj}^{(m_2)}  G_2^{(I)}(Q;\beta+m_1,\beta+m_2,A^{(m_1)}_i,A^{(m_2)}_j). \nonumber \\
\end{eqnarray}
We have already calculated $G_2^{(R)}(1,Q;\alpha_1,\alpha_2,A^{(m_1)}_i,A^{(m_2)}_j)$ as $G_2^{R \to Q}(x,\alpha_1,\alpha_2,V_1,V_2)$ in the previous sections.
Here, we calculate 
\begin{eqnarray}
&&G_2^{(Q)}(\beta,m_1,m_2) \equiv \nonumber \\
&&\sum_{i=1}^{p_{m_1}} \sum_{j=1}^{p_{m_2}} J_{Ai}^{(m_1)} J_{Aj}^{(m_2)}  G_2^{(I)}(Q;\beta+m_1,\beta+m_2,A^{(m_1)}_i,A^{(m_2)}_j) \nonumber \\
\label{G_Q}
\end{eqnarray}
for $Q>>1$.
\par
\paragraph{(i)When $\beta$ is a non-integer} \par
We study the asymptotic behaviour of $G^{(I)}(Q;\alpha_1,\alpha_2,V_1,V_2)$ in Eq. \ref{G_Q} for $Q>>1$ in the case of $V_1 \neq V_2$. 
Here, we use the following formulae of the asymptotic behaviour of the hypergeometric function \cite{hypergeom}. 
When $b-a$, $c-a$, $a$, $b$, and $c$ are non-integers for a large $x$, 
\begin{eqnarray}
&&{}_2F_1(a,b,c,x) \approx \nonumber \\
&&\frac{\Gamma(b-a) \Gamma(c)}{ \Gamma(b) \Gamma(c-a)}\frac{1}{(-x)^{a}} \nonumber \\ 
&&+\frac{(-1)^a a(1+a-c) \Gamma(b-a) \Gamma(c)}{(1+a-b)\Gamma(b) \Gamma(c-a)}\frac{1}{x^{a+1}} \nonumber \\ 
&&+\frac{(-1)^a a(1+a)(1+a-c)(2+a-c) \Gamma(b-a) \Gamma(c)}{2(1-a-b)(2-a+b)\Gamma(b) \Gamma(c-a)}\frac{1}{x^{a+2}} \nonumber \\ 
&&+\frac{\Gamma(a-b)\Gamma(c)}{\Gamma(a)\Gamma(c-b)}\frac{1}{(-x)^b}. \label{f_ap1}
\end{eqnarray}
When both $b-a$ and $c-a$ are integers and $c-b>0$ for a large $x$,
\begin{eqnarray}
&&{}_2F_1(a,b,c,x) \approx \nonumber \\
&&\frac{\Gamma(b-a) \Gamma(c)}{ \Gamma(b) \Gamma(c-a)}\frac{1}{(-x)^{a}} \nonumber \\ 
&&+\frac{(-1)^a a(1+a-c) \Gamma(b-a) \Gamma(c)}{(1+a-b)\Gamma(b) \Gamma(c-a)}\frac{1}{x^{a+1}} \nonumber \\ 
&&+\frac{(-1)^a a(1+a)(1+a-c)(2+a-c) \Gamma(b-a) \Gamma(c)}{2(1+a-b)(2+a-b)\Gamma(b) \Gamma(c-a)}\frac{1}{x^{a+2}} \nonumber \\
&&+(-1)^{b-a}\Gamma(c) \nonumber \\
&&\frac{(\log(-x)+\psi(b-a+1)-\psi(c-b)-\psi(b)-\gamma)}{\Gamma(a)\Gamma(b-a+1)\Gamma(c-b)}\frac{1}{(-x)^b}. \nonumber \\
\label{f_ap2}
\end{eqnarray}
when $b=a$ and $c-a$ are integers for a large $x$, 
\begin{eqnarray}
&&{}_2F_1(a,b,c,x) \approx \nonumber \\
&&\frac{ \Gamma(c)}{ \Gamma(a) \Gamma(c-a)}\frac{(\log(-z)-\psi(c-a)-\psi(a)-2\gamma)}{(-x)^{a}} \nonumber \\ 
&&+\frac{\Gamma(c)^2(-x)^{-c}}{\Gamma(a)^2(\Gamma(c-a+1))^2}. \nonumber \\
\label{f_ap3}
\end{eqnarray}
Here, $\gamma=0.5772..$ is the Euler constant. \par
By using these formulae, the hypergeometric function in Eq. \ref{GI_real} is written as \par
\begin{eqnarray}
&& {}_2F_1(\alpha_i,1-\alpha_j,2-\alpha_j,\frac{-(V_j+x)}{(V_i-V_j)}))  \nonumber \\
&& \approx  P_1 (\frac{(V_i-V_j)}{Q+V_j})^{\alpha_i}+ P_2 (\frac{(V_i-V_j)}{Q+V_j})^{\alpha_i-1} 
+ P_3 (\frac{(V_i-V_j)}{Q+V_j})^{\alpha_i-2}
\nonumber \\
&& + P_4 (\frac{(V_i-V_j) } {Q+V_j})^{1-\alpha_j}\log(Q) + P_5(V_i,V_j) (\frac{(V_i-V_j)}{Q+V_j})^{1-\alpha_j},  \nonumber \\
\end{eqnarray}
where
\begin{eqnarray}
P_1=
\begin{cases}
\frac{(1-\alpha_j)}{(1-\alpha_i-\alpha_j)} &  \text{($\alpha_i+\alpha_j \neq 1$)} \\
0 & \text{($\alpha_i+\alpha_j =1$)}
\end{cases}
\end{eqnarray}
\begin{eqnarray}
P_2=
\begin{cases}
\frac{-\alpha_i(\alpha_i+\alpha_j-1)}{(\alpha_i+\alpha_j)}P_1 & \text{($\alpha_i+\alpha_j \neq 0$)} \\
0 & \text{($\alpha_i+\alpha_j=0$)} 
\end{cases}
\end{eqnarray}
\begin{eqnarray}
P_3=
\begin{cases}
\frac{(1+\alpha_i)(\alpha_i+\alpha_j)}{2(1+\alpha_i+\alpha_j)}P_2 &  \text{($\alpha_i+\alpha_j \neq -1$)} \\
0& \text{($\alpha_i+\alpha_j=-1$)} \\
\end{cases}
\end{eqnarray}
\begin{eqnarray}
P_4=
\begin{cases}
0 & \text{($\alpha_i+\alpha_j$ is a non-integer)} \\
\frac{\Gamma(2-\alpha_j)}{\Gamma(\alpha_i)}  \frac{(-1)^{1-\alpha_i-\alpha_j}}{\Gamma(2-\alpha_i-\alpha_j)} & \text{($\alpha_i+\alpha_j=0,-1,1$)}
\end{cases}
\end{eqnarray}
\begin{eqnarray}
&&P_5(V_i,V_j)=\frac{\Gamma(2-\alpha_j)}{\Gamma(\alpha_i)} \times \nonumber \\  
&&\begin{cases}
\text{($\alpha_i+\alpha_j$ is a non-integer)}\\
\Gamma(\alpha_i+\alpha_j-1) \\
\text{($\alpha_i+\alpha_j=-1,0,1$)} \\
\frac{(-1)^{1-\alpha_i-\alpha_j}(-\log(V_i-V_j)+\psi(2-\alpha_i-\alpha_j)-\psi(1-\alpha_j))}{\Gamma(2-\alpha_i-\alpha_j)}. \\ 
\end{cases}
\end{eqnarray}
Substituting these results, we can obtain the following approximation, 
\begin{eqnarray}
&&G_2^{(I)}(Q;\alpha_1,\alpha_2,V_1,V_2) \approx  \frac{1}{1-\alpha_j} \left \{ \right .  \nonumber  \\
&& P_1 (Q+V_j)^{1-\alpha_i-\alpha_j}+ P_2 (V_i-V_j)(Q+V_j)^{-\alpha_i-\alpha_j} 
\nonumber \\
&&+ P_3 (V_i-V_j)^2 (Q+V_j)^{-\alpha_i-\alpha_j-1} \nonumber \\ 
&& + P_4 (V_i-V_j)^{1-\alpha_i-\alpha_j}\log(Q) \nonumber \\ 
&& \left. + P_5(V_i,V_j)\times (V_i-V_j)^{1-\alpha_i-\alpha_j} \right\}, \nonumber \\
\end{eqnarray}
where $i={\arg \max}_{k \in \{1,2\}}\{V_k\}$, $j={\arg \min}_{k \in \{1,2\}}\{V_k\}$  (under this condition, $G_2^{(I)}$ takes a real number). \par
In addition, using the asymptotic behaviour of the power law function,  
\begin{equation}
(Q+V)^{\alpha} \approx Q^{\alpha}+(\alpha)VQ^{\alpha-1}+\frac{(\alpha)(\alpha-1)}{2}V^2 Q^{\alpha-2}, \label{pow_a}
\end{equation}
we can obtain $G_2^{(I)}$ as a sum of the power law function of $Q$, 
\begin{eqnarray}
&&G_2^{(I)}(Q;\alpha_1,\alpha_2,V_1,V_2) \approx  \frac{1}{1-\alpha_j} \left \{ \right .  \nonumber  \\
&& P_1 Q^{1-\alpha_i-\alpha_j}+ (P_1(-\alpha_i-\alpha_j+1)V_j+P_2(V_i-V_j))Q^{-\alpha_i-\alpha_j} 
\nonumber \\
&&+ (\frac{P_1}{2}(-\alpha_i-\alpha_j+1)(-\alpha_i-\alpha_j)V_j^2 \nonumber \\
&&+P_2(-\alpha_i-\alpha_j)(V_i-V_j)V_j \nonumber \\
&&+P_3 (V_i-V_j)^2) Q^{-\alpha_i-\alpha_j-1} \nonumber \\ 
&& + P_4 (V_i-V_j)^{1-\alpha_i-\alpha_j}\log(Q) \nonumber \\ 
&& \left. + P_5(V_i,V_j)\times (V_i-V_j)^{1-\alpha_i-\alpha_j} \right\}. \nonumber \\
\end{eqnarray}
For $V_1=V_2$, using Eq.\ref{pow_a}, Eq. \ref{GI_AA} is approximated as
\begin{eqnarray}
&&G_2^{(I)}(x;\alpha_1,\alpha_2,V_1,V_2) \approx  \nonumber \\
&&\frac{1}{1-\alpha_1-\alpha_2}(Q+V_2)^{-\alpha_1-\alpha_2+1} \nonumber \\
&&\approx \frac{1}{1-\alpha_1-\alpha_2} (Q^{-\alpha_1-\alpha_2+1}+(-\alpha_1-\alpha_2+1)V_2Q^{-\alpha_1-\alpha_2} \nonumber \\
&&+\frac{(-\alpha_1-\alpha_2)(-\alpha_1-\alpha_2+1)}{2}V_2^2 Q^{-\alpha_1-\alpha_2-1}).
\end{eqnarray}
Hence, because $Q^{a} \to 0$ ($a<0$, $Q>>1$), 
\begin{eqnarray}
&&G_2^{(Q)}(\beta,m_1,m_2)  \approx \nonumber \\
&&\sum_{i=1}^{p_{m_1}} \sum_{j=1}^{p_{m_2}} J_{Ai}^{(m_1)} J_{Aj}^{(m_2)}  G^{(I)}(Q;\beta+m_1,\beta+m_2,A^{(m_1)}_i,A^{(m_2)}_j) \nonumber \\
&&\approx \sum_{i=1}^{p_{m_1}} \sum_{j=1}^{p_{m_2}} J_{Ai}^{(m_1)} J_{Aj}^{(m_2)} \left[ \right. \nonumber \\
&&
\begin{cases}
\frac{ P_5(V_q,V_r) \times (V_q-V_r)^{1-2\beta-m_1-m_2}}{1-\beta-m_s} & \text{($V_q \neq V_r$)} \\
0 & \text{($V_q = V_r$)} \\
\end{cases} 
\nonumber \\
&&+
\begin{cases}
0 & \text{($2\beta+m_1+m_2 \neq 0$)} \\ 
V_r & \text{($2\beta+m_1+m_2 = 0$)} \\ 
\end{cases}
\left. \right], 
\end{eqnarray}
where $V_q=\max\{A^{(m_1)}_i,A^{(m_2)}_j\}$, $V_r=\min\{A^{(m_1)}_i,A^{(m_2)}_j\}$, $s=\arg \min_{\{u=1,2\}}\{A_{q_u}^{(m_u)}\}$ and $(q_1,q_2)=(i,j)$. \par

\paragraph{Case of $\beta=1$} \par
We use the approximation formulas for $x>>1$, 
\begin{eqnarray}
Li_2(x) \approx -\frac{1}{2}\log(-x)^2-\frac{\pi^2}{6} \label{li2_app}
\end{eqnarray}
\begin{eqnarray}
\log(x+A) \approx \log(x)+\frac{A}{x} \label{log_ap}. 
\end{eqnarray}
From Eq. \ref{GI_00AB} and Eq. \ref{GI_00AA},  we can obtain 
\begin{eqnarray}
&&G_2^{(Q)}(1,-1,-1) \approx  \nonumber \\
&&\sum_{i=1}^{p_{-1}} \sum_{j=1}^{p_{-1}} J_{Ai}^{(-1)} J_{Aj}^{(-1)}  G^{(I)}(Q;0,0,A^{(-1)}_i,A^{(-1)}_j) \nonumber \\
&&\approx \sum_{i=1}^{p_{-1}} \sum_{j=1}^{p_{-1}} J_{Ai}^{(-1)} J_{Aj}^{(-1)} G_{(-1,-1)}^{(Q)}(A^{(-1)}_i,A^{(-1)}_j), 
\end{eqnarray}
where
\begin{eqnarray}
&&G_{(-1,-1)}^{(Q)}(V_1,V_2) = \nonumber \\
&&\begin{cases}
(V_2-V_1)(-\frac{1}{2}\log(V_2-V_1)^2-\frac{\pi}{6})+V_1    & (V_1 \neq V_2) \\
V_1 & (V_1=V_2).\\
\end{cases} \nonumber \\
\end{eqnarray}
\par
Similarly, from Eq. \ref{GI_01AB} and Eq. \ref{GI_01AA}, 
\begin{eqnarray}
&&G_2^{(Q)}(1,-1,0) \approx  \nonumber \\
&&\sum_{i=1}^{p_{-1}} \sum_{j=1}^{p_{0}} J_{Ai}^{(-1)} J_{Aj}^{(0)}  G^{(I)}(Q;0,0,A^{(-1)}_i,A^{(0)}_j) \nonumber \\
&&\approx \sum_{i=1}^{p_{-1}} \sum_{j=1}^{p_{0}} J_{Ai}^{(-1)} J_{Aj}^{(0)} G_{-1,0}^{(Q)}(A_i^{(-1)},A^{(0)}_j), \nonumber \\
\end{eqnarray}
where 
\begin{eqnarray}
G_{(-1,0)}^{(Q)}(V_1,V_2) =
\begin{cases}
-\frac{1}{2}\log(V_2-V_1)^2  & (V_1 \neq V_2) \\
0 & (V_1=V_2)\\
\end{cases}.
\end{eqnarray}
Because orders of Eq. \ref{GI_02AB} and Eq. \ref{GI_02AA} is $\log(Q)/Q$, 
\begin{eqnarray}
G_2^{(Q)}(1,-1,2) \approx 0.
\end{eqnarray}
In addition, from Eq. \ref{GI_int}, 
\begin{eqnarray}
G_2^{(Q)}(1,m_1,m_2) \approx 0 \quad (m_1=0,1,m_2=0,1) .
\end{eqnarray}
Note that in  Eq. \ref{GI_int} because $h_1^{(1)}=-h_1^{(2)}$, the terms of $\log(Q)$ approach zero for $Q>>1$ by cancelling each other out．
\par
\paragraph{When $\beta$ is an integer and $\beta \neq 1$ } \par 
The terms of $\log(Q)$ in Eq. \ref{GI_int} approach zero for $Q>>1$ by cancelling each other out because $h_1^{(1)}=-h_1^{(2)}$.
We use the following result, 
\begin{equation}
G_2^{(I)}(Q;\alpha_1,\alpha_2,V_1,V_2)  \approx 0. 
\end{equation}
Thus, 
\begin{equation}
G_2^{(Q)}(\beta,m_1,m_2) \approx 0. 
\end{equation}
We summarise $G_2^{(Q)}$
\begin{eqnarray}
&&G_2^{(Q)}(\beta,m_1,m_2)  \nonumber \\
&&\approx \sum_{i=1}^{p_{m_1}} \sum_{j=1}^{p_{m_2}} J_{Ai}^{(m_1)} J_{Aj}^{(m_2)}G_2^{(I;Q)}(\beta,m_1,m_2,A^{(m_1)}_i,A^{(m_2)}_j),  \nonumber \\
\end{eqnarray}
where
\begin{eqnarray}
&&G_2^{(I;Q)}(\beta,m_1,m_2,A^{(m_1)}_i,A^{(m_2)}_j)=  \nonumber \\
&&\begin{cases}
\text{($2\beta+m_1+m_2 \neq 1, 0,-1$, $A^{(m_1)}_i \neq A^{(m_2)}_j$)} \\
\frac{\Gamma(2\beta+m_1+m_2-1)\Gamma(2-\beta-m_{s})}{(1-\beta-m_{s}) \Gamma(\beta+m_{k})} (A_{q}-A_{r})^{1-2\beta-m_1-m_2}   \\
\text{($\beta$ is non-integer $2\beta+m_1+m_2=-1,1$)} \\
 \frac{(-1)^{1-2\beta-m_1-m_2} \Gamma(2-\beta-m_s)}{(1-\beta-m_s)\Gamma(\beta+m_k)} (A_{q}-A_{r})^{1-2\beta-m_1-m_2} \times \\
 \frac{(-\log(A_{q}-A_{r})+\psi(2-2\beta-m_1-m_2)-\psi(1-\beta-m_s))}{\Gamma(2-2\beta-m_1-m_2)}   \\
\text{($\beta$ is non-integer, $2\beta+m_1+m_2=0$)} \\
 \frac{(-1)^{1-2\beta-m_1-m_2} \Gamma(2-\beta-m_s)}{(1-\beta-m_s)\Gamma(\beta+m_k)} (A_{q}-A_{r})^{1-2\beta-m_1-m_2} \times \\
 \frac{(-\log(A_{q}-A_{r})+\psi(2-2\beta-m_1-m_2)-\psi(1-\beta-m_s))}{\Gamma(2-2\beta-m_1-m_2)}   \\
+A_{r} \\
\text{($\beta=1$, $m_1=-1,m_2=-1$, $A^{(m_1)}_i \neq A^{(m_2)}_j$)} \\
(A^{(m_2)}_j-A^{(m_1)}_i)(-\frac{1}{2}\log(A^{(m_2)}_j-A^{(m_1)}_i)^2-\frac{\pi}{6})+A^{(m_1)}_i  \\
\text{($\beta=1$, $m_1=-1,m_2=-1$, $A^{(m_1)}_i = A^{(m_2)}_j$)} \\
A^{(m_1)}_i \\
\text{($\beta=1$, $(m_1,m_2) \in \{(-1,0),(0,-1)\}$, $A^{(m_1)}_i \neq A^{(m_2)}_j$）} \\ 
-\frac{1}{2}\log(A^{(m_2)}_j-A^{(m_1)}_i)^2 \\
\text{($\beta=1$ and [$m_1=1$ or $m_2=1$])} \\
0 \\
\text{($\beta=2,3,4 \cdots$)} \\
0\\
\text{($A^{(m_1)}_i = A^{(m_2)}_j$, $2\beta+m_1+m_2 \neq 0$)}\\
0 \\ 
\text{($A^{(m_1)}_i = A^{(m_2)}_j$, $2\beta+m_1+m_2=0$, $\beta \neq 1$)}\\
A_{r}. \\ 
\end{cases} \nonumber \\
\end{eqnarray}
Here, $A_q=\max\{A^{(m_1)}_i,A^{(m_2)}_j\}$，$A_r=\min\{A^{(m_1)}_i,A^{(m_2)}_j\}$, $t=\arg \max_{\{u=1,2\}}\{A_{q_u}^{(m_u)}\}$, $s=\arg \min_{\{u=1,2\}}\{A_{q_u}^{(m_u)}\}$ and $(q_1,q_2)=(i,j)$.  The reason why and we change the suffix to avoid a constant of the integrations from becoming a complex number.  \par
Consequently, $U_1$ is obtained by 
\begin{eqnarray}
&&U_1(\beta,L) \approx 
\sum^{+1}_{m_1=-1}\sum^{+1}_{m_2=-1} \sum_{i=1}^{p^{(A)}_{m_1}} \sum_{j=1}^{p^{(A)}_{m_2}}  J_{A_i}^{(m_1)}J_{A_j}^{(m_2)} \{ \nonumber \\
&&G_2^{(R;Q)}(1,\beta+m_1,\beta+m_2; A_i^{(m_1)},A_j^{(m_2)}) \nonumber \\
&&+G_2^{(I;Q)}(\beta,m_1,m_2,A_i^{(m_1)},A_j^{(m_2)}) \nonumber \\
&&-G_2^{(I)}(1,\beta+m_1,\beta+m_2; A_i^{(m_1)},A_j^{(m_2)}) \}. 
 \nonumber \label{U_1_ans} \\
\end{eqnarray}

\subsection{Calculation of $U_2(\beta,L)$}
$U_2$ is defined by Eq. \ref{U_2}, 
\begin{equation}
U_2(\beta,L)=\sum^{L(I+1)}_{t=LI+2}\frac{ Z(\beta)^2}{\check{\eta}^2} (\theta_3'(t)-\theta_2'(t))^2. \label{U2_app}
\end{equation}
Substituting Eq. \ref{theta_2d} and Eq. \ref{theta_3d} into Eq. \ref{U2_app}, 
\begin{eqnarray}
&&U_2(\beta,L) \nonumber \\
&&=\sum^{L(I+1)}_{t=LI+2}\frac{ Z(\beta)^2}{\check{\eta}^2} (\sum^{L-1}_{k=0}\theta(k+L(I+1)+1-t) \nonumber \\
&&-\sum^{L(I+1)-t}_{k=0}\theta(k))^2. \nonumber \\
\end{eqnarray}
Using a shifted index $t'=t+LI+2$, $U_2$ is written as  
\begin{eqnarray}
&&U_2=\sum^{L-2}_{t'=0}\frac{ Z(\beta)^2}{\check{\eta}^2} (\sum^{L-1}_{k=0}\theta(k+L-1-t') \nonumber \\
&&-\sum^{L-2-t'}_{k=0}\theta(k))^2.  \label{U2_app2}
\end{eqnarray}
Using the Euler-Maclaurin formula in Eq. \ref{euler}, we can obtain 
\begin{eqnarray}
&&U_2 \approx  \nonumber \\
&&\frac{ Z(\beta)^2}{\check{\eta}^2} \sum^{L-3}_{t'=0} \left\{\theta(L-1-t')+\frac{\theta(L-t')+\theta(2L-2-t')}{2} \right. \nonumber  \\
&+&\frac{\theta(2L-2-t')-\theta'(L-1-t')}{12}  \nonumber \\
&+&\int^{L-1}_{1}\theta(k+L-1-t')dk \nonumber \\
&-&  \theta(0)-\frac{\theta(1)+\theta(L-2-t)}{2} \nonumber \\ 
&-&\left. \frac{\theta'(L-2-t)-\theta'(1)}{12}-\int^{L-2-t}_{1} \theta(k)  dk \right\}^2 \nonumber \\
&+&u_2^{(0)}(\beta)^2, 
\end{eqnarray}
where $u_2^{(0)}(\beta)^2$ is given by 
\begin{eqnarray}
&&u_2^{(0)}(\beta)=  \nonumber \\
&& (1+a)^{-\beta}+\frac{1}{2}((2+a)^{-\beta}+(L+a)^{-\beta}) \nonumber \\
&&+(-\beta)\frac{1}{12}((L+a)^{-\beta-1}-(2+a)^{-\beta-1})+ \nonumber \\
&&\begin{cases}
\frac{1}{-\beta+1}((L+a)^{-\beta+1}-(2+a)^{-\beta+1} & \text{($\beta \neq 1$)} \\
\log(L+a)-\log(2+a)  & \text{($\beta=1$)}\\
\end{cases}. \nonumber \\
\end{eqnarray}
Here, $u_2^{(0)}(\beta)$ corresponds to the term of $t'=L-2$ in Eq. \ref{U2_app2}. We separated and directly calculated this term in order to improve the accuracy. 
\par
Substituting Eq. \ref{theta} into $\theta(t')$ and taking some calculations, 
for $\beta>0$ and $\beta \neq 1$, we obtain
\begin{eqnarray}
&&U_2 \approx \nonumber \\
&&\sum^{L-3}_{t=0}\left\{B_0+  \sum^{p_B^{(-1)}}_{k=1}J^{(-1)}_{B,k} (-t+B^{(-1)}_k)^{-(\beta-1)} \right. \nonumber \\
&&+\left. \sum^{p_B^{(0)}}_{k=1}J^{(0)}_{B,k} ((-t+B^{(0)}_k)^{-\beta}+ \sum^{p_B^{(+1)}}_{k=1}J^{(+1)}_{B,k}(-t+B^{(+1)}_k)^{-(\beta+1)}\right\}^2 \nonumber \\
&&+u_2^{(0)}(\beta)^2, 
\end{eqnarray}
and for $\beta=1$, 
\begin{eqnarray}
&&U_2 \approx \nonumber \\
&&\sum^{L-3}_{t=0}\left\{B_0+ \sum^{p_B^{(-1)}}_{k=1}J^{(-1)}_{B,k} \log(-t+B^{(-1)}_k) \right. \nonumber \\
&&\left. + \sum^{p_B^{(0)}}_{k=1}J^{(0)}_{B,k} ((-t+B^{(0)}_k)^{-\beta}+ \sum^{p_B^{(+1)}}_{k=1}J^{+1}_{B,k}(-t+B^{(+1)}_k)^{-(\beta+1)}\right\}^2 \nonumber \\
&&+u_2^{(0)}(\beta)^2.
\end{eqnarray}
Combining these results, $U_2$ is written by  
\begin{eqnarray}
&&U_2 \approx \nonumber \\
&&\sum^{L-3}_{t=0}\left\{B_0+\sum^{+1}_{m=-1} \sum_{k=1}^{p^{(m)}_B}J^{(m)}_{B,k} f_A(-t,\beta+m,B^{(m)}_k) \right\}^2 \nonumber \\
&&+u_2^{(0)}(\beta)^2, 
\end{eqnarray}
\par
%
%
%
%
%
where 
\begin{eqnarray}
&&B_1^{(-1)}=B_1^{(0)}=B_1^{(+1)}=2L-2-a \\
&&B_2^{(-1)}=B_2^{(0)}=B_2^{(+1)}=L+a \\
&&B_3^{(-1)}=B_4^{(0)}=B_3^{(+1)}=L-2+a \\
&&B_3^{(0)}=L-1+a, 
\end{eqnarray}
$p^{(-1)}_B=3$, $p^{(0)}_B=4$, $p^{(+1)}_B=3$, 
\begin{eqnarray}
B_0=
\begin{cases}
-a^{-\beta}-\frac{1}{2}(1+a)^{-\beta}-\frac{\beta}{12}(1+a)^{-\beta-1}+\frac{(a+1)^{-\beta+1}}{-\beta+1}\\
-a^{-\beta}-\frac{1}{2}(1+a)^{-\beta}-\frac{\beta}{12}(1+a)^{-\beta-1}+\log(a+1), \\
\end{cases} \nonumber \\
\end{eqnarray}
\begin{eqnarray}
&&J_{B,1}^{(-1)}=
\begin{cases}
1/(-\beta+1)& (\beta \neq 1) \\
1 & (\beta = 1) 
\end{cases} \\
&&J_{B,2}^{(-1)}=J_{B,3}^{(-1)}=
\begin{cases}
-1/(-\beta+1)& (\beta \neq 1) \\
-1 & (\beta = 1) 
\end{cases}, 
\end{eqnarray}
\begin{eqnarray}
&&J_{B,1}^{(0)}=J_{B,3}^{(0)}=1/2 \\
&&J_{B,2}^{(0)}=1 \\
&&J_{B,4}^{(0)}=-1/2,  
\end{eqnarray}
and 
\begin{eqnarray}
&&J_{B,1}^{(+1)}=-\beta/12\\
&&J_{B,2}^{(+1)}=J_{B,2}^{(+1)}=\beta/12.
\end{eqnarray}
 \par
Expanding the squared term, 
\begin{eqnarray}
&&U_2 \approx \nonumber \\
&&\sum^{L-3}_{t'=0}\left\{B_0+\sum^{+1}_{m=-1} \sum_{k=1}^{p_m}J^{(m)}_{B,k} f_A(t,\alpha+m,B^{(m)}_k) \right\}^2= \nonumber \\
&&B_0^2(L-2)+2B_0\sum^{+1}_{m=-1}U^{(1)}_2(-t;\beta,m) \nonumber \\
&&+\sum^{+1}_{m_1=-1}\sum^{+1}_{m_2=-1}U^{(2)}_2(t;\beta,m_1,m_2) \nonumber \\
&&+u_2^{(0)}(\beta)^2, \label{u2_expand}
\end{eqnarray}
\par
where 
\begin{eqnarray}
&&U^{(1)}_2(t;\beta,m_1)=  \sum_{i=1}^{p^{(B)}_{m_1}}  J_{Bi}^{(m_1)}  
  \sum^{L-2}_{t=0}f_A(-t,\beta+m_i,B^{(m)}_i), 
 \nonumber \\
\end{eqnarray}
and 
\begin{eqnarray}
&&U^{(2)}_2(t;\beta,m_1,m_2)= \nonumber \\
&& \sum_{i=1}^{p^{(B)}_{m_1}} \sum_{j=1}^{p^{(B)}_{m_2}} J_{Bi}^{(m_1)} J_{Bj}^{(m_2)} \nonumber \\
&& \sum^{L-2}_{t=0}f_A(-t,\beta+m_i,B^{(m)}_i)f_A(-t,\beta+m_j,B^{(m)}_j).
 \nonumber \\
\end{eqnarray}
%
Using the Euler-Maclaurin formula in Eq. \ref{euler}, we can approximate the sum in $U^{(1)}_2(t;\beta,m_1,m_2))$ as 
for $L \geq 3$, 
\begin{eqnarray}
&&\sum^{L-3}_{t=0}f_A(t,\alpha_1,V_1) \\
&&\approx G_1^{(R)}(3-L,\max\{-1,3-L\};\alpha_1,V_1) \nonumber \\
&&+\int^{\max\{L-3,1\}}_1 (-t+V_1)^{-\alpha_1} dt \nonumber \\
&&\approx G_1^{(R)}(3-L,\max\{-1,3-L\},;\alpha_1,V_1) \nonumber \\
&&-G_1^{(I)}(\min\{-L+3,-1\};\alpha_1,V_1) \nonumber \\
&&+G_1^{(I)}(-1;\alpha_1,V_1), \nonumber \\
\end{eqnarray}
where $G_1^{(R)}(x_1,x_2;\alpha_1,V_1)$ is given by 
for $\alpha_1 \neq 1$ and $\alpha_2 \neq 1$, 
\begin{eqnarray}
&&G_1^{(R)}(x_1,x_2;\alpha_1,V_1)=(V_1)^{-\alpha_1}+ \nonumber \\
&&\frac{1}{2}\{(V_1+x_1)^{-\alpha_1}+(V_1+x_2)^{-\alpha_1}\}  \nonumber \\
&&-\frac{1}{12}\{(-\alpha_1)(V_1+x_2)^{-\alpha_1-1} \nonumber \\
&&-(-\alpha_1)(V_1+x_1)^{-\alpha_1-1}\},  \nonumber \\  \label{u21_ans1}
\end{eqnarray}
and for $x_1=0$, $x_2=0$ we define
\begin{eqnarray}
&&G_1^{(R)}(x_1,x_2;\alpha_1,V_1)=V_1^{-\alpha_1}.
\end{eqnarray}
For $\alpha_1 = 0$, the corresponding term is also written as 
\begin{eqnarray}
&&G_1^{(R)}(x_1,x_2;\alpha_1,V_1)= \nonumber \\
&&=\log(V_1)+ \frac{1}{2}\{\log(V_1+x_1)+\log(V_1+x_2)\}  \nonumber \\
&&-\frac{1}{12}\{(V_1+x_2)^{-1} \nonumber \\
&&-(V_1+x_1)^{-1}\},  \nonumber \\ 
\end{eqnarray}
and for $x_1=0$, $x_2=0$ we define
\begin{eqnarray}
&&G_1^{(R)}(x_1,x_2;\alpha_1,V_1)=\log(V_1).
\end{eqnarray}
In addition, $G_1^{(I)}(x;\alpha,V)$ is calculated as 
\begin{eqnarray}
&&G_1^{(I)}(x;\alpha,V)= \nonumber \\
&&
\begin{cases}
\text{($\alpha \neq 0$, $\alpha \neq 1$)} \\
\int (x+V)^{-\alpha} dx = \frac{1}{-\alpha+1} (x+V)^{-\alpha+1} \\
\text{($\alpha=1$)} \\
\int (x+V)^{-1} dx= \log(x+V)  \\
 \text{($\alpha=0$)} \\
\int \log(x+V) dx= (x+V)\log(x+V)-x.  \\ 
\end{cases} \label{u21_ans2}
\end{eqnarray}
Here, we omit a constant of integration.  \par
Next, we calculate $U^{(2)}_2(t;\beta,m_1,m_2))$ in Eq. \ref{u2_expand}. 
Using the Euler-Maclaurin formula in Eq. \ref{euler}, we can approximate the sum in $U^{(2)}_2(t;\beta,m_1,m_2))$ as 
\begin{eqnarray}
&&\sum^{L-3}_{t=0}f_A(-t,\alpha_1,V_1)f_A(-t,\alpha_2,V_2) \\
&&\approx G_2^{(R)}(3-L,\max\{-1,3-L\};\alpha_1,\alpha_2,V_1,V_2) \nonumber \\
&&+\int^{\max\{L-3,1\}}_1 (-t+V_1)^{-\alpha_1}(-t+V_2)^{-\alpha_2} dt \nonumber \\
&&\approx G_2^{(R)}(3-L,\max\{-1,3-L\},;\alpha_1,\alpha_2,V_1,V_2) \nonumber \\
&&-G_2^{(I)}(\min\{3-L,1\};\alpha_1,\alpha_2,V_1,V_2) \nonumber \\
&&+G_2^{(I)}(-1;\alpha_1,\alpha_2,V_1,V_2) . \label{u22_ans}
\end{eqnarray}
Thus, substituting Eqs. \ref{u21_ans1}, \ref{u21_ans2} and \ref{u22_ans} into Eq. \ref{u2_expand}, for $L>3$,  
\begin{eqnarray}
&&U_2 \approx 
(L-2)B_0^2 \nonumber \\
&&+2B_0\sum^{+1}_{m=-1}\sum_{i=1}^{p^{(B)}_{m}}  J_{Bi}^{(m)} \{ 
G_1^{(R)}(-1,3-L,\beta+m,B_i^{(m)}) \nonumber \\
&&-G_1^{(I)}(-L+3,\beta+m,B_i^{(m)})  \nonumber \\
&&+G_1^{(I)}(-1,\beta+m,B_i^{(m)}) \} \nonumber \\
&&+\sum^{+1}_{m_1=-1}\sum^{+1}_{m_2=-1} \sum_{i=1}^{p^{(B)}_{m_1}} \sum_{j=1}^{p^{(B)}_{m_2}}  J_{Bi}^{(m_1)}J_{Bj}^{(m_2)} \{ \nonumber \\
&&G_2^{(R)}(3-L,-1,\beta+m_1,\beta+m_2; B_i^{(m_1)},B_j^{(m_2)}) \nonumber \\
&&-G_2^{(I)}(3-L,\beta+m_1,\beta+m_2; B_i^{(m_1)},B_j^{(m_2)}) \nonumber \\
&&+G_2^{(I)}(-1,\beta+m_1,\beta+m_2; B_i^{(m_1)},B_j^{(m_2)}) \} \nonumber \\
&+& u^{(0)}_2(\beta)^2 \label{u2_ans}
\end{eqnarray}
for $L=3$, 
\begin{eqnarray}
&&U_2 \approx 
(L-2)B_0^2 \nonumber \\
&&+2B_0\sum^{+1}_{m=-1}\sum_{i=1}^{p^{(B)}_{m}}  J_{Bi}^{(m)} 
G_1^{(R)}(0,0,\beta+m,B_i^{(m)}) \nonumber \\
&&+\sum^{+1}_{m_1=-1}\sum^{+1}_{m_2=-1} \sum_{i=1}^{p^{(B)}_{m_1}} \sum_{j=1}^{p^{(B)}_{m_2}}  J_{Bi}^{(m_1)}J_{Bj}^{(m_2)} \{ \nonumber \\
&&
G_2^{(R)}(0,0,\beta+m_1,\beta+m_2; B_i^{(m_1)},B_j^{(m_2)}) \} \nonumber \\
&+& u^{(0)}_2(\beta)^2. 
\end{eqnarray}

For $L=2$, from Eq. \ref{u21_ans2}, 
\begin{eqnarray}
&&U_2 \approx  u^{(0)}_2(\beta)^2. 
\end{eqnarray}
Note that because we cannot use the integral approximation method, we calculated directly from the sums for $L=2$.
\par
\subsection{Calculation of $U_3(\beta,L)$}
$U_3$ is defined by Eq. \ref{U_3},
\begin{equation}
U_3= \frac{ Z(\beta)^2}{\check{\eta}^2} \sum^{L(I+2)}_{t=L(I+1)+1}(\theta_4'(t))^2, \label{U3_app} 
\end{equation}
Substituting Eq. \ref{theta_2d} and Eq. \ref{theta_3d} into Eq. \ref{U3_app}, 
\begin{eqnarray}
U_3&=&  \sum^{L(I+2)}_{t=L(I+1)+1}\sum^{L-(t-L(I+1))}_{k=0}(k+a)^{-\beta}.
\end{eqnarray}
Using a shifted index $t'=t-L(I+1)-1$, we can write
\begin{eqnarray}
U_3&=& \sum^{L-1}_{t'=0}(\sum^{L-(1+t')}_{k=0}(k+a)^{-\beta})^2. \\
\end{eqnarray}
Using the Euler-Maclaurin formula in Eq. \ref{euler}, we can obtain 
\begin{eqnarray}
&&U_3 \approx  \nonumber \\
&&\frac{ Z(\beta)^2}{\check{\eta}^2} \sum^{L-2}_{t'=0} \left\{\theta(0)+\frac{\theta(1)+\theta(L-1-t))}{2} \right. \nonumber  \\
&+&\left. \frac{\theta'(L-1-t)-\theta'(1)}{12} +\int^{L-1-t}_{1}\theta(k)dk \right\}^2 \nonumber \\
&+&\theta(0)^2.
\end{eqnarray}
Here, because we cannot use the integral approximation method, we calculated directly from the sums for $t'=L-1$.
\par 
Subsitituting Eq. \ref{theta} into $\theta(t')$, 
for $\beta>0$ and $\beta \neq 1$, 
\begin{eqnarray}
&&U_3 \approx \nonumber \\
&&\sum^{L-2}_{t=0}\left\{C_0+  \sum^{p_C^{(-1)}}_{k=1}J^{(-1)}_{C,k} (-t+C^{(-1)}_k)^{-(\beta-1)} \right. \nonumber \\
&&+\left. \sum^{p_C^{(0)}}_{k=1}J^{(0)}_{C,k} ((-t+C^{(0)}_k)^{-\beta}+ \sum^{p_C^{(+1)}}_{k=1}J^{(+1)}_{C,k}(-t+C^{(+1)}_k)^{-(\beta+1)}\right\}^2 \nonumber \\
&&+a^{-2\beta},  
\end{eqnarray}
and for $\beta=1$
\begin{eqnarray}
&&U_3 \approx \nonumber \\
&&\sum^{L-2}_{t=0}\left\{C_0+ \sum^{p_C^{(-1)}}_{k=1}J^{(-1)}_{C,k} \log(-t+C^{(-1)}_k) \right. \nonumber \\
&&\left. + \sum^{p_C^{(0)}}_{k=1}J^{(0)}_{C,k} ((-t+C^{(0)}_k)^{-\beta}+ \sum^{p_C^{(+1)}}_{k=1}J^{+1}_{C,k}(-t+C^{(+1)}_k)^{-(\beta+1)}\right\}^2 \nonumber \\
&&+a^{-2\beta}. 
\end{eqnarray}
Combining two equations into one,  
\begin{eqnarray}
&&U_3 \approx \nonumber \\
&&\sum^{L-2}_{t=0}\left\{C_0+\sum^{+1}_{m=-1} \sum_{k=1}^{p^{(m)}_C}J^{(m)}_{C,k} f_A(-t,\beta+m,C^{(m)}_k) \right\}^2 \nonumber \\
&&+a^{-2\beta} 
\end{eqnarray}
\par
%
%
%
%
%
where
\begin{eqnarray}
&&C_1^{(-1)}=C_1^{(0)}=C_1^{(-1)}=L-1+a, \\
\end{eqnarray}
$p^{(-1)}_C=1$, $p^{(0)}_C=1$, $p^{(+1)}_C=1$, 
\begin{eqnarray}
C_0=
\begin{cases}
a^{-\beta}+\frac{1}{2}(1+a)^{-\beta}+\frac{\beta}{12}(1+a)^{-\beta-1}-\frac{(a+1)^{-\beta+1}}{-\beta+1}\\
a^{-\beta}+\frac{1}{2}(1+a)^{-\beta}+\frac{\beta}{12}(1+a)^{-\beta-1}-\log(a+1), \\
\end{cases}
\end{eqnarray}
\begin{eqnarray}
&&J_{C,1}^{(-1)}=
\begin{cases}
1/(-\beta+1)& (\beta \neq 1) \\
1 & (\beta = 1) 
\end{cases}, 
\end{eqnarray}
\begin{eqnarray}
&&J_{C,1}^{(0)}=1/2,  
\end{eqnarray}
and
\begin{eqnarray}
&&J_{C,1}^{(+1)}=-\beta/12.
\end{eqnarray}
\par
Expanding the squared term, 
\begin{eqnarray}
&&U_3 \approx \nonumber \\
&&\sum^{L-2}_{t'=0}\left\{C_0+\sum^{+1}_{m=-1} \sum_{k=1}^{p_m}J^{(m)}_{B,k} f_A(t,\alpha+m,B^{(m)}_k) \right\}^2= \nonumber \\
&&C_0^2+2C_0\sum^{+1}_{m=-1}U^{(1)}_3(-t;\beta,m) \nonumber \\
&&+\sum^{+1}_{m_1=-1}\sum^{+1}_{m_2=-1}U^{(2)}_3(t;\beta,m_1,m_2) \nonumber \\
&&+a^{-2\beta}, \label{u3_expand}
\end{eqnarray}
\par
where 
\begin{eqnarray}
&&U^{(1)}_3(t;\beta,m)=   \nonumber \\ 
&& \sum_{i=1}^{p^{(C)}_{m}}  J_{Ci}^{(m)}  \sum^{L-2}_{t=0}f_A(-t,\beta+m,C^{(m)}_i)
 \nonumber \\
\end{eqnarray}
and
\begin{eqnarray}
&&U^{(2)}_3(t;\beta,m_1,m_2)= \nonumber \\
&& \sum_{i=1}^{p^{(C)}_{m_1}} \sum_{j=1}^{p^{(C)}_{m_2}} J_{Ci}^{(m_1)} J_{Cj}^{(m_2)} \nonumber \\
&& \sum^{L-2}_{t=0}f_A(-t,\beta+m_i,C^{(m)}_i)f_A(-t,\beta+m_j,C^{(m)}_j). \nonumber \\
\end{eqnarray}
%
Consequently, as with $U_2$ for $L>2$, $U_3$ is also calculated by 
\begin{eqnarray}
&&U_3 \approx 
(L-1)C_0^2 \nonumber \\
&&+2C_0\sum^{+1}_{m=-1}\sum_{i=1}^{p^{(C)}_{m}}  J_{C_i}^{(m)} \{ 
G_1^{(R)}(-1,2-L,\beta+m,C_i^{(m)}) \nonumber \\
&&-G_1^{(I)}(-L+2,\beta+m,C_i^{(m)})  \nonumber \\
&&+G_1^{(I)}(-1,\beta+m,C_i^{(m)}) \} \nonumber \\
&&+\sum^{+1}_{m_1=-1}\sum^{+1}_{m_2=-1} \sum_{i=1}^{p^{(C)}_{m_1}} \sum_{j=1}^{p^{(C)}_{m_2}}  J_{C_i}^{(m_1)}J_{C_j}^{(m_2)} \{ \nonumber \\
&&G_2^{(R)}(2-L,-1,\beta+m_1,\beta+m_2; B_i^{(m_1)},B_j^{(m_2)}) \nonumber \\
&&-G_2^{(I)}(2-L,\beta+m_1,\beta+m_2; C_i^{(m_1)},C_jj^{(m_2)}) \nonumber \\
&&+G_2^{(I)}(-1,\beta+m_1,\beta+m_2; C_i^{(m_1)},C_j^{(m_2)}) \} \label{u3_ans}
 \nonumber \\
&&+a^{-2\beta} , 
\end{eqnarray}
for $L=2$, 
\begin{eqnarray}
&&U_3 \approx 
(L-1)C_0^2 \nonumber \\
&&+2C_0\sum^{+1}_{m=-1}\sum_{i=1}^{p^{(C)}_{m}}  J_{C_i}^{(m)} 
G_1^{(R)}(0,0,\beta+m,C_i^{(m)}) \nonumber \\
&&+\sum^{+1}_{m_1=-1}\sum^{+1}_{m_2=-1} \sum_{i=1}^{p^{(C)}_{m_1}} \sum_{j=1}^{p^{(C)}_{m_2}}  J_{C_i}^{(m_1)}J_{C_j}^{(m_2)} \{ \nonumber \\
&&
G_2^{(R)}(0,0,\beta+m_1,\beta+m_2; C_i^{(m_1)},C_j^{(m_2)}) \} \nonumber \\
 \nonumber \\
&&+a^{-2\beta}. 
\end{eqnarray}
\par
\subsection{Calculation of $V[R_j^{(L)}]$}
We can obtain $U(\beta,L)$ by substituting $U_1(\beta,L)$ (Eq. \ref{U_1_ans}), $U_2(\beta,L)$ (Eq. \ref{u2_ans}), and $U_3(\beta,L)$ (Eq. \ref{u3_ans}) into Eq. \ref{u_ans}.
\renewcommand{\theequation}{E\arabic{equation}}
\renewcommand{\thefigure}{E-\arabic{figure}}
\renewcommand{\thetable}{S-E\arabic{table}}
\setcounter{figure}{0}
\setcounter{equation}{0}
\section{Asymptotic behaviour of  $V[R_j^{(L)}]$ in the case of the power-law forgetting process for $L>>1$}
\label{app_l_forget}
In this section, we calculate the asymptotic behaviour of  $V[R_j^{(L)}]$  for $L>>1$.
Because  $V[R_j^{(L)}]$ is decomposed into $U_1(\beta,L)$, $U_2(\beta,L)$ and $U_3(\beta,L)$ (Eq. \ref{u_ans}), we calculate the asymptotic behaviours of $U_1(\beta,L)$ (Eq. \ref{U_1_ans}), $U_2(\beta,L)$ (Eq. \ref{u2_ans}), and $U_3(\beta,L)$(Eq. \ref{u3_ans}), respectively.
 \par
\subsection{Asymptotic behaviour of $U_1(\beta,L)$ for $L>>1$}
The dominant terms of Eq. \ref{U_1_ans} are the cases of $m_1=-1$ and $m_2=-1$, namely, 
\begin{equation}
\sum^{p^{(A)}_{(-1)}}_{i=1} \sum^{p^{(A)}_{(-1)}}_{j=1} J^{(-1)}_{A_i}J^{(-1)}_{A_j} G_2^{(I;Q)}(\beta,-1,-1,A_i^{(-1)},A_j^{(-1)})
\end{equation}
and
\begin{equation}
\sum^{p^{(A)}_{(-1)}}_{i=1} \sum^{p^{(A)}_{(-1)}}_{j=1} J^{(-1)}_{A_i}J^{(-1)}_{A_j} G_2^{(I)}(1,\beta-1,\beta-1,A_i^{(-1)},A_j^{(-1}).
\end{equation}
Calculating these terms for $L>>1$, we can write 
\begin{eqnarray}
&&U_1(\beta,L) \approx 
u_1^{(1)}(\beta)L^{3-2\beta}
\label{u1_l}
\end{eqnarray}
for $\beta \neq 1$ and $\beta>0$, 
\begin{eqnarray}
&&u_1^{(1)}(\beta)=\frac{1}{(2-\beta)(1-\beta)^2} \times \nonumber \\
&&\biggl\{ 4 {}_2F_1(\beta-1,2-\beta,3-\beta,-1) -\frac{(2-\beta)(4+2^{3-2\beta})}{3-2\beta} \nonumber \\
&&+2 \frac{\Gamma(3-\beta)}{\Gamma(\beta-1)}(2^{3-2\beta}q_1(\beta,L)-4q_2(\beta,L)) \biggr\} \nonumber \\
\end{eqnarray}
for $\beta=1$, 
\begin{eqnarray}
&&u_1^{(1)}(\beta)= \nonumber \\
&&-\frac{4\pi^2}{6}-4\psi(2)-4\log(2)^2-2\log(-1)^2 
\end{eqnarray}
%
In addition, 
\begin{eqnarray}
&&q_1(\beta,L)= \nonumber \\
&&\begin{cases}
\frac{-log(2)+\psi(3-2\beta)+\psi(2-\beta)}{\Gamma(4-2\beta)} & \text{($\beta=0.5$)} \\
\Gamma(2\beta-3) & \text{($\beta \neq 0.5$)}\\
\end{cases}
\end{eqnarray}
and 
\begin{eqnarray}
&&q_2(\beta,L)= \nonumber \\
&&\begin{cases}
\frac{(\psi(3-2 \beta)+\psi(2-\beta))}{\Gamma(4-2\beta)} & \text{($\beta=0.5$)} \\
\Gamma(2 \beta-3) & \text{($\beta \neq 0.5$)}. \\
\end{cases}
\end{eqnarray}
\par
Here, for the calculations, we use these approximation formulas of the hypergeometric function \cite{hypergeom} 
\begin{equation}
{}_2F_1(a,b,c;x) \to 1  \quad (x \to 0),  \label{f_ap0}
\end{equation}
Eq. \ref{f_ap1}，Eq. \ref{f_ap2}, Eq. \ref{f_ap3} and the logarithmic function Eq. \ref{log_ap}．In addition, we replace $L+Const.$ with $L$．
\par 
\subsection{Asymptotic behaviour of $U_2(\beta,L)$ for L>>1} \par
As with $U_1(\beta,L)$, we can calculate the asymptotic behaviour of $U_2(\beta,L)$.
We focus on the higher order terms than $O(L)$ in Eq. \ref{u2_ans}.
The highest order terms are 
\begin{equation}
\sum^{p^{(B)}_{(-1)}}_{i=1} \sum^{p^{(B)}_{(-1)}}_{j=1} J^{(-1)}_{B_i}J^{(-1)}_{B_j} G_2^{(I;Q)}(3-L,\beta-1,\beta-1,B_i^{(-1)},B_j^{(-1)})
\end{equation}
and
\begin{equation}
\sum^{p^{(B)}_{(-1)}}_{i=1} \sum^{p^{(B)}_{(-1)}}_{j=1} J^{(-1)}_{B_i}J^{(-1)}_{B_j} G_2^{(I)}(-1,\beta-1,\beta-1,B_i^{(-1)},B_j^{(-1)}). 
\end{equation}
The second highest order terms are  
\begin{equation}
2B_0 \sum^{p^{(B)}_{(-1)}}_{i=1} J^{(-1)}_{B_i} G_1^{(I;Q)}(3-L,\beta-1,B_i^{(-1)})
\end{equation}
and
\begin{equation}
2B_0 \sum^{p^{(B)}_{(-1)}}_{i=1} J^{(-1)}_{B_i} G_1^{(I;Q)}(-1,\beta-1,B_i^{(-1)}).
\end{equation}
The term of $O(L)$ is 
\begin{equation}
(L-1)B_0^2.
\end{equation}
Calculating these terms, we can obtain 
\begin{eqnarray}
&&U_2(\beta,L) \approx \nonumber \\
&&\begin{cases}
u^{(2)}_1(\beta)L^{3-2\beta} u^{(2)}_2(\beta)L^{2-\beta}+u^{(2)}_3(\beta)L &\text{($0<\beta<1$)} \\
u^{(2)}_a \log(L)^2 L + u^{(2)}_b \log(L)L+u^{(2)}_cL &\text{($\beta=1$)} \\
u^{(2)}_3(\beta)L & \text{($\beta>1$)} \label{u2_l}, 
\end{cases}
\end{eqnarray}
where 
\begin{eqnarray}
&&u^{(2)}_1(\beta)=\frac{-4 {}_2F_1(\beta-1,2-\beta,3-\beta,-1)}{(2-\beta)(1-\beta)^2}+\frac{(2^{3-2\beta}+3)}{(3-2\beta)(1-\beta)^2}  \nonumber \\
\end{eqnarray}
\begin{eqnarray}
&&u^{(2)}_2(\beta)=2 B_0 \frac{2^{2-\beta}-3}{(2-\beta)(1-\beta)} 
\end{eqnarray}
\begin{eqnarray}
&&u^{(2)}_3(\beta)= B_0^2  
\end{eqnarray}
\begin{eqnarray}
&&u^{(2)}_a=1  
\end{eqnarray}
\begin{eqnarray}
&&u^{(2)}_b=-4\log(2)-2-2B_0
\end{eqnarray}
\begin{eqnarray}
&&u^{(2)}_c= 2+4\psi(2)-4\psi(1) \nonumber  \\
&&+\log(2)(4+2\log(2)-4\log(-1)+4B_0) .
\end{eqnarray}
Here, for the calculations, we use these approximation formulas of the hypergeometric function Eq. \ref{f_ap0}, Eq. \ref{f_ap1}，Eq. \ref{f_ap2}，Eq. \ref{f_ap3}, 
and the logarithmic function Eq. \ref{log_ap}．In addition, we replace $L+Const.$ with $L$．
\subsection{Asymptotic behaviour of $U_3(\beta,L)$ for L>>1} \par
As with $U_2(\beta,L)$, we can calculate the asymptotic behaviour of $U_2(\beta,L)$.
We focus on higher order terms than $O(L)$ in Eq. \ref{u3_ans}.
The highest order terms are 
\begin{equation}
\sum^{p^{(C)}_{(-1)}}_{i=1} \sum^{p^{(C)}_{(-1)}}_{j=1} J^{(-1)}_{C_i}J^{(-1)}_{C_j} G_2^{(I;Q)}(2-L,\beta-1,\beta-1,C_i^{(-1)},C_j^{(-1)}), 
\end{equation}
and 
\begin{equation}
\sum^{p^{(C)}_{(-1)}}_{i=1} \sum^{p^{(C)}_{(-1)}}_{j=1} J^{(-1)}_{C_i}J^{(-1)}_{C_j} G_2^{(I)}(-1,\beta-1,\beta-1,C_i^{(-1)},C_j^{(-1)}).
\end{equation}
The second highest order terms are  
\begin{equation}
2C_0 \sum^{p^{(C)}_{(-1)}}_{i=1} J^{(-1)}_{C_i} G_1^{(I;Q)}(2-L,\beta-1,C_i^{(-1)}), 
\end{equation}
and
\begin{equation}
2C_0 \sum^{p^{(C)}_{(-1)}}_{i=1} J^{(-1)}_{C_i} G_1^{(I;Q)}(-1,\beta-1,C_i^{(-1)}). 
\end{equation}
The term of O(L) is 
\begin{equation}
(L-1)C_0^2.
\end{equation}
Calculating these terms, we can obtain 
\begin{eqnarray}
&&U_3(\beta,L) \approx \nonumber \\
&&\begin{cases}
u^{(3)}_1(\beta)L^{3-2\beta} u^{(3)}_2(\beta)L^{2-\beta}+u^{(3)}_3(\beta)L &\text{($0<\beta<1$)} \\
u^{(3)}_a \log(L)^2 L + u^{(3)}_b \log(L)L+u^{(3)}_cL &\text{($\beta=1$)} \\
u^{(3)}_3(\beta)L & \text{($\beta>1$)} \label{u3_l} 
\end{cases}, 
\end{eqnarray}
where 
\begin{eqnarray}
&&u^{(3)}_1(\beta)=\frac{1}{(1-\beta)^2 (3-2\beta)}  \nonumber \\
\end{eqnarray}
\begin{eqnarray}
&&u^{(3)}_2(\beta)=2 C_0 \frac{2-\beta}{1-\beta}
\end{eqnarray}
\begin{eqnarray}
&&u^{(3)}_3(\beta)= C_0^2  
\end{eqnarray}
\begin{eqnarray}
&&u^{(3)}_a=1  
\end{eqnarray}
\begin{eqnarray}
&&u^{(3)}_b=(-2+2C_0)
\end{eqnarray}
\begin{eqnarray}
&&u^{(3)}_c= -1+C_0^2+2C_0. 
\end{eqnarray}
Here, for the calculations, we use the approximation formulas of the hypergeometric functions in Eq. \ref{f_ap0}, Eq. \ref{f_ap1}，Eq. \ref{f_ap2}，Eq. \ref{f_ap3}, 
and the logarithmic function in Eq. \ref{log_ap}．In addition, we replace $L+Const.$ with $L$． \par
\subsection{Asymptotic behaviour of $V[R_j^{(L)}]$ for $L>>1$}
Substituting Eq. \ref{u1_l}, Eq. \ref{u2_l}, and Eq. \ref{u3_l} into \ref{u_ans}, we can obtain
\begin{eqnarray}
&&V[\delta R_j^{(L)}] \approx \frac{\check{\eta}^2}{Z(\beta)^2} \times  \nonumber \\
&&\begin{cases}
u_1(\beta)L^{1-2\beta}+u_2(\beta)L^{-\beta}+u_3(\beta)L^{-1} &\text{($0<\beta<1$)} \\
u_a \log(L)^2 L^{-1} + u_b \log(L) L^{-1}+u_c L^{-1} &\text{($\beta=1$)} \\
u_3(\beta)L^{-1} & \text{($\beta>1$)} 
\end{cases} \label{u_ans_l}
\end{eqnarray}
\begin{eqnarray}
&&u_1(\beta)=u_1^{(1)}(\beta)+u_1^{(2)}(\beta)+u_1^{(3)}(\beta)  \nonumber \\
\end{eqnarray}
\begin{eqnarray}
&&u_2(\beta)=u_2^{(2)}(\beta)+u_2^{(3)}(\beta)
\end{eqnarray}
\begin{eqnarray}
&&u_3(\beta)= u_3^{(2)}(\beta)+u_3^{(3)}(\beta) 
\end{eqnarray}
\begin{eqnarray}
&&u_a=1  
\end{eqnarray}
\begin{eqnarray}
&&u_b=(-2+2C_0)
\end{eqnarray}
\begin{eqnarray}
&&u_c= -1+C_0^2+2C_0. \label{u_ans_l_end}
\end{eqnarray}
\par
Consequently, the highest order term is obtained by  
\begin{eqnarray}
V[\delta R_j^{(L)}] \propto 
\begin{cases}
L^{1-2\beta} &\text{($0<\beta<1$)} \\
\log(L)^2 L^{-1} &\text{($\beta=1$)} \\
L^{-1} & \text{($\beta>1$)}. 
\end{cases}
\end{eqnarray}
\renewcommand{\theequation}{F\arabic{equation}}
\renewcommand{\thefigure}{F-\arabic{figure}}
\renewcommand{\thetable}{S-F\arabic{table}}
\setcounter{figure}{0}
\setcounter{equation}{0}
\section{Mean square displacement of power-law forgetting process}
\label{MSD_forget}
We calculate the MSD of the following power-law forgetting process given by Eq. \ref{r_forget}, 
\begin{equation}
r(t)=\sum_{s=0}^{\infty} \theta(s) \cdot \eta(t-s), 
\end{equation}
where 
\begin{equation} 
\theta(s)=\frac{(s+a(\beta))^{-\beta}}{Z(\beta)}, \label{theta}
\end{equation}
and
\begin{equation}
a \equiv a(\beta) \equiv  Z(\beta)^{-1/\beta}, 
\end{equation}
$Z(\beta)>0$ is an arbitrary constant.
The MSD can be calculated as 
\begin{eqnarray}
&&<(r(t+L)-r(t))^2> \nonumber \\
&=&<(\sum^{\infty}_{s=0} \theta(s)  \eta(t+L-s) - \sum^{\infty}_{s'=0}\theta(s')  \eta(t-s'))^2> \\
&=&<\left\{\sum^{-1}_{s=-L} \theta(s+L) \eta(t-s) +\sum^{\infty}_{s=0} (\theta(s+L) \right. \nonumber \\
&-&\left. \theta(s) ) \eta(t-s) \right\}^2> \\ 
&=&\check{\eta}^2  (\sum^{-1}_{s=-L} \theta(s+L)^2 +\sum^{\infty}_{s=0} (\theta(s+L) - \theta(s) )^2)  \\ 
&=&\check{\eta}^2  (S_1+S_2),  
\label{ss}
\end{eqnarray}
where 
\begin{eqnarray}
S_1&=&\sum^{-1}_{s=-L} \theta(s+L)^2 \\
&=&\begin{cases}
\frac{1}{Z(\beta)^2}(\xi(2\beta,a)-\xi(2\beta,a+L)) & (\beta \neq 1/2, \beta \geq 0) \\ 
\frac{1}{Z(\beta)^2}(\psi^{(0)}(a+L)-\psi^{(0)}(a)) & (\beta=1/2) \\ 
\end{cases},   \nonumber \\
\label{s1_ans}
\end{eqnarray}
and
\begin{eqnarray}
S_2&=&\sum^{\infty}_{s=0} (\theta(s+L) - \theta(s) )^2 \nonumber \\
&=&\lim_{N \to \infty}\sum^{N}_{s=0} (\theta(s+L) - \theta(s) )^2 \\
&=&\lim_{N \to \infty}\sum^{N}_{s=0} \theta(s+L)^2+\sum^{N}_{s=0}  \theta(s)^2  -2 \sum^{N}_{s=0} \theta(s+L) \theta(s). \nonumber \\ \label{S_2}
\end{eqnarray}
\par
We calculate three terms in Eq. \ref{S_2}, respectively. 
The first term of Eq. \ref{S_2} is given by 
\begin{eqnarray}
&&\sum^{N}_{s=0} \theta(s+L)^2=  \nonumber \\
&&\begin{cases}
\frac{1}{Z(\beta)^2}(\zeta(2\beta,a+L)-\zeta(2\beta,a+L+N+1))  & \text{($\beta \neq 1/2$)}\\
\frac{1}{Z(\beta)^2}(\psi^{(0)}(a+L+N+1)-\psi^{(0)}(a+L)) & \text{($\beta = 1/2$)}
\end{cases} \nonumber \\
\end{eqnarray}
where $\zeta(\alpha,x)$ is the Hurwitz zeta function $\zeta(s,q)=\sum^{\infty}_{n=0}1/(q+n)^s$, and $\psi^{(0)}(x)$ is the digamma function, $\psi^{(0)}(x)=d/dx \log(\Gamma(x))$. 
The second term of Eq. \ref{S_2} is given by 
\begin{eqnarray}
&&\sum^{N}_{s=0} \theta(s)^2  \nonumber \\
&&=\begin{cases}
\frac{1}{Z(\beta)^2}(\zeta(2\beta,a)-\zeta(2\beta,a+N+1)) & \text{($\beta \neq 1/2$)}\\
\frac{1}{Z(\beta)^2}(\psi^{(0)}(a+N+1)-\psi^{(0)}(a)) & \text{($\beta = 1/2$)} 
\end{cases} \nonumber \\
\end{eqnarray}
For $N>>1$, using the general formula $\zeta(x,\alpha)$ for $x>>1$
\begin{equation}
\zeta(x,\alpha) \approx \frac{1}{\alpha-1}x^{-\alpha+1}+ \frac{1}{2} x^{-\alpha}+\cdots
\end{equation}
and
\begin{equation}
\psi^{(0)}(x) \approx \log(x)-\frac{1}{2x} + \cdots.
\end{equation}
we can obtain the approximation 
\begin{eqnarray}
&&\sum^{N}_{s=0} \theta(s+L)^2  \approx  \nonumber \\
&&
\begin{cases}
\frac{1}{Z(\beta)^2}(\zeta(2\beta,a+L)+\frac{(a+L+N+1)^{1-2\beta}}{(1-2\beta)})  & \text{($\beta \neq 1/2$)}\\
\frac{1}{Z(\beta)^2}(\log(a+L+N+1)-\psi^{(0)}(a+L)) & \text{($\beta = 1/2$)} 
\end{cases} \nonumber \\  \label{u3_1}
\end{eqnarray}
and
\begin{eqnarray}
\sum^{N}_{s=0} \theta(s)^2 
 \approx \begin{cases}
\frac{1}{Z(\beta)^2}(\zeta(2\beta,a)+\frac{(a+N+1)^{1-2\beta}}{(1-2\beta)}) & \text{($\beta \neq 1/2$)}\\
\frac{1}{Z(\beta)^2}(\log(a+N+1)-\psi^{(0)}(a)) & \text{($\beta = 1/2$)}
\end{cases} \nonumber \\ \label{u3_2} 
\end{eqnarray}
. \par
Lastly, we calculate the third term of Eq. \ref{S_2}.
Using the Euler-Maclaurin formula in  Eq. \ref{euler} and $(x+N)^{-\beta} \to 0$，$(x+N)^{-\beta-1} \to 0$ for $N>>1$, we can obtain
\begin{eqnarray}
&&\sum^{N}_{s=0} \theta(s+L) \theta(s) \approx \nonumber \\
&& \frac{1}{Z(\beta)^2}\{a^{-\beta} (a+L)^{-\beta} \nonumber \\
&& \frac{1}{Z(\beta)^2}\{\frac{1}{2} (a+1)^{-\beta} (a+1+L)^{-\beta} \nonumber \\
&&-\frac{(-\beta)}{12}((a+1)^{-\beta}(a+1+L)^{-\beta-1}+(a+1)^{-\beta-1}(a+1+L)^{-\beta}) \nonumber \\ 
&&+\int^{N}_{1}(x+a)^{-\beta} (x+a+L)^{-\beta}dx \}. \label{S2_0}
\end{eqnarray}
Here, we separate the term of s=0 from the summation to improve the accuracy.  \par
We calculate the integration term of Eq. \ref{S2_0}, 
\begin{eqnarray}
&&\int^{N}_{1}(x+a)^{-\beta} (x+a+L)^{-\beta}dx \\
&&=L^{-2\beta+1} \int^{\frac{N+a}{L}}_{\frac{a+1}{L}} x^{-\beta} (x+1)^{-\beta}dx. 
\end{eqnarray}
Executing the integration, 
\begin{eqnarray}
&&G(x,\beta)  \equiv \int x^{-\beta} (x+1)^{-\beta}dx  \nonumber \\
&&\begin{cases}
\text{($\beta \geq 0$, $\beta$ is a non-integer )} \\
\frac{x^{1-\beta}{}_{2}F_{1}(1-\beta,\beta,2-\beta,-x)}{1-\beta} \\
\text{$(\beta=1,2,3,4)$} \\ 
p^{(\beta)}_1\log(\frac{x+1}{x}) \\
+\sum^{\beta}_{k=2}\frac{p^{(\beta)}_k}{-k+1}((x+1)^{-k+1}+(-1)^kx^{-k+1}) \\
\end{cases}. \nonumber \\ \label{s2_int}
\end{eqnarray}
Here, we neglect an integral constant and  ${}_{2}F_{1}(a,b,c;x)$ is the Gaussian hypergeometric function \cite{hypergeom}, 
\begin{equation}
{}_{2}F_{1}(a,b,c;x)=\sum^{\infty}_{n=1}\frac{(a)_n(b)_n}{(c)_n}\frac{x^n}{n!}, \label{hyper_geom}
\end{equation} 
where $(A)_n=\Gamma(A+n)/\Gamma(A)$.
In additon, 
\begin{equation}
p^{(n)}_i=\frac{(-1)^n \Gamma(2n-i)}{\Gamma(n-i+1)\Gamma(n)}. \label{pp}
\end{equation}
is satisfied with a partial fraction decomposition
\begin{equation}
\frac{1}{(x+1)^n \cdot x^n}=\sum^{n}_{k=1}\frac{p^{(n)}_k}{(x+1)^k}+\frac{(-1)^k p^{(n)}_k}{x^k}.
\end{equation}
\par
For $x>>1$, using the asymptotic formula of a Gaussian hypergeometric function \cite{hypergeom}, when $b-a$ and $c$ are non-integers, 
\begin{eqnarray}
{}_2F_1(a,b,c;x) \approx \frac{\Gamma(b-a) \Gamma(c) (-x)^{-a}}{\Gamma(b) \Gamma(c-a)}+\frac{\Gamma(a-b) \Gamma(c) (-x)^{-b}}{\Gamma(a) \Gamma(c-b)} \nonumber \\
\end{eqnarray}
or when $b=a$ and $c-a$ are positive integers,
\begin{eqnarray}
&&{}_2F_1(a,b,c;x) \approx  \nonumber \\
&& \Gamma(c) (\log(-x)-\psi^{(0)}(c-a)-\psi^{(0)}(a)-2 \gamma)(-x)^{a}, \nonumber \\
\end{eqnarray}
for $x>>1$, the hypergeometric function in Eq. \ref{s2_int} is simplified 
\begin{eqnarray}
&&{}_2F_1(1-\beta,\beta,2-\beta,-x)= \nonumber \\ 
&&\begin{cases}
\frac{1-\beta}{1-2\beta} x^{-\beta}+\frac{\Gamma(2\beta-1)\Gamma(2-\beta)}{\Gamma(\beta)} x^{-(1-\beta)} & (0 < \beta < 1/2) \\ 
\frac{1}{2} (\log(x)+\log(4)) x^{-\beta} & (\beta=1/2) \\
\frac{\Gamma(2\beta-1)\Gamma(2-\beta)}{\Gamma(\beta)}x^{-(1-\beta)} & (\beta > 1/2)
\end{cases}. \nonumber \\
\end{eqnarray}
Substituting the results of these integrations into Eq. \ref{S2_0}, for $N>>1$, 
we can obtain
\begin{eqnarray}
&&\sum^{N}_{s=0} \theta(s+L) \theta(s) \\
&\approx&  \frac{1}{Z(\beta)^2}\{ \nonumber \\
&& a^{-\beta} (a+L)^{-\beta} \nonumber \\
&&\frac{1}{2} (a+1)^{-\beta} (a+1+L)^{-\beta} \nonumber \\
&&+\frac{\beta}{12}((a+1)^{-\beta}(a+1+L)^{-\beta-1} \nonumber \\
&&+(a+1)^{-\beta-1}(a+1+L)^{-\beta}) \nonumber \\ 
&&-L^{1-2\beta}G((a+1)/L,\beta)+G_2(N,L,\beta) \} \nonumber \\ \label{u3_3}
\end{eqnarray}
where 
\begin{eqnarray}
&&G_2(N,L,\beta) \equiv \nonumber \\ 
&&\begin{cases}
\text{($\beta \neq 1/2, \beta>0$ $\beta$ is non-integer)} \\
\frac{N^{1-2\beta}}{1-2\beta}+L^{1-2\beta}\frac{\Gamma(2\beta-1)\Gamma(2-\beta)}{(1-\beta)\Gamma(\beta)}   \\
\text{ $(\beta=1/2)$ }\\
\log(N+a)-\log(L)+\log(4)  \\
  \text{($\beta=1,2,3,\cdots$)} \\
0 \\
\end{cases}. \nonumber \\ 
\end{eqnarray}
\par
From the results of Eq. \ref{S_2}, Eq. \ref{u3_1}, Eq. \ref{u3_2}, and Eq. \ref{u3_3}, $S_2$ is calculated as 
\begin{eqnarray}
&&Z(\beta)^2 S_2 \approx   \nonumber \\
&&- 2a^{-\beta}(a+L)^{-\beta} \nonumber \\
&&- (a+1)^{-\beta} (a+1+L)^{-\beta} \nonumber \\
&&-\frac{\beta}{6}((a+1)^{-\beta}(a+1+L)^{-\beta-1} \nonumber \\
&&+(a+1)^{-\beta-1}(a+1+L)^{-\beta}) \nonumber \\ 
&&+2L^{1-2\beta} G((a+1)/L,\beta)+ \nonumber \\
&&\begin{cases}
 \text{($\beta > 0$, $\beta \neq 1/2$, $\beta$ is non-integer )} \\
-2L^{1-2\beta} \frac{\Gamma(2\beta-1)\Gamma(2-\beta)}{(1-\beta)\Gamma(\beta)}+\zeta(2\beta,a)+\zeta(2\beta,a+L)  \\
\text{($\beta=1/2$)} \\
2\log(L)-2\log(4)-\psi^{(0)}(a)-\psi^{(0)}(a+L) \\
\text{($\beta=1,2,3,4,\cdots$)} \\
\zeta(2\beta,a)+\zeta(2\beta,a+L).
\end{cases} \nonumber \\
 \label{s2_ans}
\end{eqnarray} 
 \par
Conseqently, substituting Eq. \ref{ss} into Eq. \ref{s1_ans} and Eq. \ref{s2_ans}, we can obtain the MSD, 
\begin{eqnarray}
&&<(r(t+L)-r(t))^2>  \approx \check{\eta}^2/Z(\beta)^2  [ \nonumber \\
&&- 2a^{-\beta}(a+L)^{-\beta} \nonumber \\
&&- (a+1)^{-\beta} (a+1+L)^{-\beta} \nonumber \\
&&-\frac{\beta}{6}((a+1)^{-\beta}(a+1+L)^{-\beta-1} \nonumber \\
&&+(a+1)^{-\beta-1}(a+1+L)^{-\beta}) \nonumber \\ 
&&+2L^{1-2\beta} G((a+1)/L,\beta)+ \nonumber \\
&&\begin{cases}
 \text{($\beta > 0$, $\beta \neq 1/2$, $\beta$ is non-integer)} \\
-2L^{1-2\beta} \frac{\Gamma(2\beta-1)\Gamma(2-\beta)}{(1-\beta)\Gamma(\beta)}+2\zeta(2\beta,a) \\
\text{$(\beta=1/2)$ }
 \\2\log(L)-2\log(4)-2\psi^{(0)}(a) \nonumber \\
 \text{($\beta$ is integer)} \\
 2\zeta(2\beta,a)  \\
\end{cases} \nonumber \\
&&]. \label{diff_ans}
\end{eqnarray} 
For $L>>1$, we can calculate the asymptotic form, 
\begin{eqnarray}
&&<(r(t+L)-r(t))^2> \approx \check{\eta}^2/Z(\beta)^2 \cdot \\
&&\begin{cases}
L & (\beta=0) \\
-2L^{1-2\beta} \frac{\Gamma(2\beta-1)\Gamma(2-\beta)}{(1-\beta)\Gamma(\beta)}+2 \zeta(2\beta,a) & (0<\beta<0.5) \\
2\log(L)-2\log(4)-2\phi^{(0)}(a) &(\beta=0.5)\\
2\zeta(2\beta,a) &(\beta>0.5) \\
\end{cases}, \nonumber  
\end{eqnarray}
which we use for $x<<1$ ${}_{2}F_{1}(1-\beta,\beta,2-\beta,-x) \approx 1$ and 
$x^{0.5}{}_{2}F_{1}(1-\beta,\beta,2-\beta,-x) =\log(\sqrt{x}+\sqrt{1+x}) \to 0$ $(x \to 0, \beta=0.5)$. 
 \par
Therefore, for $L>>1$, the dominant term is 
\begin{eqnarray}
<(r(t+L)-r(t))^2> \propto 
\begin{cases}
L^{1-2\beta}  & (0 \leq \beta <0.5) \\
\log(L) &(\beta=0.5)\\
O(1) &(\beta > 0.5) \\
\end{cases}.
\end{eqnarray}
\renewcommand{\theequation}{G\arabic{equation}}
\renewcommand{\thefigure}{G-\arabic{figure}}
\renewcommand{\thetable}{S-G\arabic{table}}
\setcounter{figure}{0}
\setcounter{equation}{0}
\section{Probability density function of $f_j(t+L)-f_j(t)$}
\label{app_pdf}
We calculate the probability density function (PDF) of $v_j(t;L)=f_j(t+L)-f_j(t)$ $(t=1,2,\cdots,T)$. 
$v_j$ is decomposed as 
\begin{eqnarray}
&&v_j(t)=f_j(t+L)-f_j(t)=\frac{g_j(t+L)}{m(t+L)}-\frac{g_j(t)}{m(t)}.  \label{v_def}
\end{eqnarray}
$g_j(t)$ is also decomposed as 
\begin{eqnarray}
g_j(t)=\check{c}_j r_j(t)m(t) + w(t)
\end{eqnarray}
and $g_j(t+L)$ is 
\begin{eqnarray}
g_j(t+L)=\check{c}_j (\delta^{(L)}r_j(t)+r_j(t)) m(t+L)+w(t+L), 
\end{eqnarray}
where  
\begin{eqnarray}
\delta^{(L)}r_j(t) \equiv r_j(t+L)-r_j(t). 
\end{eqnarray}
From the definition in Eq. \ref{rd_model}, 
$g(t)$
 and 
$g(t+L)$ obey a mixture of a Poisson distribution. 
Thus, using the PDF of $\delta^{(L)}r_j$, $P_{\delta r}(r(t);L,\check{\eta})$, 
 $w(t)$ and $P_w(w(t),\check{w})$, the PDF of $g$ at time $t$ conditioned by $r(t)$ is written as
\begin{eqnarray}
&&P_g(g;t,\check{w},\check{\eta}|r(t))= \nonumber  \\
&&\int P_{w}(w;\check{w}) P^{(dist)}_{poi}(g(t);\check{c}_j r_j(t)w m(t)) dw.
\end{eqnarray}
The PDF of $g$ at time $t+L$ conditioned by $r(t)$ is written as
\begin{eqnarray}
&&P_g(g;t+L,\check{w},\check{\eta}|r(t))= \nonumber \\ 
&&\int P_{w}(w;\check{w}) P_{\delta r}(\delta^{(L)}r;L,\check{\eta}) \nonumber \\ 
&&\times  P^{(dist)}_{poi}(g(t);\check{c}_j w(r_j(t)+\delta^{(L)}r)) dw d \delta^{(L)} r, 
\end{eqnarray}
where $P^{(dist)}_{poi}(x;A)$ is a PDF of a Poisson distribution with mean $A$. \par
Here, the PDF of $\delta^{(L)}r_j$ is calculated by 
\begin{eqnarray}
P_{\delta r}(r;L,\check{\eta})=\frac{1}{2\pi} \int^{-\infty}_{\infty} e^{-itx} \phi_{\delta r}(x)dx, 
\end{eqnarray}
where $\phi_{\delta r}(x)$ is the characteristic function of $\delta^{(L)}r_j$ given by
\begin{eqnarray}
\phi_{\delta r}(t)=\prod^{\infty}_{s=0}\phi_{\eta}((\theta(s+L)-\theta(s))t) \times \prod_{s=-1}^{-L}\phi_{\eta}(\theta(s+L)t), \nonumber \\ \label{phi_dr}
\end{eqnarray}
and $\phi_{\eta}(x)$ by the characteristic function of $\eta$.
Here, we calculate this characteristic function of $\phi_{\delta r}(x)$ by using the fact that $\delta^{L}r_j(t)$ is written as a linear combination of independent random variables $\{\eta(t)\}$, 
\begin{eqnarray}
&&\delta^{L}r_j=\sum^{\infty}_{s=0}(\theta(s+L)-\theta(s))\eta(t-s) \nonumber \\
&&+\sum_{s=-1}^{-L}\theta(s+L)\eta(s).
\end{eqnarray}
Note that, under the condition $\beta=0.5$, the first term of Eq. \ref{phi_dr} can be approximated by a finite sum. 
\par
In a comparison of the data, we use the following characteristic function $\phi_{\eta}(\nu)$,  
\begin{eqnarray}
&&\phi_{\eta}(t)=\frac{K_{\nu/2}(\sqrt{\nu}| \frac{ \check{\eta} t}{(\nu/(\nu-2))^{1/2})}|)(\sqrt{\nu}|\frac{ \check{\eta} t}{(\nu/(\nu-2))^{1/2}})|)^{\nu/2}}{\Gamma(\nu/2)2^{\nu/2-1}} \nonumber \\ 
&&\quad (\nu>0), 
\end{eqnarray}
This function is a characteristic function with a scaled t-distribution whose degree of freedom is $\nu$ and standard deviation is $\check{\eta}$. 
To obtain this fuction, we use the characteristic function with a t-distribution whose degree of freedom is $\nu$ \cite{hurst1995characteristic}, 
\begin{equation}
\phi_{t}(t)=\frac{K_{\nu/2}(\sqrt{\nu}|t|)(\sqrt{\nu}|t|)^{\nu/2}}{\Gamma(\nu/2)2^{\nu/2-1}} \quad (\nu>0).
\end{equation}
as well as the standard deviation of the t-distribution, $\nu/(\nu-2)$ $(\nu>2)$, 
where $K_{\nu/2}(x)$ is the  modified Bessel function of the second kind 
(We set $\nu=2.64$ in the data analysis, and $\check{\eta}$ is determined based on a word from the data.). 
\par
In addition, for a comparison of the data, we also set the PDF of $w$ as the t-distribution whose degree of freedom is $\nu$ and standard deviation is $\check{w}$
\begin{eqnarray}
P_{w}(x)=\frac{\nu/(\nu-2)}{\check{w}} \frac{\Gamma(\frac{\nu+1}{2})}{\sqrt{\nu \pi}\Gamma(\frac{\nu}{2})}(1+\frac{(\frac{\check{w} x}{(\nu/(\nu-2))^{1/2}})^2}{\nu})^{-\frac{\nu+1}{2}}. \nonumber \\
\end{eqnarray}
(We set $\nu=2.64$ in the data analysis, and $\check{w}$ is determined based on a word from the data.) 
\par
Hence, combining these functions and Eq. \ref{v_def}, we can obtain the PDF of $v_j$, 
\begin{eqnarray}
&&P_v(v;m(t),m(t+L),L,\check{w},\check{\eta}|r(t))= \nonumber \\
&&\int^{\infty}_{0}\frac{1}{m(t+L)} P_{g}((g+v)m(t+L);t+L,\check{w},\check{\eta}|r(t)) \nonumber \\
&&\times\frac{1}{m(t)} P_{g}(g m(t);t,\check{w},\check{\eta}|r(t))) dg.  \label{pdf_ans}
\end{eqnarray}
%
%
%
%
\renewcommand{\theequation}{S.H\arabic{equation}}
\renewcommand{\thefigure}{H-\arabic{figure}}
\renewcommand{\thetable}{S-H\arabic{table}}
\setcounter{figure}{0}
\setcounter{equation}{0}
\section{Estimation of scaled total number of blogs $m(t)$ from the data}
\label{app_m}
  We estimate the scaled total number of blogs $m(t)$ by using the moving median as follows: 
\textcolor{black}{
\begin{enumerate}
\item[Step 1.] We create a set $S$ consisting of indexes of words such that $c_j$ takes a value larger than the threshold $\check{c}_j(t) \geq 100$.
\item[Step 2.] We estimate $m(t)$ as the median of $\{F_j(t)/c_j:j \in S \}$ with respect to $j$.
\item[Step 3.] For $t=1,2,\cdots,T$, we calculate $m(t)$ using step 2. 
\end{enumerate}
} \par
Here, we use only words with $\check{c}_j(t) \geq 100$ in step 1 because we neglect the discreteness. In step 2, we apply the median because of its robustness to outliers.

\clearpage
\clearpage
\end{document}